\documentclass[
12pt, 
english, 
onehalfspacing, 
nolistspacing, 
liststotoc, 
headsepline, 
]{MastersDoctoralThesis} 
\usepackage{amsmath}
\usepackage{amssymb}
	\usepackage{amsfonts}
\usepackage{amsbsy}
\usepackage{mathrsfs}
\usepackage{bm}
\usepackage{enumitem}
\newcounter{numln}
\usepackage{multirow}
\usepackage{adjustbox}
\usepackage[page,toc,titletoc,title]{appendix}
\newlist{numitemise}{itemize}{2}
\setlist[numitemise]{wide}%
\setlist[numitemise, 1]{labelindent=0pt,labelwidth=2em, label=\stepcounter{numln}\makebox[2em]{\thenumln.\hfill}, leftmargin=\dimexpr\labelwidth+\labelsep\relax}%
\setlist[numitemise, 2]{labelindent=\dimexpr -2em-\labelsep\relax, labelwidth=\dimexpr 2em+\labelsep\relax, label=\stepcounter{numln}\makebox[\dimexpr\labelwidth + \labelsep\relax]{\thenumln.\hfill\textbullet}, leftmargin=\dimexpr\leftmargin+2\labelsep\relax}%

\usepackage{mathpazo} 

\RequirePackage[numbers,sort&compress]{natbib}

\usepackage{setspace}
\usepackage{graphicx}
\usepackage{subfig}
\usepackage{wasysym}
\usepackage{comment}
\usepackage{textgreek}
\usepackage{mathtools}

\usepackage{wrapfig}
\usepackage{makecell}

\thesistitle{On the effect of scalar fields on Hawking radiation and quasinormal modes of black holes} 

\supervisor{Prof. Narayan Banerjee} 

\examiner{} 

\degree{Doctor of Philosophy} 

\author{Avijit Chowdhury} 
\addresses{} 

\subject{Physical Sciences} 
\keywords{} 
\university{\href{https://www.iiserkol.ac.in/}{IISER Kolkata}} 
\department{Department of Physical Sciences} 
\def\DateSub{May 21, 2021}

\AtBeginDocument{
\hypersetup{pdftitle=\ttitle} 
\hypersetup{pdfauthor=\authorname} 
\hypersetup{pdfkeywords=\keywordnames} 
}

\begin{document}

\frontmatter 

\pagestyle{plain} 


\begin{titlepage}
\begin{center}

\textsc{\Large Doctoral Thesis}\\[0.5cm] 
\HRule \\[0.4cm] 
{\huge \bfseries \ttitle\par}\vspace{0.4cm} 
\HRule \\[1.5cm] 

\begin{center}
	\large{By\\
    Avijit Chowdhury \\
	Roll No.: 13IP001\\
    Department of Physical Sciences\\\vspace{0.0cm}
    Indian Institute of Science Education and Research Kolkata\vspace{0.0cm}}
\end{center}
\begin{center} 
	\large{\emph{Supervisor:} Prof. Narayan Banerjee}\\
\end{center}

\vspace{1.5cm}

\centering
	\includegraphics[width=0.3\textwidth]{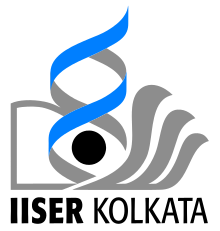}
	
\vspace{1.2cm}

\textit{ A thesis submitted in fulfilment of the requirements for 
the degree of\\ \degreename \ 
in the \deptname \ at the\\ \ Indian Institute of 
Science Education and Research Kolkata }

\vspace{0.7cm}

{\large \DateSub}\\[4cm] 

\vfill
\end{center}
\end{titlepage}

\sloppy
\begin{declaration}
\begin{flushright}
Date: {\DateSub} \\
\end{flushright}
\noindent
\par I, \textbf{Avijit Chowdhury} Registration No. \textbf{13IP001} dated 
25/07/2013, a student of Department of Physical Sciences of the 
Integrated PhD Programme of the Indian Institute of Science Education and Research 
Kolkata (IISER Kolkata), hereby declare that this thesis is my own work and, to 
the best of my knowledge, it neither contains materials previously published or 
written by any other person, nor it has been submitted for any degree/diploma or 
any other academic award anywhere before.
\par 
I also declare that all copyrighted material incorporated into this thesis is in 
compliance with the Indian Copyright (Amendment) Act, 2012 and that I have 
received written permission from the copyright owners for my use of their work.
\par 
I hereby grant permission to IISER Kolkata to store the thesis in a database 
which can be accessed by others.

\vspace{2.cm}
\begin{flushleft}
	-----------------------------------\\
	{\bf Avijit Chowdhury}\\
	Department of Physical Sciences\\
	Indian Institute of Science Education and Research Kolkata \\
	Mohanpur 741246, West Bengal, India
\end{flushleft}

\vspace{3.0cm}

\end{declaration}

\cleardoublepage

\thispagestyle{plain}
\null\vfil
{\noindent\huge\bfseries Certificate from the Supervisor \par\vspace{10pt}}

\begin{flushright}
	Date: {\DateSub} \\
\end{flushright}
\noindent
\par
{\doublespacing This is to certify that the thesis entitled \textbf{``On the effect of scalar fields on Hawking radiation and quasinormal modes of black holes''} submitted by \textbf{\authorname} Registration No. \textbf{13IP001} dated 25/07/2013, a 
student of Department of Physical Sciences of the Integrated PhD 
Programme of IISER Kolkata, is based upon his own research work under my 
supervision. This is also to certify that neither the thesis nor any part of it 
has been submitted for any degree/diploma or any other academic award anywhere 
before. In my opinion, the thesis fulfils the requirement for the award of the 
degree of Doctor of Philosophy.} \\

\vspace{2.5cm}

\begin{flushleft}
-----------------------------------\\
	{\bf Prof. Narayan Banerjee} \\
	Professor\\
	Department of Physical Sciences \\
	Indian Institute of Science Education and Research Kolkata\\
    Mohanpur 741246, West Bengal, India
\end{flushleft}

\cleardoublepage

\thispagestyle{plain}
\null\vfil
{\noindent\huge\bfseries Abstract \par\vspace{12pt}}
The present thesis attempts to study the effect of scalar fields on Hawking radiation and quasinormal modes of black holes. We selected a static, spherically symmetric electrically charged black hole with an additional scalar `hair' for our analysis. The scalar `hair' is sourced by a scalar field, conformally coupled to the Einstein-Hilbert action. The scalar field can survive even in the absence of the black hole's electric charge and is characterized as a `primary hair'. This scalar field changes the gravitational constant, and hence modifies the ADM (Arnowitt-Deser-Missner) mass of the black hole. The scalar field's strength is determined by a scalar `charge' that manifests itself as an additive correction to the square of the electric charge in the standard Reissner-Nordstr\"{o}m metric. This seemingly simple modification leads to nontrivial physical implications.

We start with an analysis of the quasinormal modes of the black hole mentioned above against perturbation by massless and massive, uncharged and charged scalar and Dirac test fields. The quasinormal modes encode a black hole's response to perturbations (either of the metric or that induced by a test field). The quasinormal modes are characterized by damped oscillations with complex frequencies called the Quasinormal frequencies. The real part of the quasinormal frequency gives the actual frequency of the wave motion, whereas the imaginary part corresponds to the damping rate. We observed that the presence of the scalar hair affects both the real part and imaginary part of the quasinormal frequency.

If one relaxes the quasinormal mode boundary conditions of purely ingoing waves at the event horizon and purely outgoing waves at spatial infinity and instead considers a charged bosonic wave incident on the black hole from spatial infinity, then the reflected wave from the event horizon may be amplified. This superradiant amplification occurs at the cost of the electrical (or rotational in case of rotating black hole) energy of the black hole. The superradiantly amplified waves can be confined and made to interact repeatedly with the black hole by a hypothetical `mirror' surrounding the black hole. This repeated reflection and superradiance may lead to an exponentially increasing amplitude and hence instability. Interestingly, the mass of the incident bosonic wavefield can effectively act as a `mirror' to reflect the low-frequency modes and lead to superradiant instability. We studied the effect of the scalar hair on the superradiant stability of the black hole.

An isolated black hole is also capable of radiating via pair-production in the form of Hawking radiation. To an asymptotic observer, the Hawking emission spectrum `coarsely resembles' a black body spectrum with a temperature inversely proportional to the black hole's mass. Though detection of Hawking radiation from astrophysical black holes is a near impossibility due to the extremely low temperature of the black holes, the study of Hawking radiation is significant in its own right, particularly in the pursuit to arrive at a quantum theory of gravity. The Hawking radiation emitted at the event horizon gets modified as it propagates through the spacetime surrounding the black hole, which acts as a filter, allowing only a fraction of the emitted radiation to reach an asymptotic observer. This fraction is referred to as the Greybody factor and measures the deviation of the Hawking emission spectrum from perfect blackbody like Planckian distribution. Another critical aspect that distinguishes Hawking radiation from blackbody radiation is its `sparsity', which measures the number of particles emitted per unit time. Whereas the sparsity of a blackbody spectrum is extremely low, suggesting an enormous number of emitted particles per unit time, the sparsity of the Hawking radiation is exceptionally high. The latter half of the thesis is devoted to studying the effect of the black hole scalar hair on Hawking radiation of scalar particles, its sparsity, and greybody factor.

\cleardoublepage


\doublespacing

\thispagestyle{plain}
\null\vfil
{\noindent\huge\bfseries Acknowledgements \par\vspace{12pt}}

First and foremost, I  wish to express my deepest gratitude to my supervisor, Professor Narayan Banerjee, for his patient guidance, continuous support, and enthusiastic encouragement throughout my Ph.D. He has always guided me with his immense knowledge, experience, and expertise both academically and personally. It was a great pleasure to work under his supervision. I could not have wished for a better mentor.

I am grateful to Dr. Golam Mortuza Hossain, Dr. Rajesh Kumble Nayak, and Dr. Ananda Dasgupta for their insightful comments and suggestions. I would also like to extend my sincere gratitude to the office staff of the Department of Physical Sciences, IISER Kolkata, and Assistant Librarian, IISER-K Library, Dr. Siladitya Jana, for their help and cooperation.

I want to thank my seniors Nandan Roy, Ankan Mukherjee, Soumya Chakrabarti, Gopal Sardar, Santanu Tripathi, Nivedita Bhadra, Biswarup Ash, Swati Sen, Rafiqul Rahman, Subhajit Barman, Chiranjeeb Singha, Anushree Datta, Supriya Pan and Souvik Pramanik for their help and support. I am thankful to my friends and colleagues Anurag, Ankit, Sachin, Souraj, Sajal, Sayak, Soumik,  Purba, Tanima, and Shibendu for making my stay at IISER Kolkata a memorable experience. Special thanks to my dear friend and colleague Srijita Sinha for her immense help, unswerving support, motivation, and encouragement.

Finally, I want to thank my parents for their unconditional love, support, encouragement, and inspiration.

\cleardoublepage
\thispagestyle{plain}
\null\vfil
{\noindent\huge\bfseries List of Publications \par \vspace{12pt}}

\onehalfspacing

\paragraph*{Publications included in this thesis:}
  \begin{numitemise}
  
        \item {\textbf{Avijit Chowdhury} and Narayan Banerjee, “Quasinormal modes of a charged spherical black hole with scalar hair for scalar and Dirac perturbations”, Eur.\ Phys.\ J.\ C \textbf{78}, 594 (2018); arXiv:1807.09559 [gr-qc].\\(Chapter 2)}
        
        \item {\textbf{Avijit Chowdhury} and Narayan Banerjee, “Superradiant stability of mutated Reissner–Nordström black holes”, Gen.\ Rel.\ Grav.\  \textbf{51}, 99 (2019); arXiv:1906.09796 [gr-qc].\\(Chapter 3)}

     	\item {\textbf{Avijit Chowdhury}, “Hawking emission of charged particles from an electrically charged spherical black hole with scalar hair”, Eur.\ Phys.\ J.\ C \textbf{ 79}, 928 (2019);  arXiv:1911.00302 [gr-qc].\\(Chapter 4)}

        \item {\textbf{Avijit Chowdhury} and Narayan Banerjee, “Greybody factor and sparsity of Hawking radiation from a charged spherical black hole with scalar hair”, Phys.\ Lett.\ B \textbf{805}, (2020) 135417; arXiv:2002.03630 [gr-qc].\\(Chapter 5)}
  	
\end{numitemise}

\paragraph*{Publications not included in this thesis:}

\begin{numitemise}

    \item {\textbf{Avijit Chowdhury} and Narayan Banerjee, “Echoes from a Singularity”, Phys.\ Rev.\ D \textbf{102}, 124051 (2020); arXiv:2006.16522 [gr-qc] }

\end{numitemise}


\tableofcontents 

\listoffigures 

\listoftables 

\dedicatory{To my loving parents and teachers \ldots} 

\mainmatter 

\pagestyle{thesis} 


\def\bh{black hole\xspace}
\newcommand{\sech}{\mathrm{sech} \,} 
\makeatletter
\def\blfootnote{\gdef\@thefnmark{}\@footnotetext}
\makeatother

\chapter{Introduction} 
\label{chap1}


\section{General Relativity and black holes}\label{ch1:sec_bh}
The theory of General Relativity, as proposed by Albert Einstein in 1915~\cite{einstein_1914,einstein_SPAWB_1915_2,einstein_SPAWB_1915_1,einstein_SPAWB_1915_3}, is undoubtedly one of the most elegant physical theories ever created. General Relativity is by far the most widely accepted theory of gravity and its predictions have been verified in the weak field regime as well as in the strong field regime. The essence of General Relativity is summarised in Einstein's field equations, which brilliantly connects geometry with matter.

Einstein envisioned spacetime as a four-dimensional differentiable manifold with a metric that measures the infinitesimal separation between two neighbouring points, $x^\mu$ and $x^\mu+dx^\mu$ (say) as
\begin{equation}
ds^2=g_{\mu\nu}dx^\mu dx^\nu,
\end{equation}
where $g_{\mu\nu}$ is a symmetric rank two tensor, with negative definite determinant, known as the metric tensor.\\
To arrive at Einstein's field equations, one starts from the Einstein-Hilbert action,
\begin{equation}\label{ch1:eq_E-H}
S_H=\int  R \sqrt{-g} ~d^4 x,
\end{equation}
where $g$ is the determinant of the metric tensor and $R$ is the Ricci scalar, obtained by the contraction of the Ricci tensor $R_{\mu \nu}$,
\begin{equation}
R=g^{\mu \nu} R_{\mu \nu}~.
\end{equation}
Varying the action~\eqref{ch1:eq_E-H} with respect to the inverse metric $g^{\mu\nu}$ and demanding that this variation is zero in the first order,
\begin{equation}
\delta S_H=0,
\end{equation} one arrives at the vacuum Einstein's equation,
\begin{equation}
R_{\mu \nu}-\frac{1}{2}R g_{\mu \nu}=0.
\end{equation}
In the presence of matter, the total action contains an additional contribution due to the matter sector,
\begin{equation}\label{ch1:eq_totalS}
S=\frac{1}{2 \kappa_0}S_H + S_M ~,
\end{equation}
where $\kappa_0=8 \pi G$ (in units where the speed of light $c=1$), $G$ is the Newton's gravitational constant and 
\begin{equation}
S_M=\int  \mathcal{L}_M \sqrt{-g} ~d^4 x
\end{equation} 
is the action for the matter sector with Lagrangian density $\mathcal{L}_M$. Proceeding as before with the action~\eqref{ch1:eq_totalS}, one arrives at the Einstein equation for `non-vacuum' spacetime,
\begin{equation}
R_{\mu \nu}-\frac{1}{2}R g_{\mu \nu}=\kappa_0 T_{\mu\nu}~,
\end{equation}
where
\begin{equation}
T_{\mu\nu}=-\frac{2}{\sqrt{-g}}\left[ \frac{\partial \left( \sqrt{-g} \mathcal{L}_M \right)}
{\partial g^{\mu \nu}} - \left[ \frac{\partial \left( \sqrt{-g} \mathcal{L}_M \right)}
{\partial g^{\mu \nu}_{,\lambda}} \right]_{,\lambda} \right]
\end{equation}
is the is the energy-momentum tensor associated with $\mathcal{L}_M$. For detailed discussion on the properties of Einstein's equation, {we} refer to~\cite{MTW}. Henceforth, unless otherwise mentioned, we will use units in which $G=c=k_B=\hbar=1$.


A particularly simple yet immensely important application of Einstein's equations is to study the external gravitational field produced by a spherically symmetric distribution of matter. In General Relativity, according to the Birkhoff's theorem~\cite{birkhoff}, the Schwarzschild solution~\cite{schwarzschild_1916} is the only, static, spherically symmetric vacuum solution to Einstein's equation. The Schwarzschild solution is described by the line element,
\begin{equation}\label{ch1:eq_sch}
ds^2=-f(r)dt^2+f(r)^{-1}dr^2+r^2\left( d\theta^2+\sin^2\theta d\phi^2 \right)
\end{equation}
where $f(r)=(1-2M/r)$ and $M$ is the mass of the gravitating object. The Schwarzschild spacetime is asymptotically flat. The metric\eqref{ch1:eq_sch} has {apparently} two singularities at $r=0$ and $r=2M$. The radius $r=r_h=2M$ is referred to as the Schwarzschild radius. For an object of 1 solar-mass, the Schwarzschild radius is about 3 km, whereas the solar-radius $\sim$ 700000 km. It is important to note that the singularity at $r_h$ is just an artefact of the chosen coordinate system, since the curvature scalar for Schwarzschild metric,
\begin{equation}
I=R_{\mu \nu \alpha \beta}R^{\mu \nu \alpha \beta}= \frac{48M^2}{r^2},
\end{equation}
is finite at $r_h$. However, the singularity at $r = 0$ is a true singularity since the curvature diverges at $r=0$.
%
To show that the coordinate singularity at $r=r_h$ is merely due to a bad choice of coordinates, one can transform to a new set of coordinates namely, the Kruskal-Szekeres coordinates~\cite{MTW},
\begin{equation}
\begin{rcases}
&X=(r/r_h-1)^{1/2}e^{r/2r_h}\cosh(t/2r_h)\\
&T=(r/r_h-1)^{1/2}e^{r/2r_h}\sinh(t/2r_h)
\end{rcases}
\mbox{~for $r>r_h$}~,
\end{equation}
\begin{equation}
\begin{rcases}
&X=(1-r/r_h)^{1/2}e^{r/2r_h}\sinh(t/2r_h)\\
&T=(1-r/r_h)^{1/2}e^{r/2r_h}\cosh(t/2r_h)
\end{rcases}
\mbox{~for $r<r_h$}~.
\end{equation}
The Schwarzschild metric in the Kruskal-Szekeres coordinates takes the form,
\begin{equation}
ds^2=-\frac{4r_h^3}{r}e^{-r/r_h}\left(dT^2-dX^2\right)+r^2\left( d\theta^2+\sin^2\theta d\phi^2 \right),
\end{equation}
where $r$ is an implicit function of $X$ and $T$, given by,
\begin{equation}
\left(\frac{r}{r_h}-1\right)e^{r/r_h}=X^2-T^2~.
\end{equation}
\begin{figure}[!t]
\begin{center}
  \includegraphics[width=0.8\textwidth, keepaspectratio]{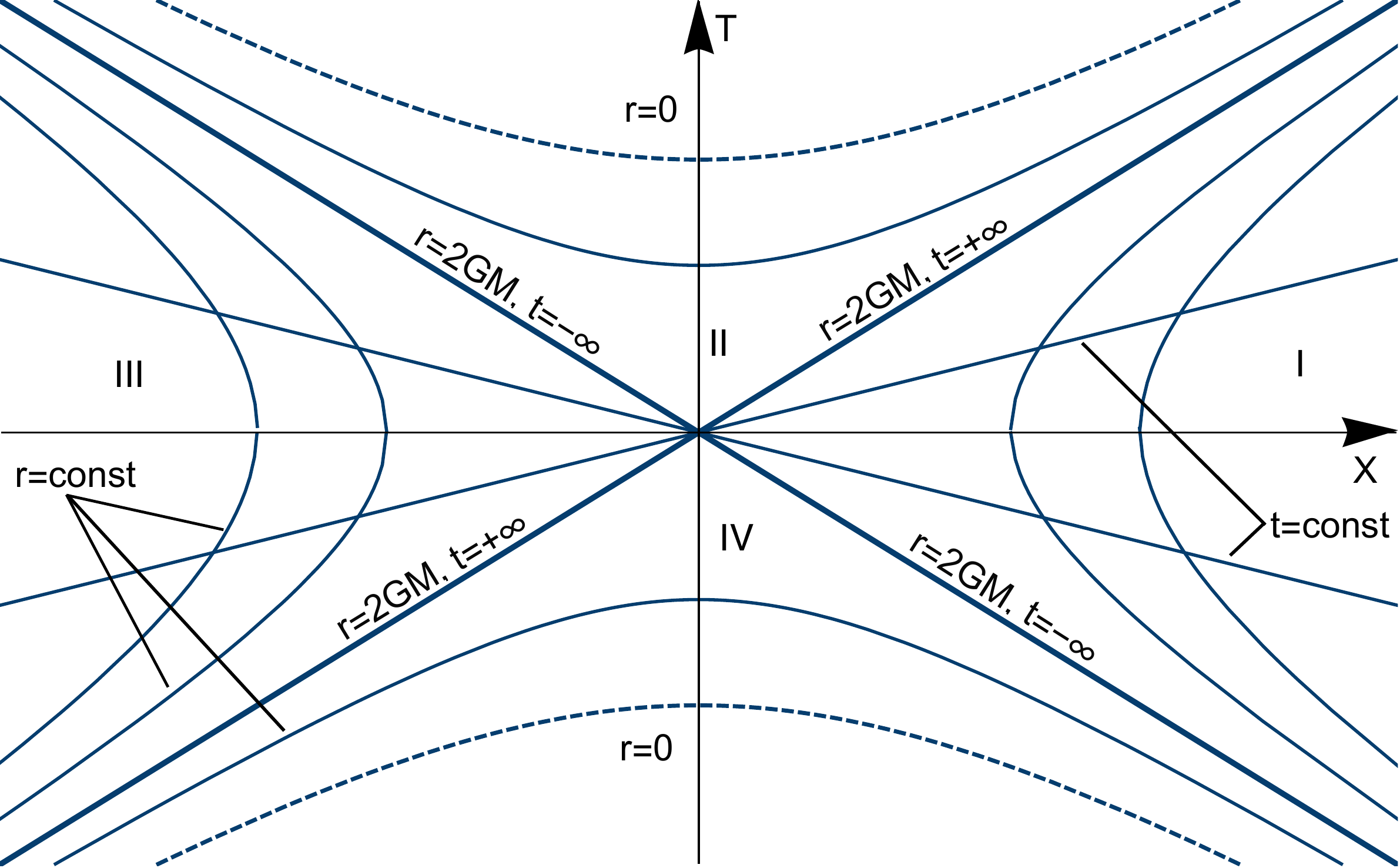}
\caption{Spacetime diagram of the Schwarzschild solution in the Kruskal-Szekeres coordinates.} 
\label{ch1:fig_kruskal}
\end{center}
\end{figure}
Figure~\ref{ch1:fig_kruskal} shows the Schwarzschild spacetime in the Kruskal-Szekeres coordinates. A radial null ray $\left( d\theta^2=d\phi^2=ds^2=0 \right)$ in this coordinates satisfies 
\begin{equation}
dT=\pm dX~. 
\end{equation}
Since the $\theta, \phi$ coordinates are suppressed, each point in the figure represents a two-sphere. The lines of constant $r$  satisfy
\begin{equation}
T^2-X^2=constant,
\end{equation}
and are represented by hyperbolae in the X-T plane. The hyperbola corresponding to $r=0$ represents the true singularity. The lines of constant $t$ are given by
\begin{equation}
\frac{T}{R}=\tanh\left(t/2r_h\right)~,
\end{equation}
which are straight lines of slope $\tanh\left(t/2r_h\right)$ passing through the origin. The line $X=T$ simultaneously corresponds to $r=r_h$ and $t=\infty$. Since the light cones are made up of 45$^\circ$ lines, all timelike curves (confined within the light cone) in the region $r<r_h$ (region II) end up at the singularity. The region $r<r_h$ (region II) cannot send any signal to the region $r>r_h$ (region I). The null hypersurface $r=r_h$, thus, acts as a one-way membrane, forming a causal boundary between the two regions of the spacetime (regions I and II). It is also interesting to note that while an in-falling particle will cross all the $t=constant$ lines in finite proper time, an asymptotic observer (with $t$ as proper time) will observe that it takes an infinite time for the falling object to reach the horizon.  The hypersurface at $r=r_h$ is referred to as the event horizon and the spacetime (region II) is referred to as a black hole. The Kruskal diagram also depicts two additional region of the spacetime -- region III and region IV. Region IV is the time reverse of region II with a singularity in the past from which anything can escape to us (region I). Region IV can never be reached from region I. The boundary of region IV is the past event horizon whereas that of region II is the future event horizon. This region (IV) of spacetime is referred to as a white hole. Region III is geometrically identical to region I, however, one cannot reach region III from region I or vice versa either forward or backward in time.
\section{Quasinormal modes}\label{ch1:sec_QNM}
Isolated black holes are idealised objects. Black holes in nature are surrounded by complex matter distribution, for example, supermassive black holes at the centre of galaxies, and intermediate-mass black holes surrounded by accretion disks. Black holes always interact with their surroundings, even in the absence of matter distribution, for example, black holes interact with the surrounding vacuum, creating particle pairs and evaporating due to Hawking radiation\cite{hawking_CMP_1975}. Thus, black holes in nature are always found in a perturbed state. The study of black hole perturbations is essential to the understanding of the emission of gravitational waves and the stability of {a} black hole spacetime. 

When perturbed, a black hole emits gravitational waves. The time evolution of these gravitational waves can be classified into three stages: (I) a brief period of initial outburst of radiation, (II) a relatively longer period of damped oscillations which are dominated by the Quasinormal Modes (QNMs), as proposed by Vishveshwara~\cite{vishveshwara_PRD_1970, vishveshwara_Nature_1970}, (III) at late time the damped oscillation give way to power-law or exponential tail~\cite{price_PRD_1972_1,price_PRD_1972_2}. 
\begin{figure}[!h]
\begin{center}
  \includegraphics[width=1\textwidth, keepaspectratio]{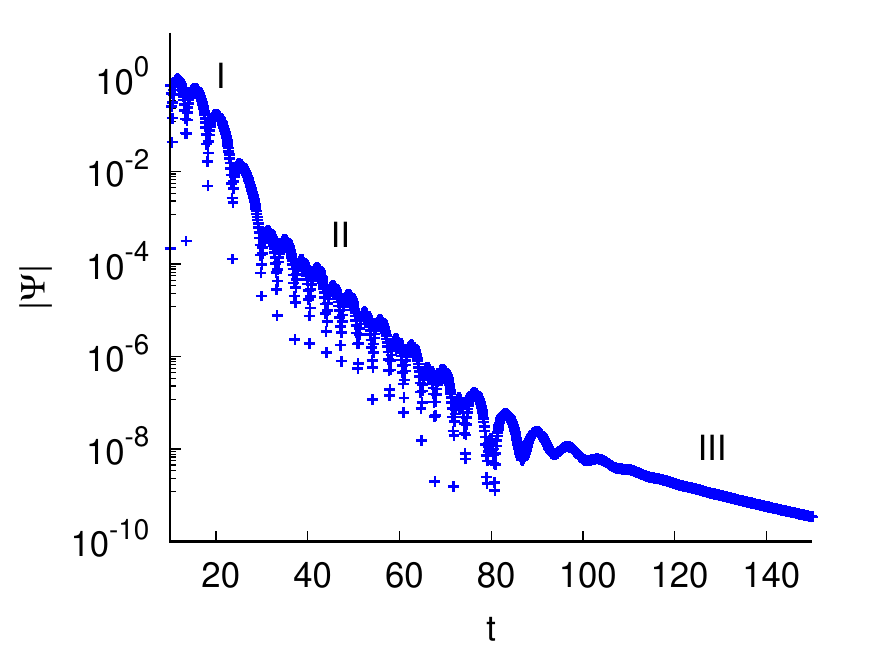}
\caption{An example of the time-domain profiles for $l=2$ mode of the axial gravitational perturbation of a Schwarzschild black hole at $r_*=10r_h$.} 
\label{ch1:fig_td_sch}
\end{center}
\end{figure}
Theoretically, a black hole can be perturbed either by adding small perturbation to the background metric~\cite{regge_PR_1957,zerilli_PRL_1970,zerilli_PRD_1970} or by adding fields to the black hole spacetime~\cite{,davis_PRL_1971,davis_PRD_1972}. In the linear approximation, when the field does not back-react on the spacetime, the latter reduces to the study of the propagation of the field in the black hole  background. In either case, one arrives at a second-order partial differential equation of the form
\begin{equation}\label{ch1:eq_evolution}
\frac{\partial^2 \Psi}{\partial r_*^2} - \frac{\partial^2 \Psi}{\partial t^2}-V\left(r\left(r_*\right)\right) \Psi=0,
\end{equation}
where $r_*$ is a spatial variable, ranging from $-\infty$ to $\infty$ and $t$ is the time. For spherically symmetric black hole solutions, $r_*$ is defined by $dr_*=\sqrt{-g_{11}/g_{00}}~dr$, mapping the event horizon to $-\infty$. $r_*$ is known as the tortoise coordinate and $r$ is an implicit function of $r_*$. $V(r)$ is the effective potential that $\Psi$ experiences. $V(r)$ usually vanishes as $r\rightarrow \{r_h ,\infty\}$. For example, let us consider the perturbation of a Schwarzschild black hole, induced by a massless test scalar field $\zeta$. The dynamics of the field is governed by the Klein-Gordon equation~\cite{landau_book},
\begin{equation}
\label{ch1:eq_KG_sch}
\nabla^\nu \nabla_\nu \zeta=0~,
\end{equation}
where $\nabla_\nu$ is the covariant derivative. Equation~(\ref{ch1:eq_KG_sch}) can be explicitly written as 
\begin{equation}
\frac{1}{\sqrt{-g}}\partial_\nu\left(g^{\mu \nu}\sqrt{-g}\partial_\mu \zeta\right)=0
\end{equation}
Using the Schwarzschild metric given in Eq.~(\ref{ch1:eq_sch}) and separating the angular dependence of $\zeta$ as 
\begin{equation}
\zeta(t,r,\theta,\phi) = Y^m_l(\theta,\phi)\Psi(t,r),
\end{equation}
where $ Y^m_l(\theta,\phi)$ are the spherical harmonics, one obtains an equation of the form (\ref{ch1:eq_evolution}). Choosing the time dependence of $\Psi$ as $e^{-i \omega t}$, 
\begin{equation}
\Psi(t,r)= \psi(r) e^{-i \omega t}~,
\end{equation}
Eq.\eqref{ch1:eq_evolution} reduces to
\begin{equation}\label{ch1:eq_radial_KG_sch}
\frac{d^2 \psi}{d r_*^2} +\left(\omega^2-V\left(r\left(r_*\right)\right)\right)\psi=0~.
\end{equation}
{Equation~\eqref{ch1:eq_radial_KG_sch} is of the form of the one-dimensional time-independent Schr\"{o}dinger equation~\cite{landau_QM_book} for a particle (of mass m=1/2 in the chosen units) with total energy $\omega^2$ { in a potential} $V$. Hence, the term $V(r)$ in Eq.~\eqref{ch1:eq_radial_KG_sch} is referred to as the effective potential.}

The tortoise coordinate for a Schwarzschild black hole is given by
\begin{equation}
r_*=r+2M \ln\left(\frac{r}{2M}-1\right)
\end{equation}
and the effective potential $V(r)$ is given by 
\begin{equation}
V(r)=\left(1-\frac{2M}{r}\right)\left(\frac{l(l+1)}{r^2}+\frac{2 M(1-s^2)}{r^3}\right)~,
\end{equation}
where $s$ is the spin of the perturbation field, $s=0,1,2$ for scalar, electromagnetic and gravitational fields~\cite{regge_PR_1957} respectively.
\begin{figure}[!h]
\begin{center}
  \includegraphics[width=1\textwidth, keepaspectratio]{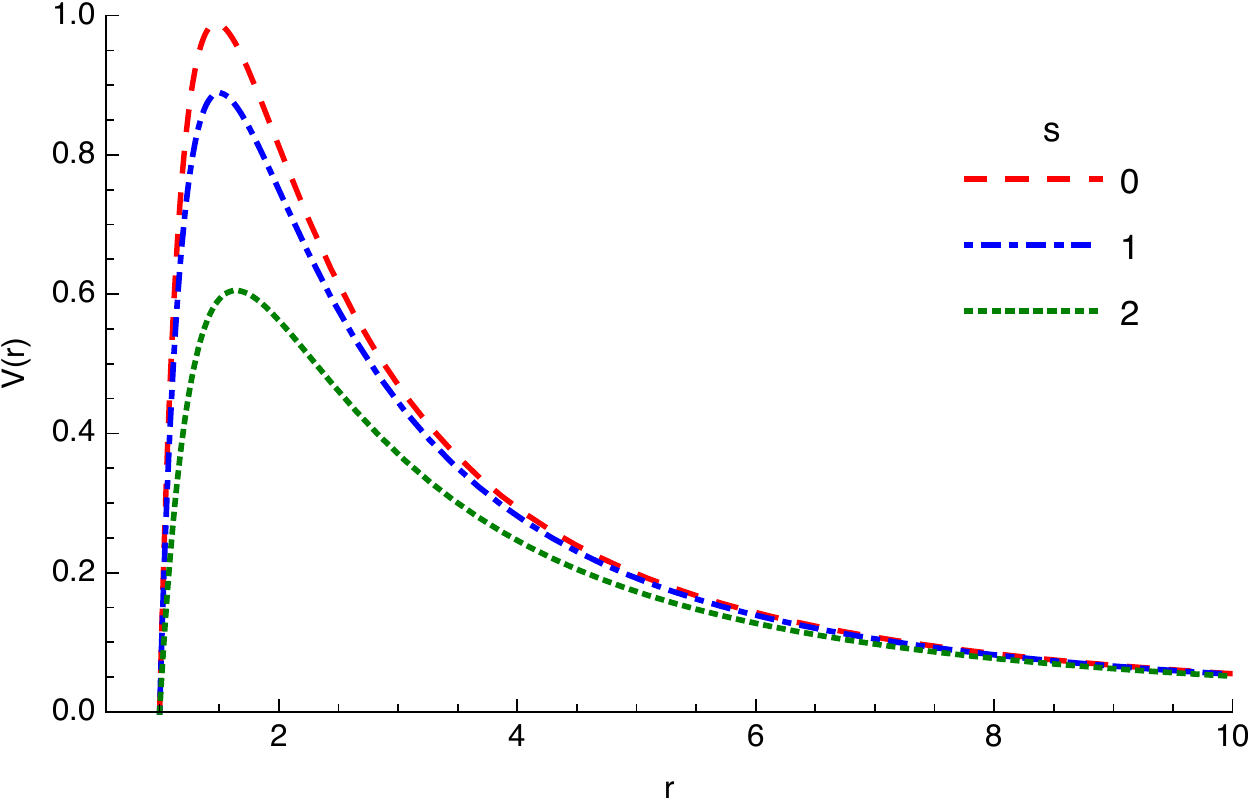}
\caption{Plot of the effective potentials for the $l=2$ mode of scalar ($s=0$), electromagnetic ($s=1$) and axial gravitational ($s=2$) perturbations of a Schwarzschild black hole with event horizon at $r_h=2M=1$.} 
\label{ch1:fig_pot_sch}
\end{center}
\end{figure}

The quasinormal modes correspond to the discrete set of solutions of Eq.~\eqref{ch1:eq_radial_KG_sch} subject to the condition of purely ingoing waves at the horizon and purely outgoing waves at $\infty$,
\begin{eqnarray}
&\psi(r_*\rightarrow -\infty)\sim e^{-i \omega r_*}\\
& \psi(r_*\rightarrow +\infty)\sim e^{+i \omega r_*}~.
\end{eqnarray}
The discrete set of complex frequencies $\omega$ corresponding to the quasinormal modes are called the quasinormal frequencies (QNFs). The real part of $\omega$ represents the actual oscillation frequency, whereas the imaginary part is responsible for the decay rate. A negative imaginary part of the quasinormal frequency thus implies the corresponding quasinormal mode decays exponentially in time. Physically, this suggests that a black hole, when perturbed, continually loses energy by emitting gravitational waves. A negative imaginary part of the quasinormal frequency also implies the boundedness of the perturbation, and hence the stability of the spacetime~\cite{vishveshwara_PRD_1970, vishveshwara_Nature_1970}. The exponential decay of the quasinormal modes also suggests that during the quasinormal ringing of the black hole, only the mode with the lowest imaginary part of the quasinormal frequency dominates the signal. This mode is referred to as the fundamental mode.

The completeness of the quasinormal modes is another subtle issue~\cite{pal_IJMPD_2015}. The quasinormal modes, in general, do not form a complete set and as such the gravitational wave signal cannot be represented by a sum of the quasinormal modes at all times~\cite{ching_RMP_1999,nollert_JMP_1998,horowitz_PRD_2000}. This is clearly manifested in the late time behaviour, when the signal is dominated by power law or exponential tail. However, in case of black holes in asymptotically anti-de Sitter (ADS) spacetime, the lack of outgoing waves at spatial infinity allows for the confinement of the perturbation close to the black hole. This in-turn removes the `late-time-tail' of the perturbation~\cite{horowitz_PRD_2000}, and thus the signal can always be represented by a sum of quasinormal modes.

The most important property of quasinormal frequencies is that they depend only on the parameters of the black hole such as the mass, electric charge, angular momentum (or the scalar charge as will be discussed in chapter~\ref{chap2}), and not on the way by which the black hole or the field around it was perturbed. Thus, accurate determination of the quasinormal frequencies is crucial to estimate the black hole parameters. Starting from the work of Chandrasekhar and Detweiler~\cite{chandrasekhar_PRSA_1975}, there are numerous numerical and semi-analytic methods to compute the quasinormal frequencies, for examples, Schutz and Will~\cite{schutz_APJL_1985} used the semi-analytic WKB technique at the first order to determine the quasinormal frequeencies of black holes. The WKB technique was extended to 3rd order by Iyer and Will~\cite{iyer_PRD_1987_1}. This was further extended to sixth order~\cite{konoplya_PRD_2003_1,konoplya_PRD_2003_2} and thirteenth order~\cite{matyjasek_PRD_2017} for better accuracy. However, the most accurate numerical determination of the quasinormal frequencies was provided by Leaver~\cite{leaver_PRSLA_1985}, using a continued fraction approach to Eq.\eqref{ch1:eq_radial_KG_sch}. Alternatively, one can study the time-evolution of the perturbation $\Psi$ in Eq.~\eqref{ch1:eq_evolution} numerically using suitable discretization scheme~\cite{gundlach_PRD_1994} with appropriate boundary condition to obtain the time profile of the perturbation such the one given in Fig.~\ref{ch1:fig_td_sch}. The quasinormal ringing phase (stage II in Fig.~\ref{ch1:fig_td_sch}) of the time-domain profile can then be fitted  with a superposition of damped exponentials with excitation factors to extract the quasinormal frequencies. For a complete and detailed discussion of black hole perturbation, quasinormal modes and a detailed account of the different techniques to compute the quasinormal frequencies, we refer to the excellent reviews~\cite{nollert_CQG_1999,kokkotas_LRR_1999,cardoso_thesis,berti_CQG_2009,konoplya_RMP_2011} (see also~\cite{glampedakis_PRD_2019, silva_PRD_2020}).

For a basic understanding of the estimation of black hole parameters from the quasinormal frequencies, let us consider a  particular quasinormal mode of a black hole. Now, the strain measured by a gravitational wave detector is given by~\cite{echeverria_PRD_1989,finn_PRD_1992},
\begin{equation}\label{ch1:eq_strain}
h(t)=\frac{1}{V^{1/3}} e^{-\pi f \frac{t-T}{Q}} \sin 2\pi f \left(t-T\right),
\end{equation}
where $f$ is the oscillation frequency, $\frac{1}{V^{1/3}}$ is the amplitude of the waveform depending on the distance from the source, size of the perturbation and relative orientation of the detector with the source. $T$ is the time at which the perturbation starts. $Q(=\pi f \tau)$ is the quality factor of the oscillation depending on the oscillation, frequency $f$ and the damping time $\tau$. $Q$ and $f$ depend on the real and imaginary parts of the quasinormal frequency. Thus, one starts with an analytical waveform depending on the black hole parameters that match with the data and knowing the real and imaginary parts of the quasinormal frequencies one can, in principle, evaluate the black hole parameters with the help of Eq.~\eqref{ch1:eq_strain}. For a detailed discussion on the gravitational wave and parameter estimation from gravitational wave observation, we refer to~\cite{maggiore_book1_2007, barack_CQG_2019}.
 
\section{Black hole superradiance}\label{ch1:sec_superrad}

As discussed in Sec.~\ref{ch1:sec_bh}, the event horizon, classically, acts as a one-way membrane, allowing matter and radiation to fall in (towards the singularity) but prevents them from going out. This one-way nature of the event horizon leads to the curious phenomenon of Black hole Superradiance.

Superradiance is a process of radiation enhancement in dissipative systems. The term `superradiance' was coined by Dicke~\cite{dicke_PR_1954} in 1954 to denote a class of phenomena dealing with the enhancement of radiation by the coherence of emitters. Later, in 1971, Zel'dovich observed that scattered monochromatic radiation from a rotating dissipative system is amplified when the frequency of the incident radiation $\omega$ satisfies the condition~\cite{zeldovich_JETPL_1971,zeldovich_JETP_1972},
\begin{equation}
\omega< m \Omega~,
\end{equation}
where $m$ is the azimuthal index and $\Omega$ is the angular velocity of rotation. This phenomenon is known as `rotational superradiance'. 

For Kerr black holes, incident bosonic waves satisfying the above condition are amplified at the cost of the rotational energy of the black hole. Superradiance can also occur in an electrically charged black hole when a charged bosonic wave scattered by the black hole is amplified by the electric field of the black hole~\cite{bekenstein_PRD_1973}. We consider a scalar field $\Psi$ with electric charge $q$ incident on a static spherically symmetric charged (Reissner-Nordstr\"{o}m or RN) black hole of mass $M$ and charge $e$. The Reissner-Nordstr\"{o}m black bole is defined by the metric~\eqref{ch1:eq_sch} with $f(r)=\left(1-2M/r+e^2/r^2\right)$ and is characterised by an event horizon at $r_+$ and an inner Cauchy horizon at $r_-$,
\begin{equation}
r_{\pm}=M \pm \sqrt{M^2-e^2}~.
\end{equation}
 To study the dynamics of the scalar field, one starts with the Klein-Gordon equation~\cite{landau_book},
\begin{equation}\label{ch1:eq_KG}
\left(\nabla^\nu-iqA^\nu\right)\left(\nabla_\nu-iqA_\nu\right)\Psi=0,
\end{equation}
where $\nabla_\nu$ is the covariant  derivative with respect to the coordinate $x^\nu$ and $A_\nu=-\delta^0_\nu e/r$ is the electromagnetic potential of the black hole.
Using the ansatz,
\begin{equation}\label{ch1:eq_decom}
\Psi\left(t,r,\theta,\phi\right)=e^{-i\omega t}Y^m_l\left(\theta, \phi \right)\psi\left(r\right).
\end{equation}
one arrives at an equation of the form,
\begin{equation}
\frac{d^2\psi }{d r_*^2}+W\left(\omega,r \right)\psi=0~,
\end{equation}
where $Y^m_l\left(\theta, \phi \right)$ are the spherical harmonics, $r_*$ is the tortoise coordinate for the Reissner-Nordstr\"{o}m black hole given by $dr_*=dr/f(r)$ and 
\begin{equation}
W(r,\omega)=\left(\omega -\frac{e q}{r} \right)^2-\frac{f(r)}{r^2} l (l+1)-\frac{f(r)}{r}\frac{d f(r)}{d r}~.
\end{equation}
$W(r,\omega)$ is constant at the horizon $r_+ ,~(r_*\rightarrow -\infty)$, and vanishes as $r\rightarrow \infty$.  
The boundary condition for the scattering problem corresponds to an incident wave of amplitude $\mathcal{I}$ from spatial infinity, which gives rise to a reflected wave of amplitude $\mathcal{R}$ near infinity and a transmitted wave of amplitude $\mathcal{T}$ at the horizon,  
\begin{equation}\label{ch1:eq_BC_superrad}
\psi \sim 
\begin{cases}
\mathcal{I}e^{-i\omega r_*}+\mathcal{R}e^{i\omega r_*} &\mbox{as} \quad r_*\rightarrow \infty\\
 \mathcal{T}e^{-i \left(\omega-\frac{e q}{r_{+}} \right) r_*}&\mbox{as} \quad  r_* \rightarrow -\infty. 
\end{cases}
\end{equation}

Reversing the sign of $t$ and $\omega$ leads to a second linearly independent solution $\psi^{*}$ which satisfies the complex conjugate boundary conditions. The Wronskian of the two solutions being independent of $r_{*}$, the Wronskian evaluated at the horizon, $W_h= 2i\left(\omega-\frac{e q}{r_{+}} \right)|\mathcal{T}|^2$, must be equal to that evaluated near infinity, $W_\infty=-2i\sqrt{\omega^2-\mu^2}\left(|\mathcal{R}|^2-|\mathcal{I}|^2\right)$. This results in,
\begin{equation}\label{ch1:eq_R^2}
|\mathcal{R}|^2=|\mathcal{I}|^2-\frac{\omega-\frac{e q}{r_+}}{\omega}|\mathcal{T}|^2~.
\end{equation}
Thus, the reflected wave will be superradiantly amplified $\left( |\mathcal{R}|^2>|\mathcal{I}|^2 \right)$,  provided~\cite{zouros_AP_1979}
\begin{equation}\label{ch1:eq_sprrdnc}
\omega<\frac{e q}{r_+}~.
\end{equation}

To highlight the importance of dissipation in superradiance, one may consider a situation in which the event horizon in the previous example is replaced by a regular two-way membrane. In that case, the boundary condition at $r=r_h$ will contain an additional contribution of an outgoing wave of amplitude $\mathcal{O}$,
\begin{equation}\label{ch1:eq_chrg_dissip}
\psi \sim  \mathcal{T}e^{-i \left(\omega-\frac{e q}{r_{+}} \right) r_*} + \mathcal{O}e^{i \left(\omega-\frac{e q}{r_{+}} \right) r_*} \mbox{as} \quad  r_*\rightarrow -\infty.
\end{equation}
In view of the above boundary condition Eq.~\eqref{ch1:eq_R^2} becomes
\begin{equation}
\label{ch1:eq_R^2_no_diss}
|\mathcal{R}|^2=|\mathcal{I}|^2-\frac{\omega-\frac{e q}{r_+}}{\omega}\left(|\mathcal{T}|^2-\mathcal{O}|^2\right)~.
\end{equation}
If there is no loss of energy or dissipation, energy conservation would demand that the outgoing energy flux, $|\mathcal{O}|^2$, across the surface at $r_h$, be equal to the transmitted energy flux, $|\mathcal{T}|^2$. Thus,  from Eq.~\eqref{ch1:eq_R^2_no_diss} one obtains $|\mathcal{R}|^2 = |\mathcal{I}|^2$, implying no superradiance. In general, dissipation is absolutely crucial in any system for superradiance to occur.

It can be further shown that in the case of fermions $\mathcal{R}\leq \mathcal{I}$ for any $\omega$ and hence no superradiance is possible~\cite{unruh_PRL_1973,toth_CQG_2016,chandrasekhar_PRSLA_1976,iyer_PRD_1978,dolan_CQG_2015}. This is related to the intrinsic difference between bosons and fermions. 
The current density for bosons can change sign as they are partially transmitted with a transmission amplitude that can be negative, $-\infty<\frac{\omega-\frac{e q}{r_+}}{\omega}|\mathcal{T}|^2\leq |\mathcal{I}|^2$. Fermions, on the other hand have positive definite current densities and a transmission amplitude that is bounded from below, $0\leq|\mathcal{T}|^2\leq|\mathcal{I}|^2$.

If the superradiantly amplified waves is confined by some mechanism and made to interact with the back hole, they will be further amplified. A recursion of this process will ultimately lead to an instability of the system~\cite{damour_LNC_1976,detweiler_PRD_1980,zouros_AP_1979}. This is known as superradiant instability, and will be the main topic of discussion of chapter~\ref{chap3}.

As discussed in~\cite{manogue_AP_1988} superradiance is related to spontaneous creation of particle antiparticle pairs in the presence of strong fields. However, spontaneous particle pair production can occur even when superradiance is not allowed, for example, in case of fermions and even in the vacuum as in Hawking radiation.\\
For a detailed discussion on the vast topic of superradiance, we refer to the excellent book by Brito, Cardoso and Pani~\cite{brito_Springer_2015}.

\section{Hawking radiation}
In 1973, Bardeen, Carter and Hawking~\cite{dewitt_LH_1973, bardeen_CMP_1973} formulated a set of four laws governing the behaviour of black holes. These laws were formulated in analogy with standard thermodynamics with the thermodynamic variables like temperature etc. suitably defined in context of black holes.

The zeroth law of black hole mechanics states that the surface gravity of a stationary black hole is always constant over the entire event horizon~\cite{dewitt_LH_1973, bardeen_CMP_1973}. This is equivalent to the zeroth law of thermodynamics  which states that when a system is in thermal equilibrium, it has a constant temperature $T$.

The first law of black hole mechanics states that the variations in the mass $M$, area $A$, angular momentum $J$, and charge $e$ of a black hole obey~\cite{wald_1993}
\begin{equation}
\label{ch1:eq_1stlaw}
\delta M=\frac{1}{8\pi}\kappa \delta A + \Omega \delta J+ \Phi \delta e
\end{equation}
where $\kappa$, $\Omega$ and $\Phi$ are respectively the surface gravity, angular velocity and electrostatic potential at the event horizon. In a static spherically symmetric spacetime of the form given in Eq.~\eqref{ch1:eq_sch}, the surface gravity is defined to be the force required of an observer at infinity to keep a particle of unit mass stationary at the event horizon, $\kappa=\frac{1}{2}f'\left(r_h\right)$. In case of a Schwarzschild black hole of mass $M$, $\kappa=1/4M$, which is equal to the Newtonian acceleration due to gravity at the event horizon.  Equation~\eqref{ch1:eq_1stlaw} is essentially analogous to the first law of thermodynamics.

The second law of black hole mechanics is the Hawking's area theorem~\cite{hawking_PRL_1971} stating that the area $A$ of the event horizon can never decrease,
\begin{equation}
\delta A\geq 0~.
\end{equation}
The second law assumes that the spacetime is globally hyperbolic and the energy condition, $R_{\mu \nu} k^\mu k^\nu\geq0$ for any null-like four vector $k^\mu$. The second law is comparable to the second law of thermodynamics, which states that the entropy $S$ of a closed system can never decrease.

The third law of black hole mechanics states the surface gravity of the event horizon cannot be reduced to zero by any finite number of steps~\cite{israel_PRL_1986}. This is comparable to the Nearnst (weaker) form of the third law of thermodynamics which states that the temperature of a thermodynamic system cannot be reduced to zero by any finite number of operations.

The analogy between the four laws of  black hole mechanics and the corresponding laws of thermodynamics highlights that some multiple of the horizon area $A$ must behave as entropy, whereas the temperature must be {equivalent} to some multiple of the surfacegravity $\kappa$. This would imply that the term $\kappa \frac{\delta A}{8 \pi}$  in the first law of mechanics becomes $T \delta S$. It is important to mention that Bekenstein~\cite{bekenstein_LNC_1972,bekenstein_PRD_1973_1,bekenstein_PRD_1974} had proposed that the black hole entropy is equal to some multiple of the horizon area, prior to the formulation of the laws of black hole mechanics. Later, in 1974, from semi-classical considerations (quantum fields in a fixed black hole background) Hawking concluded that a black hole would spontaneously emit particles~\cite{hawking_Nature_1974,hawking_CMP_1975}. This radiation, dubbed as `Hawking radiation', has a characteristic thermal nature corresponding to a temperature of $T=\kappa/2\pi$ . This yields the celebrated Bekenstein-Hawking formula for the black hole entropy,
\begin{equation}
S_{BH}=A/4~.
\end{equation}
There are two essential ingredients in understanding Hawking radiation~\cite{carlip_LNP_2008}. \\
First, the quantum mechanical vacuum is filled with virtual particle-antiparticle pairs with opposite energies that pop in and out of existence. In a stable quantum field theory, negative energy states are usually forbidden, and the vacuum state is the lowest energy state. However, energy has a quantum mechanical uncertainty, so a virtual particle-antiparticle pair with energies $\pm E$ can fluctuate in and out of existence over a time period $t\sim \hbar /E$. It may be noted that the particle-antiparticle pair may have energies $\mp E $ as well.
\\
Second, in General Relativity, energy and particularly its signature can be frame dependent. The Hamiltonian is the generator of time-translations and thus depends on the choice of the time coordinate.
 In case of a Schwarzschild black hole given in Eq.~\eqref{ch1:eq_sch} with $f(r)=(1-r_h/r)$, as one crosses the horizon at $r_h$, $f(r)$ changes sign and with it the nature of the $r$ and $t$ coordinates gets interchanged in their role as space and time coordinates. Inside the event horizon, the direction of decreasing $r$ becomes the direction of increasing time. This is also evident from the Kruskal diagram in Fig.~\ref{ch1:fig_td_sch}. Let us the consider a virtual particle-antiparticle pair that is created just outside the event horizon with energies $\pm E$ relative to an observer at infinity. The component of the virtual particle-antiparticle pair that has negative energy relative to an asymptotic observer may cross the event horizon in time less than $\hbar/E$. 
 Thus, the positive energy component of the pair that is outside the horizon can materialize and reach an asymptotic observer. 
 
The heuristic explanation provided above captures the essence of Hawking radiation~\cite{hawking_CMP_1975}. For a historical account of the formulation of Hawking radiation, we refer to the excellent article by Page~\cite{page_NJP_2004}.

\subsection{Sparsity and Greybody factor}
\label{ch1:sec_sparsity}
One curious feature of Hawking radiation is its 'sparsity'. The sparsity of Hawking radiation is used to distinguish it from blackbody radiation~\cite{gray_CQG_2016, hod_PLB_2016, hod_EPJC_2015, miao_PLB_2017,schuster_thesis, bibhas_IJMPA_2017}. Loosely speaking, sparsity is a measure of the number of particles  emitted per unit time in a radiation process. It depends on the frequency of the emitted radiation as well as on rate of emission. \\
To understand the essence of 'sparsity', let us consider the emission of massless scalar particles from a black body in flat Minkowski space. The number of particles emitted by a black body at a temperature $T$ in unit time, in the momentum range $d^3\vec{k}$ into a surface element $d\vec{A}$ with unit normal $\hat{n}$ is given by,
\begin{equation}
dN_n=\frac{1}{(2\pi)^3}\frac{1}{e^{ k/ T}-1}d^3\vec{k} \cdot d\vec{A}.
\end{equation}
 Thus, the total number of particles emitted by a blackbody at a temperature $T$ in unit time, assuming finite surface area $A$ is given by
\begin{eqnarray}
N_n&=&\int_{0}^{A} \int_{0}^{\infty} \int_{0}^{2\pi} \int_{0}^{\pi/2}\frac{1}{(2\pi)^3}\frac{\hat{n}\cdot \hat{k}}{e^{ k/ T}-1} k^2 \sin{\theta} d\theta d\phi dk dA~,\\
&=&\frac{\zeta(3)}{4 \pi^2} T^3 A~,
\end{eqnarray}
where $\zeta(\alpha)=\frac{1}{\Gamma(\alpha)}\int_0^\infty \frac{x^{\alpha-1}}{e^x -1}dx$
is the Riemann Zeta function and $\Gamma(\alpha)=\int_0^\infty x^{\alpha-1}e^{-\alpha}dx$ is the Gamma function (with $\Gamma(n+1)=n!$). 
The inverse of $N_n$ gives the time gap between the emission of two consecutive quanta,
\begin{equation}\label{ch1:eq_tgap}
\tau_{gap}=\frac{1}{N_n}~.
\end{equation}
 The definition~\eqref{ch1:eq_tgap} for the time interval between consecutive emission is applicable only when the detector is of the shape of a sphere surrounding the blackbody, otherwise $\tau_{gap}<1/N_n$. 

To obtain the sparsity of radiation, one usually compares the timescale $\tau_{gap}$ with a localisation timescale $\tau_{localisation}$~, which is the time taken by the emitted wave field to complete one oscillation cycle~\cite{gray_CQG_2016}. For simplicity, we choose the localisation time scale to be,
\begin{equation}
\tau_{localisation,n}=\frac{2\pi}{\omega_{peak,n}},
\end{equation}
where $\omega_{peak,n}$ refers to the value of the angular frequency $\omega$ at which the peak of the number density spectrum occurs, i.e., the position of the maximum of  $k^2/\left(e^{k/T}-1\right)$. Other relevant choices for localisation time include~\cite{gray_CQG_2016},  $2\pi/\omega_{average,n}$ which is related to the position of the average of the number density spectrum; $2\pi/\omega_{peak,E}$ and $2\pi/\omega_{average,E}$ which are related to the position of the peak and average of the energy density spectrum, $dN_E\propto \frac{k^3}{e^{k/T}-1}dk$, respectively.

Using Lambert $W$-function (defined as $W(x) e^{W(x)}=1$) we get
\begin{equation}
\omega_{peak,n}=T\left(2+W\left(-2e^-2\right)\right).
\end{equation}
Thus, we define the sparsity of radiation as,
\begin{equation}\label{ch1:eq_sparsity}
\eta=\frac{\tau_{gap}}{\tau_{localisation,n}}=\frac{\left(2+W\left(-2e^-2\right)\right)}{2\pi\zeta(3)}\frac{\lambda_{thermal}^2}{A},
\end{equation}
where, $\lambda_{thermal}=\frac{2\pi \hbar c}{k_B T}$ (explicitely).

For standard (laboratory and astrophysical) blackbodies, the thermal wavelength is much smaller than their surface area, and hence $\lambda_{thermal} ^2<<A$ which implies $\eta<<1$. This implies that the sparsity of a blackbody emission spectrum is extremely low, suggesting that a large number of particles are emitted by a blackbody per unit time per unit time. 
However, for black holes, $\eta>>1$ and the Hawking emission spectrum is extremely sparse~\cite {gray_CQG_2016, hod_PLB_2016, hod_EPJC_2015, miao_PLB_2017,schuster_thesis, bibhas_IJMPA_2017}.

In case of Hawking emission of massless scalar particles from black holes, the definition of sparsity is modified by the inclusion of the greybody factor as discussed in chapter~\ref{chap5}. The definition of $\tau_{gap}$ and $\tau_{localisation}$ used in chapter~\ref{chap5} is based on the energy flux rather than the number flux considered here, however, the basic philosophy remains the same and the Hawking emission spectrum is shown to be extremely sparse.

The Hawking radiation emitted at the event horizon gets modified as it propagates through the spacetime outside the black hole. The geometry outside the event horizon gives rise to a potential barrier that transmits only a part of the Hawking radiation while reflecting the rest back to the black hole. Thus, an asymptotic observer receives only a fraction of the Hawking radiation emitted at the event horizon. This fraction is referred to as the greybody factor. Hence, the greybody factor is the transmission probability through the potential barrier and measures the deviation of the Hawking emission spectrum, as observed by an asymptotic observer from a perfect black body spectrum~\cite{page_PRD_1976,page_PRD2_1976}.\\
{The Hawking emission rate of a black hole at the event horizon in a mode with frequency $\omega$ is given by
\begin{equation}
\Gamma(\omega)=\frac{1}{e^{\beta \omega}\pm 1}~\frac{d^3 k}{{2\pi}^3}~,
\end{equation}
where the Hawking temperature is the inverse of $\beta$ and the plus and minus sign corresponds to the emission of fermions and bosons respectively. However, an asymptotic observer would measure the emission rate to be,
\begin{equation}
\Gamma(\omega)=\frac{T(\omega)}{e^{\beta \omega}\pm 1}~\frac{d^3 k}{{2\pi}^3}~,
\end{equation}
where $T(\omega)$ is the frequency dependent greybody factor.}
\section{No-hair conjecture}
One of the fundamental questions regarding a black hole is the nature of the information that an external observer may get from the exterior gravitational field of a black hole.
The answer to this is usually given in terms of the No-hair Conjecture.

To understand the no-hair conjecture one must recollect the black hole uniqueness theorems~\cite{chrusciel_LRR_2012,israel_PR_1967}, which in a nutshell { states that the most general stationary black hole solution to the Einstein-Maxwell equations is given by} the Kerr-Newman metric, which depends only on the mass $M$, angular momentum $J$ and electric charge $e$ of the black hole. This led Ruffini and Wheeler~\cite{ruffini_PT_1971} to propose the No-hair conjecture. The conjecture claims that the most general result of a gravitational collapse in the presence of any form of matter-energy is a Ker-Newman black hole~\cite{MTW}. The term `hair' in this context refers to the physical parameters other than the mass, charge and angular momentum that can be measured by an external observer. For further discussion on the topic, we refer to~\cite{bekenstein_Sharkov_1996,herdeiro_IJMPD_2015,volkov_MG_2016,barcelo_CQG_2019}.
\subsection{Quest for hairy black holes}
Unlike the standard black holes that strictly obey the No Hair Conjecture, there were numerous attempts to the find so-called ``hairy black holes" which have other nontrivial physical parameters (hair). Black hole hair can be classified into two broad categories viz. the primary hair and the secondary hair~\cite{herdeiro_IJMPD_2015}. A primary hair is one that can exist on its own without depending on any other black hole parameters viz. mass, charge and angular momentum. A secondary hair, on the other hand, is the one that cannot exist on its own. It depends on some other parameter (primary hair), such that if the primary hair vanishes the secondary hair also ceases to exist. 

Scalar fields are one of the simplest matter fields considered by physicists. The existence of scalar field in nature has been experimentally verified by the Large Hadron Collider at CERN in the form of the standard model Higgs boson~\cite{aad_PLB_2012, chatrchyan_PLB_2012}. Scalar fields has also been considered in different phenomenoogical models in gravity and particularly in cosmology~\cite{faraoni_book} to model dark energy and dark matter.  {This is because scalar fields can be used as a substitute for realistic matter. Canonical scalar fields can mimic a perfect fluid with some equation of state~\cite{faraoni_PRD_2012}.} Scalar fields are also well motivated by beyond standard model particle physics. So, one naturally enquires about the validity of the no-hair concept in the context of scalar fields coupled to gravity. In this context, Chase~\cite{chase_CMP_1970}, following earlier suggestions by Penney~\cite{penney_PR_1968}, showed that asymptotically flat static black hole spacetimes cannot support a massless scalar field in equilibrium with it. Later, Bekenstein~\cite{bekenstein_PRL_1972,bekenstein_PRD_1972_1,bekenstein_PRD_1972_2,bekenstein_PRD_1995} put forward a no-go theorem for the existence of massive scalar hair and extended it to higher spin fields as well. Bekenstein's theorem has also been extended for other theories of gravity for example Lanczos-Lovelock gravity~\cite{skakala_CQG_2014}. Recently, no-hair theorem in the context of  noncanonical self-gravitating static multiple scalar fields in spherically symmetric spacetimes has been studied in~\cite{doneva_PRD_2020}. For a detailed review on the status of the no-scalar hair theorems, we refer to Herdeiro and Radu~\cite{herdeiro_IJMPD_2015} and Winstanley~\cite{winstanley_FP_2002}. 
 
In the regime of non-minimally coupled scalar field theories, a case of particular interest is when the scalar field is conformally-coupled to gravity. The Bocharova–Bronnikov–Melnikov–Bekenstein (BBMB)~\cite{bocharova_1970,bekenstein_AP_1974,bekenstein_AP_1975} black hole solution of General Relativity coupled conformally to a massless scalar field was the first counter-example of the no-hair conjecture~\cite{ruffini_PT_1971}.  The action for the theory with conformal scalar-vacuum reads as,
\begin{equation}\label{ch1:eq_BBMB_action}
I=\frac{1}{16 \pi}\int d^4 x\sqrt{-g}\left[R-8\pi\left( \bigtriangledown_\mu \psi \bigtriangledown^\mu \psi-\frac{1}{6} R \psi^2 \right)\right].
\end{equation}
The scalar field equation, $\bigtriangledown_\mu \bigtriangledown^\mu \psi-R \psi /6=0$,
remains invariant under local conformal transformation, $g_{\mu\nu}\rightarrow \bar{g}_{\mu\nu}=\Omega^2 g_{\mu\nu} $~. The BBMB black hole is defined by the line element,
\begin{equation}
ds^2=-\left(1-\frac{M}{r}\right)^2 dt^2+\left(1-\frac{M}{r}\right)^{-2} dr^2+r^2\left(d\theta^2 +\sin^2\theta d\phi^2\right)~,
\end{equation}
with the scalar field $\psi$ given by,
\begin{equation}
\psi=\frac{\sqrt{3} M}{r-M}~,
\end{equation}
where $M$ is the total mass.
For vanishing $M$ the metric corresponds to a standard Minkowski black hole whereas for non-zero  mass, it corresponds to an extremal Reissner-Nordstr\"{o}m black hole with $|e|=M$.
However, the BBMB black hole solution is plagued with several disturbing anomalies, for example, the scalar field diverges at the horizon and the BBMB metric does not have a continuous limit to a Schwarzschild solution. Bronnikov and Kireev~\cite{bronnikov_PLA_1978} showed that the BBMB black hole is unstable under linear perturbation, although McFadden and Turok~\cite{mcfadden_PRD_2005} have concluded otherwise.  Further, the scalar field vanishes in the absence of the $M$, and hence the scalar hair is a secondary hair. Astorino~\cite{astorino_PRD_2013} considered a charged generalization of the BBMB solution and removed the anomalies to generate a new class of scalar hairy black holes with conformal matter coupling.\\
For some more examples and discussion on hairy black holes, we refer to \cite{tahamtan_PRD_2015, tahamtan_PRD_2016,sen_Pramana_2001, antoniou_PRL_2018,antoniou_PRD_2018,nucamendi_PRD_2003,canate_CQG_2015} and references therein.
\subsection{A Charged black hole with scalar hair}\label{ch1:sec_shRN}
{ Starting from the action~\eqref{ch1:eq_BBMB_action}, Astorino~\cite{astorino_PRD_2013} arrived at a static spherically symmetric charged black hole solution endowed with a primary scalar hair. This solution is dubbed as the scalar-hairy Reissner-Nordstr\"{o}m black hole or the scalar-hairy RN black hole and will be considered in this thesis.} The scalar-hairy RN spacetime is defined by the line element,
\begin{equation}\label{ch1:eq_shRN}
ds^2=-f(r)dt^2+\frac{dr^2}{f(r)}+r^2\left(d\theta^2 +\sin^2\theta d\phi^2\right)~,
\end{equation}
where
\begin{equation}\label{ch1:eq_f(r)}
f(r)=1-\frac{2M}{r}+\frac{e^2 + s}{r^2}~.
\end{equation}
 The scalar charge $s$ occurs as an additive correction term to the square of the electric charge $e$. The scalar hairy black hole is characterised by an event horizon at $r_+$ and a Cauchy horizon at $r_-$, where
\begin{equation}\label{ch1:eq_r_+-}
r_{\pm}=M\pm\sqrt{M^2-e^2-s}~.
\end{equation}
The associated scalar field,
\begin{equation}\label{ch1:eq_scalarfield}
\psi =\pm \sqrt{\frac{6}{8 \pi G}}\sqrt{\frac{s}{s+e^2}}~,
\end{equation}
can exist independently and hence constitutes a primary scalar hair.
The net energy-momentum tensor is given by,
\begin{equation}\label{ch1:eq_stressenergy}
T^\mu_\nu=\frac{e^2+s}{r^4} diag\left(-1,-1,1,1\right) . 
\end{equation}
The energy-momentum tensor of the scalar-hairy RN spacetime satisfies both the dominant and strong energy conditions for $s>-e^2$. This is to say that for $s>-e^2$ and for any future directed normalized timelike vector $v^\mu$, the vector $-T^\mu_\nu v^\nu$ is a future directed timelike vector field (dominant energy condition) and 
\begin{equation}
\left(T_{\mu\nu}-\frac{1}{2}T g_{\mu\nu}\right) v^\mu v^\nu \geq 0 \qquad \mbox{(strong energy condition)},
\end{equation}
where $T$ is the trace of $T^\mu_\nu$. The strong energy condition coupled to the fact that the $T^\mu_\nu$ is traceless implies that the weak energy condition is also satisfied for $s>-e^2$,
\begin{equation}
T_{\mu\nu}v^\mu v^\nu \geq 0~.
\end{equation}
It is important to note that the energy-momentum tensor due to the scalar field alone is also traceless. According to a theorem given by Banerjee and Sen~\cite{narayan_pramana_2015}, for an asymptotically flat spherical black hole to have a non-trivial hair, the trace of the  energy-momentum tensor of the corresponding field must either vanish or fall off at least as rapidly as $1/r^3$ with $r$ being the proper radius. Thus, the existence of this scalar hair is completely consistent with the above theorem. It must also be stressed that unlike the BBMB solution the scalar field $\psi$ in Eq.~\eqref{ch1:eq_scalarfield} of the scalar-hairy RN black hole~\cite{astorino_PRD_2013} does not diverge at the event horizon.\\
The scalar-hairy RN solution~\eqref{ch1:eq_shRN}  is similar to the black hole solution given by Bekenstein~\cite{bekenstein_AP_1974} for a conformally invariant scalar field. In the form (\ref{ch1:eq_shRN}) as provided by Astorino~\cite{astorino_PRD_2013}, the solution opens a new possibility. \\
{With $s<-e^2$, the coefficient of $1/r^2$ in $f(r)$ in Eq.~\eqref{ch1:eq_f(r)} becomes negative and the Astorino solution mimics a mutated Reissner-Nordstr\"{o}m metric~(\ref{ch1:eq_shRN}), where $f = (1-\frac{2m}{r} - \frac{\epsilon^2}{r^2})$. This mutated Reissner-Nordstr\"{o}m metric leads to an Einstein-Rosen bridge or a wormhole, that connects two causally  disconnected spacetime~\cite{rosen_PR_1935}.
}

The scalar-hairy Reissner Nordstr\"{o}m black hole evades the no-hair theorem since in the absence of electromagnetic charge, it cannot be connected to the Einstein frame by means of conformal transformations~\cite{astorino_PRD_2013}.

It is interesting to note that a metric, identical to the mutated Reissner-Nordstr\"{o}m solution has been obtained earlier by Dadhich et al.~\cite{dadhich_PLB_2000} as an exact static, spherically symmetric black hole solution on 3 - brane in five-dimensional gravity in the Randall-Sundrum scenario. The parameter, $q= -(e^2+s)$ with $s<-e^2$ being interpreted as a tidal charge parameter. Echoes in the gravitational wave signals from such braneworld black holes has been recently studied in~\cite{dey_PRD_2020}\\
The Penrose diagrams of the scalar-hairy RN black hole (for $s>-e^2$) are discussed in Appendix~\ref{AppendixA}.
\section{Outline of the thesis}
The motivation of the present thesis is to study the effect of scalar fields on black hole radiation of different forms. We look for various aspects of the quasinormal modes arising out of the perturbation by test field of a charged spherical black hole with a scalar hair. We also look for the superradiant stability of such black holes under perturbation by massive charged test scalar field. Our second object of interest is the effect of the scalar hair on the Hawking emission of charged particles from these black holes and the associated grey body factor. We pick up the scalar-hairy black hole solution put forward by Astorino~\cite{astorino_PRD_2013}, given by Eqs.~\eqref{ch1:eq_sch} and \eqref{ch1:eq_f(r)} for our investigation.

In Chapter~\ref{chap2}, we study the quasinormal modes of the scalar-hairy RN black hole due to perturbation by massless electrically charged test scalar and Dirac fields for both positive and negative values of the black hole scalar charge $s$. We also study the effect of the mass of the test scalar field on the quasinormal mode spectrum. We employed the Leaver's method of  continued fraction~\cite{leaver_PRSLA_1985,leaver_PRD_1990,leaver_PRD_1991} to evaluate the quasinormal frequencies numerically. We observe that the presence of the scalar hair  changes the quasinormal mode spectrum from that of a standard Reissner-Nordstr\"{o}m black hole. The change is much more perceptible for $s<-e^2$. When the test field has non-zero electric charge, the damping rate of the fundamental quasinormal mode decreases monotonically with $s$ for $s<-e^2$, whereas, for $s\geq -e^2$ the damping rate { is not monotonic and} maximizes for a particular value of the scalar charge. In the case of the uncharged test field, the spacetime behaves similar to a standard Reissner-Nordstr\"{o}m black hole with an effective electric charge $Q=\sqrt{e^2+s}$, provided $s$ lies in $M^2-e^2\geq s> -e^2$. {Similar results are also obtained for charged Dirac fields.}

In Chapter~\ref{chap3}, we analyze  the superradiant stability of the scalar-hairy Reissner-Nordstr\"{o}m black hole, primarily for $s<-e^2$. We study the evolution of massive charged test scalar fields in the background of the scalar-hairy RN black hole. If the frequency of the {incident scalar wave satisfies} the superradiance condition as given in Eq.~\eqref{ch1:eq_sprrdnc}, then the reflected wave gets amplified. {The mass, $\mu$ of the test field in principle, acts a natural `mirror' (effective potential barrier for the superradiant modes) to reflect the superradiantly amplified waves with frequency $\omega<\mu$ (bound state condition) back into the black hole which are reamplified. }This back and forth reflection of the superradiant modes may lead to an instability of the black hole. However, we observe that much like the Reissner-Nordstr\"{o}m black hole, the scalar-hairy RN superradiant modes do not satisfy the bound state condition and hence the scalar-hairy RN black hole is superradiantly stable for both positive and negative values of the scalar charge.

In Chapter~\ref{chap4}, we study the Hawking emission of charged particles from the scalar-hairy Reissner-Nordstr\"{o}m black hole using the tunnelling method, developed by Parikh and Wilczek~\cite{parikh_PRL_2000}. The tunnelling method critically hinges on the principle of conservation of energy. Let us consider a Schwarzschild black hole of mass $M$. The radius of its event horizon is $r_h=2M$. Let us consider a virtual particle pair produced inside the event horizon. Now, if the outgoing particle has an energy $E$, then the radius of the event horizon shrinks to $r_h=2(M-E)$. Thus, the outgoing particle has to tunnel through the classically forbidden region, $r=2M$ to $r=2(M-E)$.

We evaluate the Hawking emission rate of the scalar-hairy RN black hole and observe that the total entropy of the black hole contains an energy-dependent contribution, which vanishes with the scalar charge. We also evaluate the upper bound on the charge-mass ratio of the emitted particles and observe that the charge-mass ratio decreases with the black hole scalar charge.

Chapter~\ref{chap5} is devoted to the study of the grey body factor and the sparsity of the Hawking emission of massless uncharged scalar particles from the scalar-hairy RN black hole. We provide semi-analytic bounds on the grey body factor. We observe that the scalar and electric charges of the black hole contribute oppositely to the grey body factor and the sparsity Hawking emission cascade. The grey body factor increases with the scalar charge, whereas it decreases with the electric charge of the black hole. Also, the sparsity of the Hawking emission flow decreases with the scalar charge of the black hole, whereas it increases with the black hole electric charge.

We notice that as the black hole continues to Hawking radiate, its ADM mass decreases, which in turn raises its temperature and enhances the Hawking emission power. The lowering of the ADM mass of the black hole also lowers the grey body factor and makes the Hawking radiation even more sparse.

Finally, in Chapter~\ref{chap6}, we conclude with a brief summary of the work presented in this thesis.


\chapter{Quasinormal modes of a charged spherical black hole with scalar hair for scalar and Dirac perturbations}\blfootnote{\begin{flushleft} The work presented in this chapter is based on “Quasinormal modes of a charged spherical black hole with scalar hair for scalar and Dirac perturbations”, \textbf{Avijit Chowdhury} and Narayan Banerjee, Eur.\ Phys.\ J.\ C \textbf{78}, 594 (2018).\end{flushleft}}  
\label{chap2}
\chaptermark{QNMs of a charged spherical black hole with scalar hair \ldots}
\section{Introduction}
After the dream of detecting gravity waves came true, and that too from a merger of two black holes~\cite{ligo_PRL_2016,abott_PRL_2017, abott_GWTC_2018}, the importance of a thorough investigation of the quasinormal modes in connection with the black hole perturbations cannot perhaps be exaggerated. These investigations started a long way back, through the work of Regge and Wheeler~\cite{regge_PR_1957} and  Vishveshwara~\cite{vishveshwara_PRD_1970, vishveshwara_Nature_1970}. As mentioned in section~\ref{ch1:sec_QNM}, the response of a black hole to a perturbation of an external field or the perturbation of the metric is manifested in the form of a damped waves, characterized by complex frequencies, called the quasinormal frequencies. The real part of the frequency corresponds to the actual frequency of the wave motion while the imaginary part takes care of the damping factor. . 

Quasinormal modes (QNM) for a Schwarzchild black hole has been studied by Vishveshwara~\cite{vishveshwara_Nature_1970} and also by Davis, Ruffini, Press and Price~\cite{davis_PRL_1971}. QNMs for a Reissner-Nordstr\"{o}m black hole was first investigated by Gunter~\cite{gunter_PTRSLA_1980}. Investigations regarding QNMs for various kind of black holes are already there in the literature. Dreyer discussed the QNMs, area spectrum and entropy of a black hole and also fixed the value of the Immirizi parameter which arises in Loop quantum gravity~\cite{dreyer_PRL_2003}. Cardoso and Lemos discussed the QNMs of a BTZ black hole~\cite{cardoso_PRD_2001_1} and also Schwarzchild-AdS black holes~\cite{cardoso_PRD_2001_2}. The latter had been discussed by Horowitz and Hubeny~\cite{horowitz_PRD_2000} also. QNMs for a  near extremal black hole has been investgated by Starinets~\cite{starinets_PRD_2002} and by Cardoso and Lemos~\cite{cardoso_PRD_2003}. QNMs for a Gauss-Bonnet black hole has been discussed by Chakrabarti~\cite{sayan_GRG_2007}.

In this chapter, we will investigate the QNMs of the scalar-hairy Reissner-Nordstr\"{o}m (scalar-hairy RN) black hole.  We will mostly concentrate on the negative values of the scalar charge $s$, specifically in the regime $s<-e^2$, $e$ being the electric charge of the scalar-hairy RN black hole. As discussed in section~\ref{ch1:sec_shRN}, the scalar-hairy RN black hole in this regime acts as a mutated Reissner-Nordstr\"{o}m black hole. The perturbation of massless and massive uncharged/charged scalar particles and massless charged Dirac particles and the QNMs generated by the perturbations in the vicinity of a mutated Reissner-Nordstr\"{o}m black hole {will be} discussed in the present chapter. The continued fraction method (see Refs.~\cite{leaver_PRSLA_1985,leaver_PRD_1990,leaver_PRD_1991}) has been adopted. The fundamental mode is the dominating one in the signal and only that mode is dealt with.

In almost all the cases both the frequency and the damping rate decrease with the magnitude of the negative scalar charge. For perturbation by a massive scalar field, the damping rate falls off sharply compared to the massless case, whereas the real frequency falls off at a much slower rate. For charged fields, the oscillation frequency and the damping rate is more for higher values of the field charge. Similar results are also obtained for massless charged Dirac fields as well.

\section{Review of the scalar-hairy RN Spacetime}
Though the scalar-hairy RN black hole~\cite{astorino_PRD_2013} has already been described in section~\ref{ch1:sec_shRN}, here we will again briefly highlight some of its relevant properties.\\
The scalar-hairy RN black hole is described by the line element,
\begin{equation}\label{ch2:eq_metric}
ds^2=-f\left(r \right)dt^2+{f\left(r \right)}^{-1}dr^2+r^2 \left( d\theta^2+\sin^2{\theta} d\phi^2 \right),
\end{equation}
with
\begin{equation}
\label{ch2:eq_f(r)}
f\left( r \right)=\left(1-\frac{2M}{r}+\frac{e^2+s}{r^2}\right),
\end{equation}
where $M$ is the mass of the black hole and $e$ and $s$ are respectively the electric and scalar charges of the black hole.

The scalar-hairy RN black hole, is characterised by an inner Cauchy horizon $\left(r_{-} \right)$ and an outer event horizon $\left(r_{+} \right)$. The horizons of the scalar-hairy RN black hole are located at
\begin{eqnarray}
r_{+}&=&M+\sqrt{M^2-e^2-s},\\
r_{-}&=&M-\sqrt{M^2-e^2-s}.
\end{eqnarray}
The maximum value of the scalar (or electric) charge is determined by the extremality condition,
\begin{equation}\label{ch2:eq_extremality}
\sqrt{M^2-e^2-s} = 0.
\end{equation}
For $s<-e^2$, $r_{-}$ is negative and of no physical significance and hence the mutated RN spacetime is characterised by a single event horizon.

\section{Massive scalar field around a charged black hole with scalar hair}
In this section we discuss the dynamics of a massive charged scalar field in the background of a scalar-hairy RN black hole and study the fundamental $(n=0)$ mode of the  quasinormal spectrum of the field around the black hole.
\subsection{Field dynamics}\label{ch2:sec_msvsclrfld}
The dynamics of a massive charged test scalar field $\Phi$ of mass $\mu$ and electric charge $q$ in the background~(\ref{ch2:eq_metric}) is governed by the Klein-Gordon equation,
\begin{equation}\label{ch2:eq_KG}
[\left(\nabla^\nu-iqA^\nu\right)\left(\nabla_\nu-iqA_\nu\right)-\mu^2]\Phi=0
\end{equation}
where $A_\nu=-\delta^0_\nu e/r$ is the electromagnetic vector potential of the black hole.
We can decompose the field $\Phi$ as
\begin{equation}\label{ch2:eq_decom}
\Phi_{lm}\left(t,r,\theta,\phi\right)=e^{-i\omega t}Y^m_{l}\left(\theta\right)R_{lm}\left(r\right),
\end{equation}
{where $\omega$ is the conserved frequency, $l$ is the spherical harmonic index, $m$ ($-l\leq m\leq l$) is the azimuthal harmonic index, $Y^m_l(\theta,\phi)$ are the spherical harmonics and $R_{lm}(r)$ is the radial component of $\Phi$}. Hereafter, we will drop the subscripts $l$ and $m$ for brevity.

With the decomposition~(\ref{ch2:eq_decom}) one can separate the Klein-Gordon equation~(\ref{ch2:eq_KG}) into a radial and an angular equation with the separation constant $K_l=l\left( l+1 \right)$. The radial Klein-Gordon equation is given by
\begin{equation}\label{ch2:eq_radialKG}
\frac{d}{dr}\left(\Delta \frac{dR}{dr}\right)+\frac{U}{\Delta} R=0,
\end{equation}
where $\Delta=r^2 f\left(r\right)$ and 
\begin{equation}
U=\left( \omega r^2-e q r \right)^2-\Delta \left[ \mu ^2 r^2+l \left( l+1 \right)\right].
\end{equation}
If we define a new radial function $\zeta=r R$ and adopt the tortoise coordinate $r_{*}$ $\left( \right.$defined by, $d r_{*}=dr/f\left( r \right)\left.\right)$, mapping the semi infinite region $\left[r_{+},\infty\right)$ into $(-\infty,\infty)$, then the radial Klein-Gordon equation~(\ref{ch2:eq_radialKG}) becomes
\begin{equation}\label{ch2:eq_trts}
\frac{d^2 \zeta}{d r_{*}^2}+W\left(\omega, r\right) \zeta=0,
\end{equation}
where
\begin{equation}
\begin{split}
W\left(\omega, r\right)&=\left(\omega -\frac{e q}{r}\right)^2\\
& -f(r) \left(-\frac{2 \left(e^2+s\right)}{r^4}+\frac{2 M}{r^3}+\frac{(l+1) l}{r^2}+\mu ^2\right).
\end{split}
\end{equation}
In the asymptotic limit equation~(\ref{ch2:eq_trts}) can be solved analytically with the quasinormal mode boundary conditions of purely ingoing waves at the horizon $\left(r_{*}\rightarrow-\infty\right)$ and purely outgoing waves at spatial infinity $\left( r_{*}\rightarrow \infty \right)$,
\begin{equation}\label{ch2:eq_bc}
\zeta \approx 
\begin{cases}
 e^{-i \left(\omega-\frac{e q}{r_{+}} \right) r_*}&\mbox{ as \hspace*{2mm}}r_* \rightarrow -\infty \\
r_{*}^{-i e q} e^{i\Omega r_{*}}&\mbox{ as \hspace*{2mm}}r_* \rightarrow \infty ,
\end{cases}
\end{equation}
where $\Omega=\sqrt{\omega^2-\mu^2}$. Equation~(\ref{ch2:eq_trts}) together with the boundary conditions~(\ref{ch2:eq_bc}) becomes an eigenvalue problem with complex eigenvalues $\omega$ representing the quasinormal frequencies.\\

\subsection{Continued Fraction technique}\label{sec_scalar_CF}
In 1985, Leaver~\cite{leaver_PRSLA_1985,leaver_PRD_1990,leaver_PRD_1991} inspired by a seminal work of Jaff\'{e}~\cite{Jaffe_Z_Phys_1934} on the computation of the electronic spectra of hydrogen molecular ion, proposed a very accurate method for finding out the QNM frequencies of black holes.

To implement Leaver's method we start with equation~(\ref{ch2:eq_radialKG}) and observe that it has two regular singularities at $r_{+}$ and $r_{-}$ and an irregular singularity as $r\rightarrow \infty$ . { We can write the radial function as the product of a function which diverges at the singularities and a series which converges in the region $r_+\leq r\leq \infty$.,

Thus, we write the series solution to equation~(\ref{ch2:eq_radialKG}) with the desired behaviour at the boundaries following the usual technique~\cite{leaver_PRSLA_1985,leaver_PRD_1990,leaver_PRD_1991} as}
\begin{equation}\label{ch2:eq_scalar_ansatz}
R=e^{i \Omega r} (r-r_{-})^\rho\sum_{n=0}^{\infty}a_n u^{n+\delta},
\end{equation}
where $u=\frac{r-r_{+}}{r-r_{-}}$,	 $\rho=\frac{i \left(i \Omega+M \left(\Omega^2+\omega^2\right)-e q \omega\right)}{\Omega}$	and $\delta=-\frac{i r_{+}^2 \left(\omega -\frac{e q}{r_{+}}\right)}{r_{+}-r_{-}}$.\\
Substituting the ansatz~(\ref{ch2:eq_scalar_ansatz}) into equation~(\ref{ch2:eq_radialKG}) we arrive at the following three term recurrence relations, satisfied by the coefficient $a_n$
\begin{eqnarray}
\alpha_0 a_1 +\beta_0 a_0 & = & 0,\\
\alpha_n a_{n+1} +\beta_n a_n+\gamma_n a_{n-1} & = & 0,
\end{eqnarray}
where $\alpha_n$, $\beta_n$ and $\gamma_n$ are given by,
\begin{equation}
\alpha_n=-\frac{\left(n+1\right)^2 r_{-}+(n+1)r_{+} \left(-2 i e q-n+2 i r_{+} \omega -1\right)}{r_{+}-r_{-}}~,
\end{equation}
\begin{equation}
\begin{split}
\beta_n=
&\frac{1}{2 \Omega (r_{-}-r_{+})}\left[ r_{+} \left\lbrace 2 \left(-2 e^2 q^2 \left(\Omega+\omega \right)+i e (2 n+1) q \left(2 \Omega+\omega \right)\right.\right.\right.\\
&\left.\left.\left.+\left(l \left(l+1\right)+2 n^2+2 n+1\right) \Omega\right)\right.\right.\\
&\left.\left.+r_{+} \left(4 \omega  \left(\Omega+\omega \right) (3 e q-2 i n-i)+3 i \mu ^2 (2 i e q+2 n+1)\right)\right.\right.\\
&\left.\left.+2 r_{+}^2 \left(\mu ^2 \left(\Omega+3 \omega \right)-4 \omega ^2 \left(\Omega+\omega \right)\right)\right\rbrace \right.\\
&\left.-2 r_{-} \left\lbrace i r_{+} \left(-2 (2 n+1) \omega ^2+\mu ^2 (i e q+4 n+2)\right)+i e (2 n+1) q \omega\right.\right. \\
&\left.\left.+\left(l \left(l+1\right)+2 n^2+2 n+1\right) \Omega +\mu ^2 r_{+}^2 \left(\Omega+\omega \right)\right\rbrace +i \mu ^2 (2 n+1) r_{-}^2\right],
\end{split}
\end{equation}
\begin{equation}
\begin{split}
\gamma_n=
&\left[\frac{i \left\lbrace e q \omega -\frac{1}{2} \left(\Omega^2+\omega^2\right) \left(r_{-}+r_{+}\right)\right\rbrace }{\Omega}+i e q+n-i \omega  (r_{-}+r_{+})\right]\\
&\Bigg[n-\frac{i}{2 \Omega \left(r_{+}-r_{-}\right)} \Bigg\lbrace-2 \left(r_{+}-r_{-}\right) \left(e q \omega -\frac{1}{2}\left(r_{-}+r_{+}\right)\left(\Omega^2+\omega ^2\right)\right)\\
&+\Omega \left(r_{-}+r_{+} \right) \left(\omega  \left(r_{-}+r_{+}\right)-2 e q\right)+\omega  \Omega \left(r_{-}-r_{+}\right)^2\Big\rbrace\Bigg].
\end{split}
\end{equation}
The convergence of the series~(\ref{ch2:eq_scalar_ansatz}) requires 
the recursion coefficients to satisfy an infinite continued fraction relation
\begin{equation}\label{ch2:eq_scalar_CF}
0=\beta_0-\frac{\alpha_0 \gamma_1}{\beta_1-} \frac{\alpha_1 \gamma_2}{\beta_2-}\cdots\frac{\alpha_n \gamma_{n+1}}{\beta_{n+1}-}\cdots
\end{equation}
The solution to this infinite continued fraction equation gives the QNM frequencies. One can invert the  continued fraction relation~(\ref{ch2:eq_scalar_CF}) any number of times. Numerically, the most stable root of the $n^{th}$ inversion of the continued fraction relation gives the $n^{th}$ quasinormal frequency,
\begin{equation}\label{ch2:eq_scalar_CF_inv}
\begin{split}
\beta_n-\frac{\alpha_{n-1}\gamma_{n}}{\beta_{n-1}-} \frac{\alpha_{n-2}\gamma_{n-1}}{\beta_{n-2}-}\cdots\frac{\alpha_{0}\gamma_{1}}{\beta_{0}}=\frac{\alpha_{n} \gamma_{n+1}}{\beta_{n+1}-} \frac{\alpha_{n+1} \gamma_{n+2}}{\beta_{n+2}-}\cdots, \\ \left(n=1,2,3,4\cdots \right).
\end{split}
\end{equation}
In practice the infinite continued fraction in equations~(\ref{ch2:eq_scalar_CF}, \ref{ch2:eq_scalar_CF_inv}) is truncated at some large truncation index, $N$. Nollert~\cite{Nollert_PRD_1993} has shown that the "error" due to truncation can be minimised and the convergence of the method can be improved by a wise choice of the "remaining" part of the infinite continued fraction , $R_N=-\frac{a_{N+1}}{a_N}$, which in turn satisfies the recurrence equation,
\begin{equation}\label{ch2:eq_Nollert}
R_N=\frac{\gamma_{N+1}}{\beta_{N+1}-\alpha{N+1}R_{N+1}}.
\end{equation}
Assuming that $R_N$ can be expanded in a power series of $N^{-1/2}$,
\begin{equation}\label{ch2:eq_nollert_series}
R_N=\sum_{k=0}^{\infty}C_{k}~N^{-k/2},
\end{equation} 
we  obtain the first three coefficients $C_k$ as,\\
$C_0=-1$, $C_1=\sqrt{2i\left(r_{-}-r_{+}\right)\left(\omega^2-\mu^2\right)^{1/2}}$ and \\$C_2=-\frac{i \left(e q \omega -\mu ^2 M\right)}{\sqrt{\omega ^2-\mu ^2}}+2 i r_{+} \sqrt{\omega ^2-\mu ^2}+\frac{3}{4}$.

\subsection{Numerical Results}
We first study the fundamental QNMs due to uncharged massive scalar field then we  add electric charge to the perturbing field and study the effect of the scalar hair on the QNMs.
For the sake of numerical simplicity we scale the mass of the black hole to unity.
\section{Uncharged massive scalar field}
For an uncharged scalar field $(q=0)$ in the scalar-hairy RN background, we assume, without any loss of generality, the constant electric charge of the black hole to be zero,$(e=0)$. The complex function $ $ appearing in equation~(\ref{ch2:eq_trts}) can now be written as $W(\omega,r)=\omega^2-V(r)$ with
\begin{equation}
V(r)=f(r) \left(-\frac{2 s}{r^4}+\frac{2 M}{r^3}+\frac{(l+1) l}{r^2}+\mu ^2\right).
\end{equation}

\begin{figure*}
\centering
\includegraphics[width=1\textwidth, keepaspectratio]{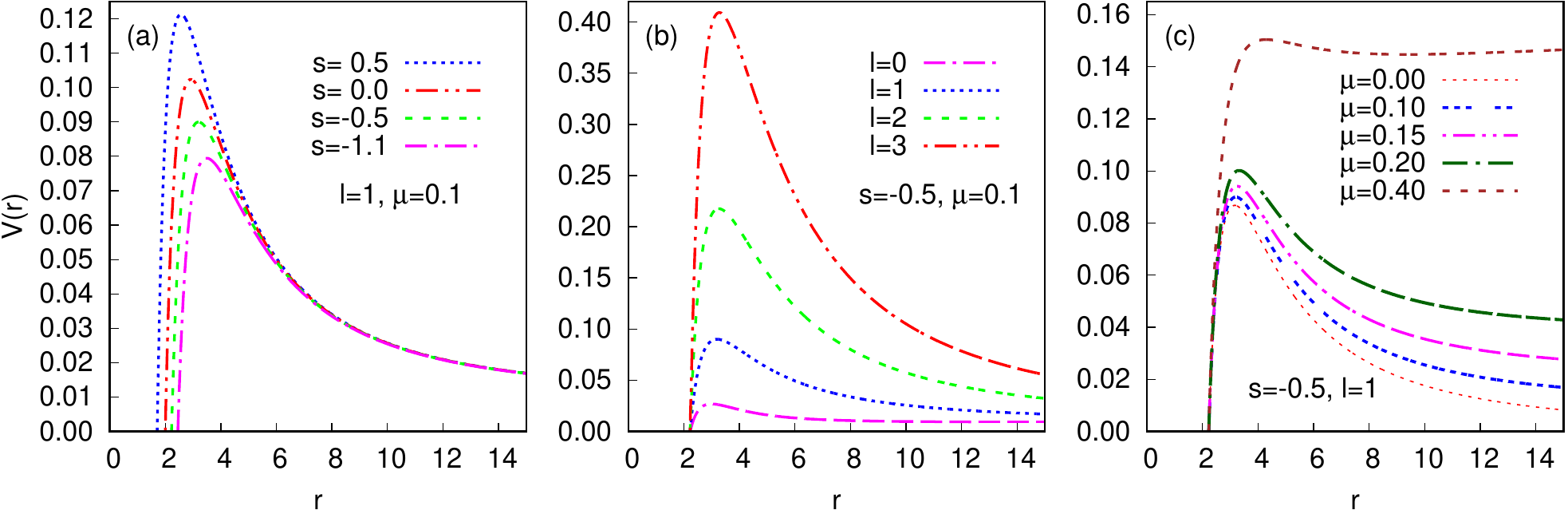}
\caption{Panel (a) shows the variation of $V(r)$ with $r$ for $l=1$ and $\mu=0.1$ for different values of $s$ as indicated. Panels (b) and (c) shows the variation of $V(r)$ with $r$ for $s=-0.5$, $\mu=0.1$ and for $s=-0.5$, $l=1$, respectively. Each curve in  (b) corresponds to a particular value of $l$ and each curve in (c) corresponds to a particular value of $\mu$ as indicated.}
\label{ch2:fig_1}
\vspace*{0.5cm}
\includegraphics[width=1\textwidth, keepaspectratio]{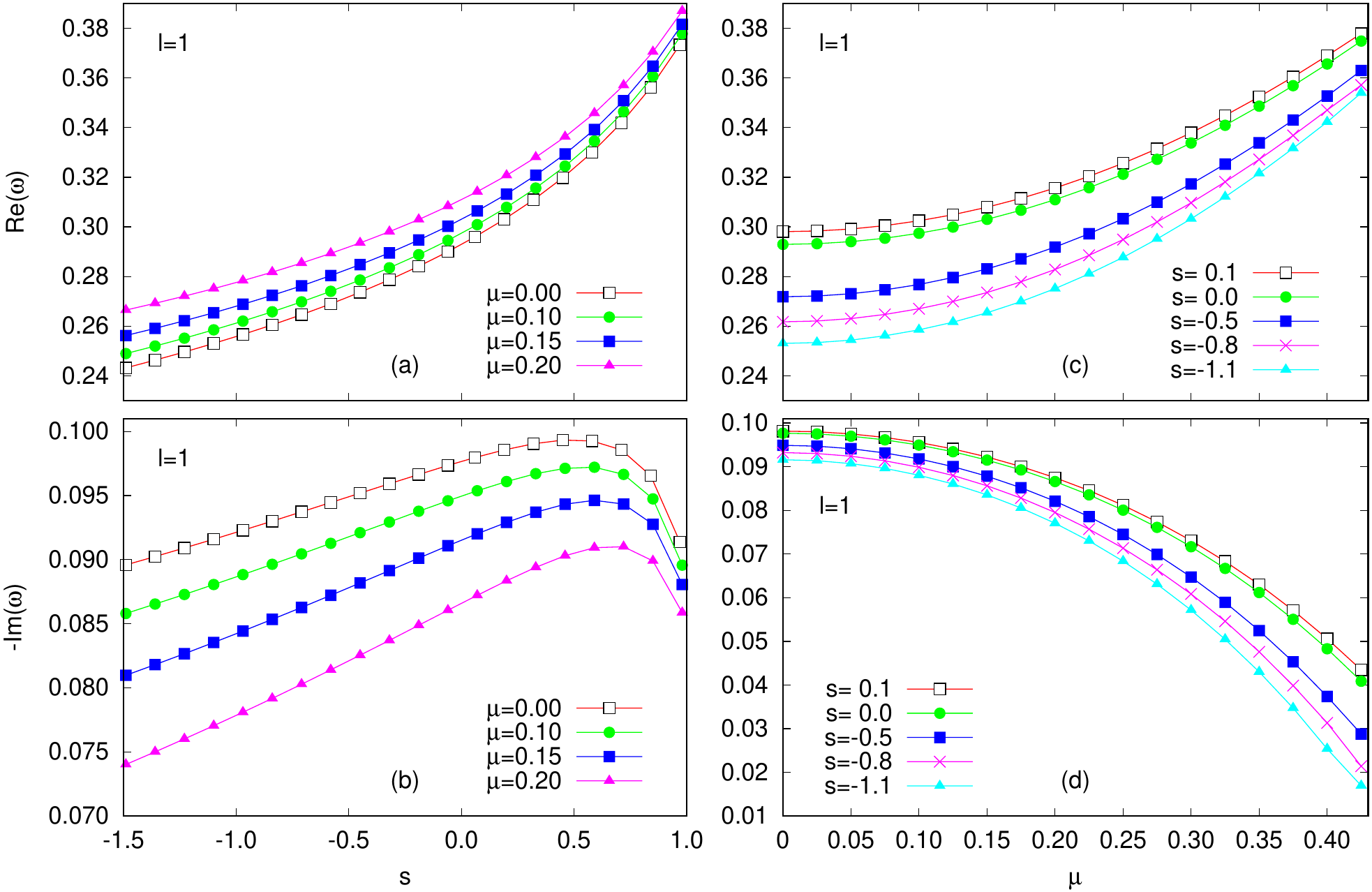}
\caption{Panels (a) and (b), respectively, show the real and imaginary parts of the fundamental (scalar) QN frequency as a function of $s$ for $l=1$ with each curve corresponding to a particular value of $\mu$ as indicated. Panels (c) and (d), respectively, show the real and imaginary parts of fundamental (scalar) QN frequency as a function of $\mu$ for $l=1$ with each curve corresponding to a particular value of s as indicated. The curve with $s=0.1$ corresponds to an RN black hole with $e\simeq 0.316$, while the curve with $s=0$ corresponds to the Schwarzschild black hole.}
\label{ch2:fig_2}
\end{figure*}

\begin{figure*}
\centering
\includegraphics[width=1\textwidth, keepaspectratio]{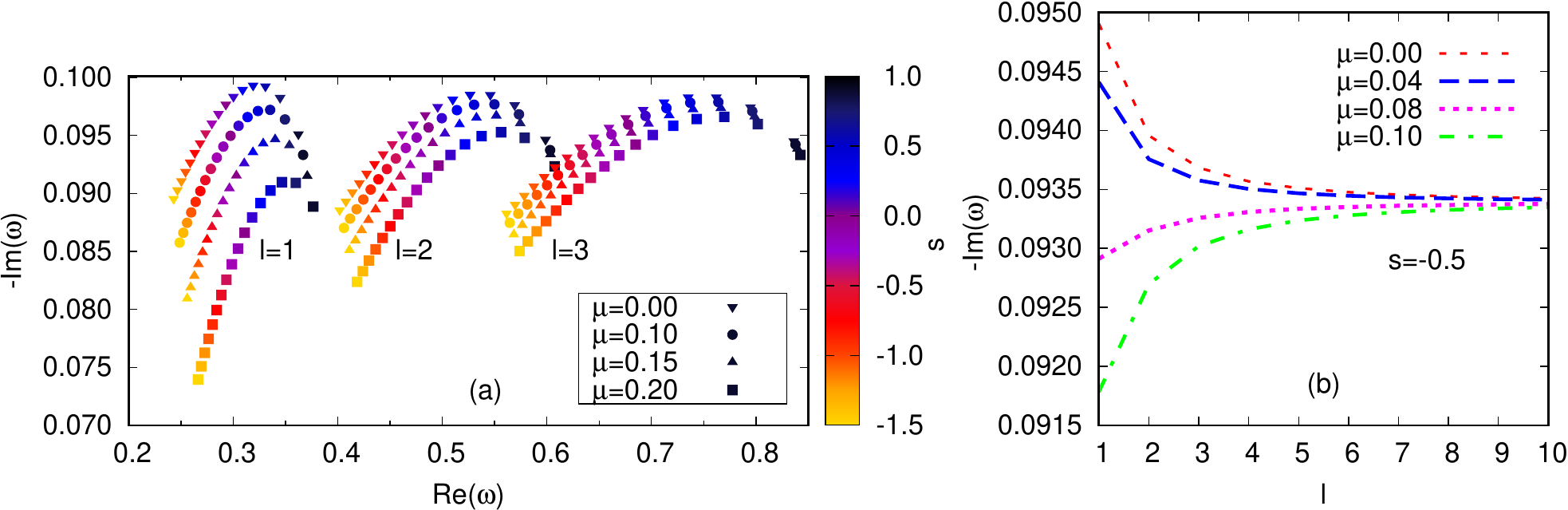}
\caption{Panel (a) shows the variation of the imaginary part of the fundamental (scalar) QN frequency with the Real part for different values of $s$ and $l$. For a given $l$, each curve corresponds to a particular value of $\mu$ as indicated. Panel (b) shows the imaginary part of the fundamental (scalar) QN frequency as a function of $l$ with $s=-0.5$ for different values of $\mu$ as indicated.}
\label{ch2:fig_3}
\vspace*{0.5cm}
\includegraphics[width=1\textwidth, keepaspectratio]{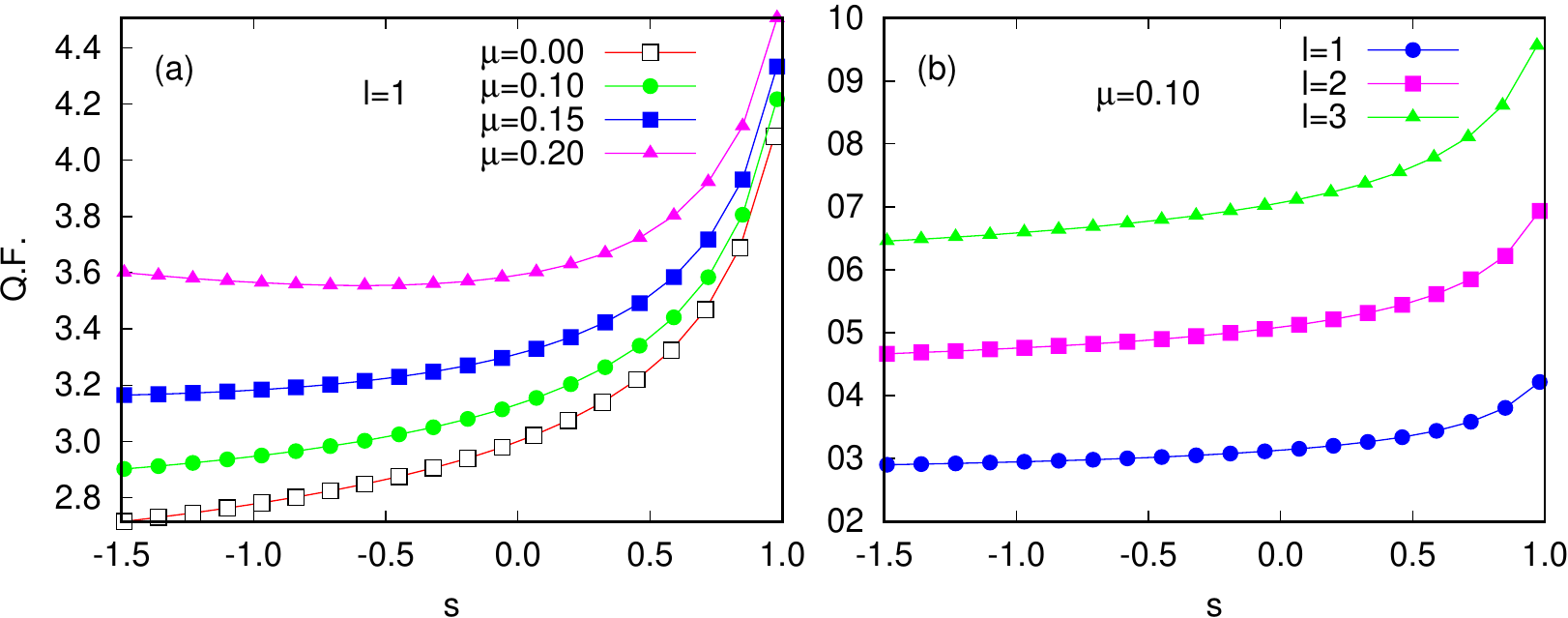}
\caption{Panels (a) and (b) show the quality factor as a function of $s$ with $l=1$ for different values of $\mu$ and with $\mu=0.1$ for different values of $l$, respectively. }
\label{ch2:fig_4}
\end{figure*}
The potential of the of the scalar-hairy RN black hole for different values of the scalar charge, field mass and multipole index are shown in Fig.~\ref{ch2:fig_1}.

In Figs.~\ref{ch2:fig_2}(a) and \ref{ch2:fig_2}(b), we show the behaviour of the real and imaginary parts of the fundamental QN frequency with the scalar charge for a particular multipole index $(l=1)$ and different field masses. We observe that for $s<0$ the magnitude of both the real and imaginary parts of the  QN frequency decrease with the absolute value of the scalar charge. This implies that the real oscillation frequency as well as the damping rate decrease with increasing magnitude of the negative scalar charge. For $s>0$, the spacetime~(\ref{ch2:eq_metric}) effectively behaves as an RN black hole of unit mass and electric charge, $e=\sqrt{s}$, showing a distinct peak in the magnitude of the imaginary part of the fundamental quasinormal frequency (see Refs.~\cite{konoplya_PRD_2002,konoplya_PLB_2002}). We also observe that for a particular value of the scalar charge the real part of the QN frequency increases with the field mass whereas the magnitude of the imaginary part decreases. This behaviour is manifested more clearly in Figs.~\ref{ch2:fig_2}(c) and \ref{ch2:fig_2}(d), where we observe that for sufficiently large field masses the imaginary part of the QN  frequency becomes vanishingly small. This results in long lived, purely real modes in the quasinormal spectrum, called {quasi-resonance modes}~\cite{ohashi_cqg_2004}.
We also note that as the scalar charge changes from positive to negative, {quasi-resonance} occurs at lower field masses with smaller real frequencies.

Fig.~\ref{ch2:fig_3}(a) shows a compact view of the behaviour of the real and imaginary parts of the QN frequency with the scalar charge for different values of the multipole number and field mass. We note that as the multipole number increases the real part of the fundamental QN frequency increases  and so does the imaginary part, but only for higher field masses. This behaviour of the imaginary part of the QN frequency can be seen more clearly in Fig.~\ref{ch2:fig_3}(b) where we note that for lower field masses the damping rate decreases with the multipole index whereas for higher field masses the damping rate increases with the multipole index. For large values of the multipole number, the damping rate is almost insensitive to the field mass.

Following Ref.~\cite{sayan_EPJC_2009} we define the {Quality Factor} as $Q.F. \sim \vert \frac{\omega_{Re}}{\omega_{Im}}\vert$. The quality factor is a measure of the product of the frequency and the ring down time of a black hole radiation, and is an important tool to figure out the black hole parameters~\cite{cardoso_thesis}. In Fig.~\ref{ch2:fig_4}(a), we observe that for a given multipole index and for large positive values of the scalar charge the quality factor decreases sharply, however for smaller values of the scalar charge it decreases very gradually. For massless field, the gradual decrease of the quality factor continues to persist for negative values of the scalar charge as well. However, beyond a certain value of the field mass $\mu$, it tends to increase for high negative values of $s$, the plot corresponding to $\mu=0.2$ in Fig.~\ref{ch2:fig_4}(a) reveals this feature. The quality factor has higher values for higher multipole indices (see Fig.~\ref{ch2:fig_4}(b)).

\section{Charged scalar field}
\begin{figure*}
\centering
\includegraphics[width=1\linewidth, keepaspectratio]{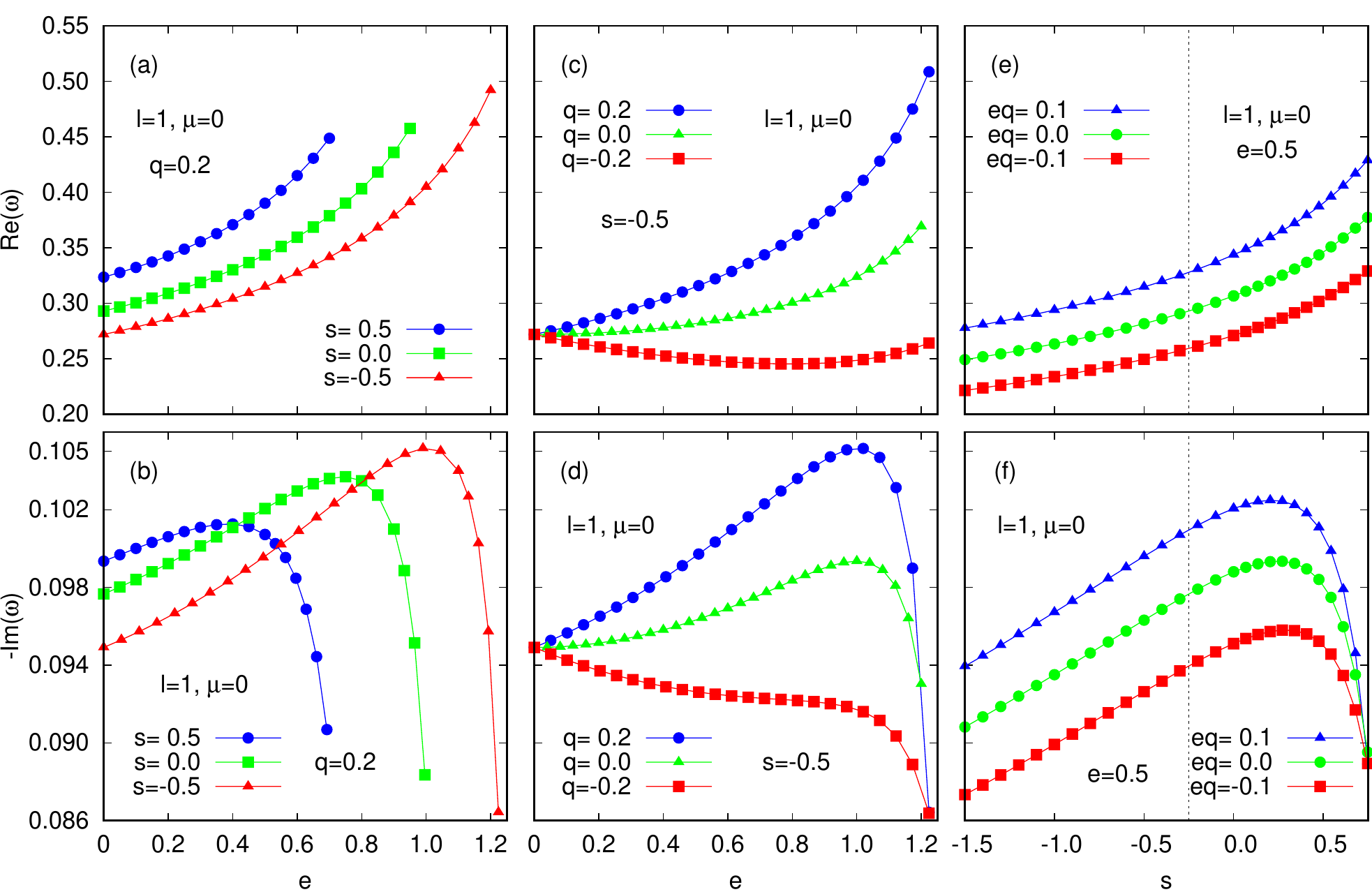}
\caption{Panels (a) and (b), respectively, show the real and imaginary parts of the fundamental (scalar) QN frequency as a function of $e$ for $l=1$, $\mu=0$ and $q=0.2$ with each curve corresponding to a particular value of $s$ as indicated. Panels (c) and (d), respectively, show the real and imaginary parts of the fundamental (scalar) QN frequency as a function of $e$ for $l=1$, $\mu=0$ and $s=-0.5$ with each curve corresponding to a particular value of $q$ as indicated. Panels (e) and (f), respectively, show the real and imaginary parts of the fundamental (scalar) QN frequency as a function of $s$ for $l=1$, $\mu=0$ and $e=0.5$ wih each curve corresponding to a particular value of $eq$ as indicated. The vertical line denotes the value of $s$ (=-0.25) below which the spacetime behaves as mutated RN.}
\label{ch2:fig_5}
\vspace*{0.5cm}
\includegraphics[width=1\textwidth, keepaspectratio]{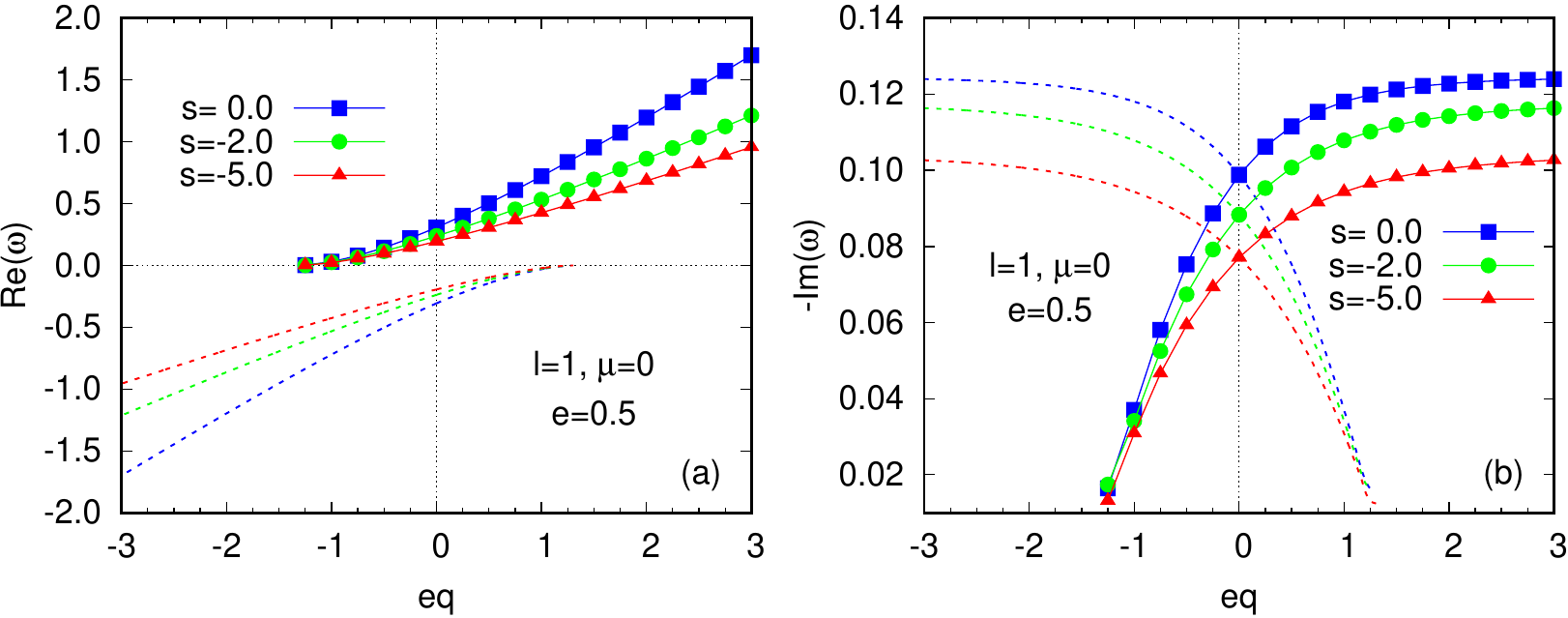}
\caption{Panels (a) and (b), respectively, show the real and imaginary parts of the fundamental (scalar) QN frequency as a function of $eq$ for $l=1$, $\mu=0$ and $e=0.5$. Each curve in each panel corresponds to a particular value of $s$ as indicated. The dashed lines represent the symmetric curves due to the simultaneous transformation $eq\rightarrow -eq$ and $\omega\rightarrow -\omega^{*}$.}
\label{ch2:fig_6}
\end{figure*}
The presence of the scalar hair changes the frequency and damping rate of the QN spectrum of  charged scalar fields as well. In Figs.~\ref{ch2:fig_5}(a) and \ref{ch2:fig_5}(b), we observe that compared to the RN black hole, for fixed non-zero values of $e$ and $q$, the magnitude of both the real and imaginary parts of the fundamental QN frequency are higher for positive values of the scalar charge and lower for negative values.

Konoplya~\cite{konoplya_PRD_2002} observed that for an RN black hole, the imaginary part of the QN frequency, for any given value of the field charge, approaches that for the uncharged field as the extremal limit is approached. Apart from a similar observation in the presence of a scalar hair (see Figs.~\ref{ch2:fig_5}(c) and \ref{ch2:fig_5}(d)) as well, we note from Fig.~\ref{ch2:fig_5}(f) that such convergence of the imaginary part of the fundamental QN frequency occurs for any given value of the black hole electric charge, as the  maximal value of the scalar charge is approached.  This maximal value is determined by the extremality condition~(\ref{ch2:eq_extremality}). We further observe that, for $s>-e^2$, the magnitude of the imaginary part of the fundamental QN frequency shows a distinct peak whereas for $s<-e^2$ it decreases monotonically with the magnitude of $s$. The corresponding behaviour of the real part of the QN frequency with the scalar charge is shown in Fig.~\ref{ch2:fig_5}(e).

The symmetry of the QNMs with respect to the transformation $\left(eq\rightarrow -eq, \omega\rightarrow -\omega^{*}\right)$ is depicted in Fig.~\ref{ch2:fig_6}.  Fig.~\ref{ch2:fig_6}(a) also highlights the existence of a critical value of $|eq|$ at which the real part of the QN frequency vanishes. However such a behaviour of the QN frequency is not new and has been previously observed for the RN black hole for charged scalar and Dirac fields (see Refs.~\cite{konoplya_PRD_2013,richartz_PRD_2014}). We note in particular, that the critical value of $|eq|$ is almost unaffected by the presence of the scalar hair and does not change with the black hole electric charge. For an RN black hole with unit multipole index, the critical value is $|eq|\approx1.3$.
\section{Charged Dirac field around a charged black hole with scalar hair}
The dynamics of a massless charged Dirac field propagating in the scalar-hairy RN spacetime is given  by the Dirac equation,
\begin{equation}
\gamma^{\mu}D_{\mu}\Psi=0,
\end{equation} 
where $\Psi$ is the Dirac four-spinor, $\gamma^{\mu}$ are the coordinate dependent Dirac four-matrices and $D_{\mu}$ is spinor covariant derivative defined by,
\begin{equation}
D_{\mu}=\partial_{\mu}-\Gamma_{\mu}-i q A_{\mu}.
\end{equation}
Here $q$ is the charge of the Dirac field and $\Gamma_{\mu}$ are the spinor connection matrices. Following Refs.~\cite{huang_PRD_2017,dolan_CQG_2015} we decompose the Dirac four-spinor as 
\begin{equation}
\Psi=\frac{1}{\sqrt{r\sqrt{\Delta}}}\left(\begin{array}{c} -Q(r)S_1(\theta) \\ -P(r)S_2(\theta) \\  P(r)S_1(\theta) \\ Q(r)S_2(\theta) \end{array}\right) e^{i (m \phi-\omega t)}
\end{equation}
 and use the canonical orthonormal (symmetric) tetrad proposed by Carter~\cite{carter_CMP_1968} to yield two pairs of coupled first order differential equation,
\begin{eqnarray}\label{ch2:eq_radial_dirac1}
\sqrt{\Delta}\left(\frac{d}{dr}-\frac{i K}{\Delta}\right) P=\lambda Q,\label{ch2:eq_radial_dirac1}\\
\sqrt{\Delta}\left(\frac{d}{dr}+\frac{i K}{\Delta}\right) Q=\lambda P,\label{ch2:eq_radial_dirac2} 
\end{eqnarray}
and
\begin{eqnarray}
\left(\frac{d}{d \theta}+\frac{1}{2}\cot{\theta}-m \csc{\theta}\right)S_1=\lambda S_2,\\
\left(\frac{d}{d \theta}+\frac{1}{2}\cot{\theta}+m \csc{\theta}\right)S_2=\lambda S_1,
\end{eqnarray}	
where, $K=\omega r^2-e q r$,  $-j\leq m\leq j$ and $\lambda=j+1/2$ $($ with $j=1/2,3/2...)$ is the separation constant.
The radial equations~(\ref{ch2:eq_radial_dirac1},\ref{ch2:eq_radial_dirac2}) can then be 
combined to yield,
\begin{equation}
\begin{split}
\sqrt{\Delta}&\frac{d}{dr}\left(\sqrt{\Delta}\frac{dP}{dr}\right)+\left(\frac{K^2+i (r-M)K}{\Delta}-2i \omega r+i e q-\lambda^2 \right)P=0.
\label{ch2:eq_combinedradial_dirac}
\end{split}
\end{equation}
If we define a new radial function, $\xi=\Delta^{-1/4}r P$, then equation~(\ref{ch2:eq_combinedradial_dirac}) can be written in a Schr\"{o}dinger like form in terms of the tortoise coordinate as
\begin{equation}
\frac{d^2\xi}{dr_{*}^2}+\tilde{W}(\omega,r)=0,
\end{equation}
where
\begin{equation}
\label{ch2:eq_trts_dirac}
\begin{split}
\tilde{W}\left(\omega, r\right)= \frac{\Delta}{r^4}\left[\frac{\left(K+\frac{i}{2}\left(r-M\right)\right)^2}{\Delta}-2i \omega r+i e q\right.
\\\left.-\lambda^2-\frac{2M}{r}+\frac{2(e^2+s)}{r^2}\right].
\end{split}
\end{equation}
In the asymptotic limits of the tortoise coordinate, equation~(\ref{ch2:eq_trts_dirac}) can be solved analytically with the QNM boundary conditions, yielding
\begin{equation}
\label{ch2:eq_bc_dirac}
\xi \approx 
\begin{cases}
 e^{\frac{1}{4}\frac{\left(r_{+}-r_{-}\right)}{r_{+}^2}r_{*}-i \left(\omega-\frac{e q}{r_{+}} \right) r_*}&\mbox{ as \hspace*{2mm}}r_* \rightarrow -\infty \\
r_{*}^{\frac{1}{2}-i e q} e^{i\omega r_{*}}&\mbox{ as \hspace*{2mm}}r_* \rightarrow \infty.
\end{cases}
\end{equation} 
Equation~(\ref{ch2:eq_combinedradial_dirac}), similar to equation~(\ref{ch2:eq_radialKG}), also has two regular singularities at $r_{+}$ and $r_{-}$ and an irregular singularity as $r\rightarrow\infty$. So proceeding as before we introduce an ansatz, consistent with the boundary conditions~(\ref{ch2:eq_bc_dirac}),
\begin{equation}\label{ch2:eq_dirac_ansatz}
P=e^{i \omega r} (r-r_{-})^{\tilde{\rho}}\sum_{n=0}^{\infty}b_n u^{n+\tilde{\delta}},
\end{equation}
where  $u=\frac{r-r_{+}}{r-r_{-}}$,	 $\tilde{\rho}=-ieq+i\omega\left(r_{+}+r_{-}\right)$ and $\tilde{\delta} = \frac{1}{2}-\frac{i r_{+}^2 \left(\omega -\frac{e q}{r_{+}}\right)}{r_{+}-r_{-}}$.
Plugging (\ref{ch2:eq_dirac_ansatz})  back into equation~(\ref{ch2:eq_combinedradial_dirac}) we again arrive at the three term recurrence relations,
\begin{eqnarray}
\tilde{\alpha}_0 b_1 +\tilde{\beta}_0 b_0 & = & 0,\\
\tilde{\alpha}_n b_{n+1} +\tilde{\beta}_n b_n+\tilde{\gamma}_n b_{n-1} & = & 0,
\end{eqnarray}
\begin{figure*}
\centering
\includegraphics[width=1\textwidth, keepaspectratio]{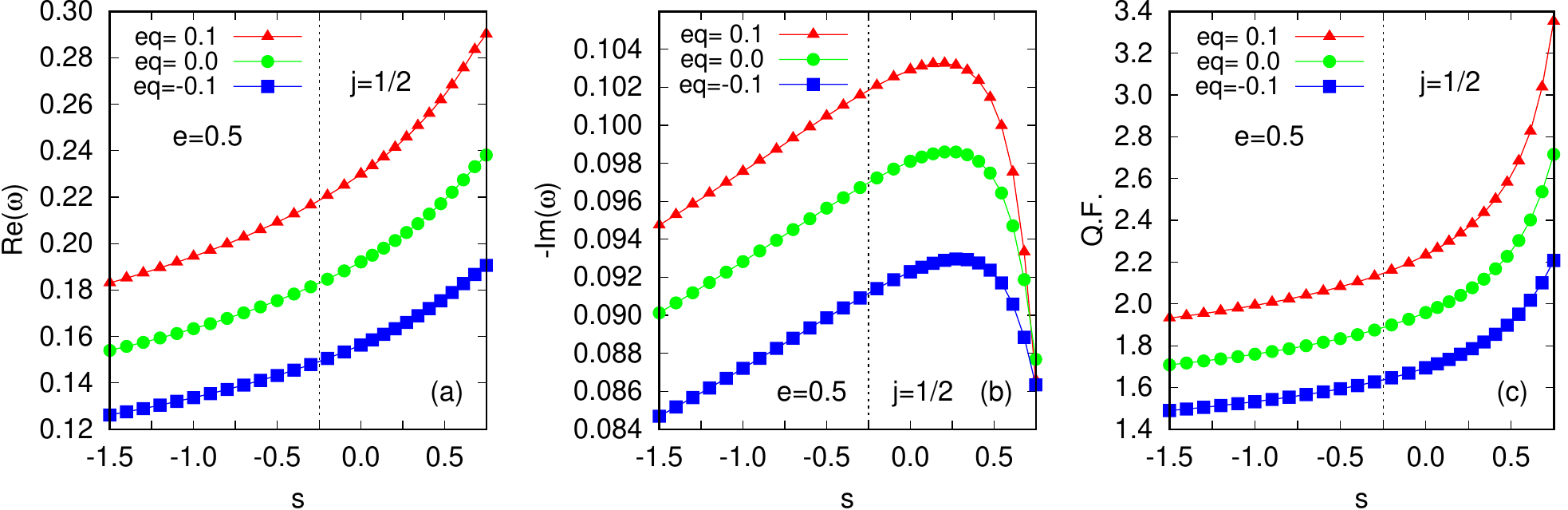}
\caption{Panels (a), (b) and (c), respectively, show the real and imaginary parts of the fundamental (Dirac) QN frequency and the quality factor as a function of $s$ for $j=1/2$ and $e=0.5$. Each curve in each panel corresponds to a particular value of $eq$ as indicated. The curve for $eq=0$ represents the perturbation by an uncharged Dirac field.  The vertical line denotes the value of $s$ (=-0.25) below which the spacetime behaves as mutated RN.}
\label{ch2:fig_7}
\vspace*{0.5cm}
\includegraphics[width=1\textwidth, keepaspectratio]{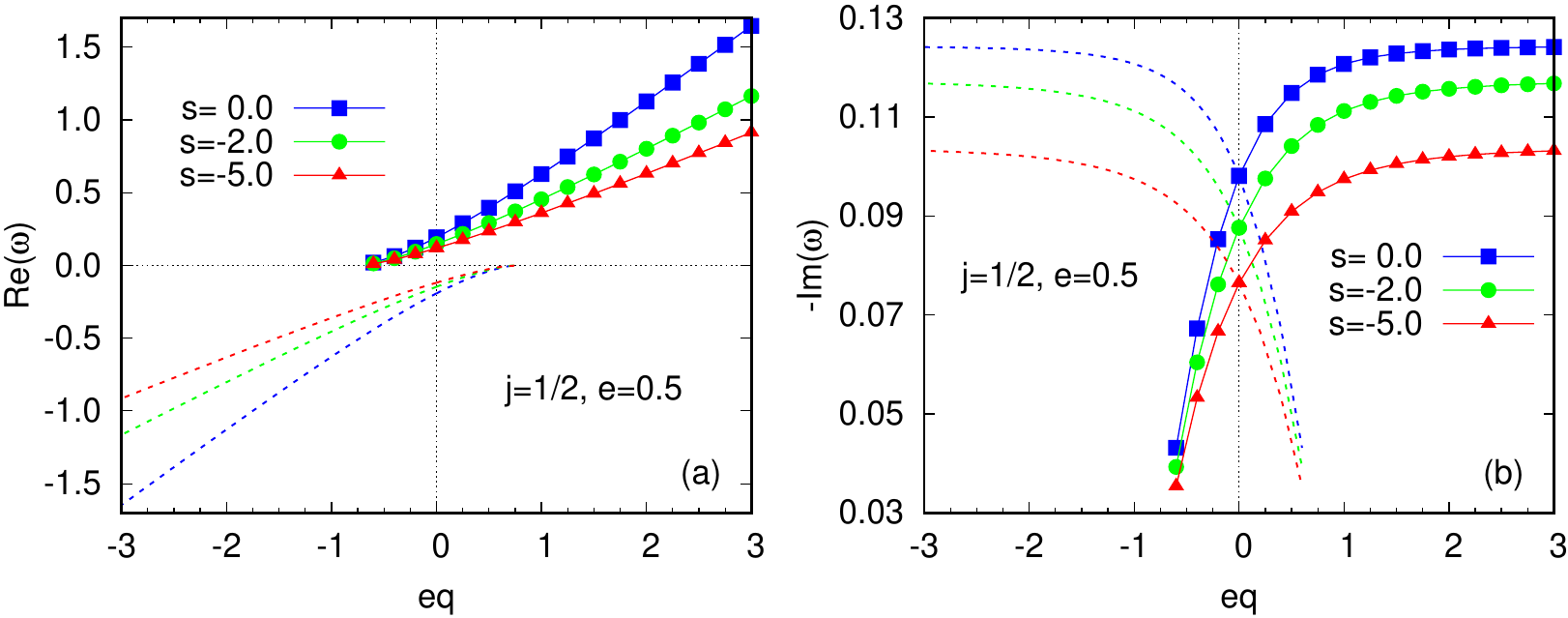}
\caption{Panels (a) and (b), respectively, show the real and imaginary parts of the fundamental (Dirac) QN frequency as a function of $eq$ for $j=1/2$ and $e=0.5$. Each curve in each panel corresponds to a particular value of $s$ as indicated. The dashed lines represent the symmetric curves due to the simultaneous transformation $eq\rightarrow -eq$ and $\omega\rightarrow -\omega^{*}$.}
\label{ch2:fig_8}
\end{figure*}
where
\begin{align}
\centering
\tilde{\alpha}_n&=(n+1) \left[\frac{1}{2} (2 n+3)+\frac{2 i r_{+} (e q-r_{+} \omega )}{r_{+}-r_{-}}\right],\\
\tilde{\beta}_n&=
\begin{aligned}[t]
&-\frac{r_{+}}{r_{+}-r_{-}} \bigg[-4 e^2 q^2-4 i r_{+} \omega  (3 i e q+2 n+1)\\
&\left.+6 i e n q+3 i e q+\lambda^2 +2 n^2+2 n-8 r_{+}^2 \omega ^2+\frac{1}{2}\right]\\
&+\frac{r_{-}\left[2 n (i e q+n+1)+i e q+\lambda^2 +\frac{1}{2}\right]-2 i (2 n+1) r_{-} r_{+} \omega}{r_{+}-r_{-}},
\end{aligned}\\[\jot]
\tilde{\gamma}_n&=
\begin{aligned}[t]
&-\frac{n+2 i (e q-\omega  (r_{-}+r_{+}))}{2 (r_{+}-r_{-})}[(2 n-1) r_{-}+r_{+} (-4 i e q-2 n+4 i r_{+} \omega +1)].
\end{aligned}
\end{align}
The convergence of the series~(\ref{ch2:eq_dirac_ansatz}) demands the recurrence  coefficients to satisfy an infinite continued fraction relation similar to equations~(\ref{ch2:eq_scalar_CF},\ref{ch2:eq_scalar_CF_inv}).\\
Applying Nollert's improvement, we now get the first few coefficients of the series~(\ref{ch2:eq_nollert_series}) as,\\
$C_0=-1$, $C_1=\sqrt{2i\omega\left(r_{-}-r_{+}\right)}$ and $C_2=\frac{5}{4}-i e q+2 i \omega r_{+}$.

\subsection{Numerical Results}
The behaviour of the real and imaginary parts of the fundamental QN frequency with the scalar charge is shown in Figs.~\ref{ch2:fig_7}(a) and \ref{ch2:fig_7}(b). As before, we observe that for a fixed  value of the black hole electric charge, the magnitude of the imaginary part of the QN frequency for a given value of the field charge $q$, increases as the scalar charge changes from negative to positive and ultimately approaches the neutral one in the extremal limit $\left(s_{extremal}={M^2-e^2}\right)$. Thus, in the extremal limit the damping rate is independent of $s$. Here also we observe a peak in the magnitude of the imaginary part of the fundamental QN frequency for $s>-e^2$ whereas for $s<-e^2$, it decreases monotonically with the magnitude of the scalar charge. The real QN frequency on the other hand continues to grow with the scalar charge. Away from the extremal value of the scalar charge this causes the quality factor to grow slowly but steadily (see Fig.~\ref{ch2:fig_7}(c)), however as the extremal value of the scalar charge is reached the growth of the quality factor becomes quite rapid. 

Similar to the scalar {perturbation}, we observe in Fig.~\ref{ch2:fig_8} that the QN frequency is symmetric with respect to the transformation $\left(eq\rightarrow -eq, \omega\rightarrow -\omega^{*}\right)$   and note that the critical value of electromagnetic interaction $(|eq|=0.7)$ at which the real part of the QN frequency vanishes, is almost unaffected by the presence of the scalar hair.

\section{Summary and Discussion}\label{ch2:sec_sum}

In the present chapter, we discussed the quasinormal mode spectrum of massless and massive uncharged as well as charged scalar field and massless charged Dirac field in the vicinity of a scalar-hairy Reissner-Nordstr\"{o}m black hole. We mainly focussed on negative values of the scalar charge with $s<-e^2$, for which the metric~(\ref{ch2:eq_metric}) represents  a "mutated RN" spacetime mimicking the Einstein-Rosen bridge. It must be noted that the perturbing scalar field $\Phi$~ in Eq.~\eqref{ch2:eq_KG} is distinct from the  scalar field $\psi$ in Eq.~\eqref{ch1:eq_scalarfield} associated with the scalar hair.

Unlike the appearance of a distinct peak in the magnitude of the imaginary part of the fundamental QN frequency of a scalar-hairy RN black hole for scalar and Dirac fields with $s>-e^2$, the mutated RN spacetime $(s<-e^2)$ is characterised by monotonically decreasing $|Im(\omega)|$ (see Figs.\ref{ch2:fig_2}(b), \ref{ch2:fig_5}(f), \ref{ch2:fig_7}(a) and \ref{ch2:fig_7}(b)). For uncharged test fields, the scalar-hairy RN black hole effectively behaves as an RN black hole with effective electric charge; $Q=\sqrt{e^2+s}$ provided $s$ lies in $M-e^2\geq s> -e^2$.

For massive scalar field, the phenomenon of {quasi-resonance}, characterised by vanishingly small $|Im(\omega)|$, is observed. We also showed the behaviour of the quality factor with the scalar charge for both the scalar and Dirac fields. As the extremal limit is approached either by increasing the electric charge for a fixed $s$ or vice-versa, we find that the imaginary  part of $\omega$ for neutral and charged scalar or Dirac perturbations to be coincident. 

In the presence of electric charge of the perturbing fields, we observe the existence of a critical value of $|eq|$, above which the real part of the QN frequency vanishes, for both the scalar and Dirac fields. This value is completely unaffected by the presence of  scalar hair.

It should be pointed out that we also calculated the area spectrum of the scalar-hairy RN black hole based on the proposals of Kunstatter~\cite{kunstatter_PRL_2003} and Maggiore~\cite{maggiore_PRL_2008}, starting from the asymptotic form of the Dirac QN frequency as suggested by Cho~\cite{cho_PRD_2006}. We obtain the area quantum as $\Delta A=8 \pi \hbar$. This being the same as that of an RN black hole~\cite{ortega_CQG_2011}, we refrain from including a detailed calculation of the same.

Recently Saleh, Thomas and Kofane~\cite{saleh_ASS_2014,saleh_ASS_2016} discussed the QN spectrum of massless uncharged scalar and Dirac fields in the vicinity of a "quantum-corrected" Schwarz\-schild black hole~\cite{kim_PLB_2012} using 3rd order WKB approximation. The metric used by them is effectively similar to that of the mutated RN spacetime discussed in the present work. The results obtained by us for the massless uncharged scalar and Dirac fields in the scalar-hairy RN background (with $s<-e^2$) using the more accurate continued fraction method, is qualitatively similar to them. The work presented in this chapter is, however, much more general as it includes charge for both the scalar and the Dirac fields and mass for the scalar field.

The QN spectrum analysis was also carried out with the 3rd order WKB approximation which is generally believed to be less accurate. The table below compares the results for one example, namely that for an uncharged massless scalar perturbation of the scalar-hairy RN black hole. The WKB approximation is known to yield more and more accurate results for higher and higher values of multipoles ($l$). The table contains the values given by the Leaver method and that by the WKB approximation for the real and the imaginary parts of the quasinormal modes for $l=1, 2, 3$. It is apparent from the table that the difference in the results given by the two methods reduces for higher values of $l$.

We point out that in the present work we have not observed any QNM with positive imaginary part indicating the stability of the scalar-hairy RN black hole under  massive (and massless)   charged (and uncharged)  scalar perturbations as well as under massless charged (and uncharged) Dirac perturbations for both $s$ in $[-e^2,M^2-e^2)$ and in the ``mutated" regime,  $s<-e^2$. This implies that the mutated RN spacetime is also stable under all the above mentioned types of perturbation.

Finally, we mention that the qualitative features for the quasinormal modes for the charged or uncharged massive scalar fields and also that for the charged Dirac field for a mutated RN (scalar-hairy RN with $s<-e^2$) background are qualitatively same as that for a usual RN black hole except for the complete monotonic behaviour of the damping (imaginary part of the QNM) in the case of  the former as opposed to the existence of a peak in the latter case, an RN black hole.
\begin{table}[!h]
\begin{center}
\caption{Fundamental QN frequencies of massless uncharged scalar fields in the background of a scalar-hairy RN black hole of mass, $M=1$ and electric charge, $e=0$ for different values of the multipole index and scalar charge. For each value of $s$, the first line is obtained using the contined fraction method with $175$ terms and the second line is obtained using the 3rd order WKB approximation.}
\begin{adjustbox}{width=\textwidth}
\begin{tabular}{ccccccc} 
\hline\hline
\multicolumn{3}{c}{\hspace{0.065\textwidth}$l=1$}                                                                                                         & \multicolumn{2}{c}{$l=2$}                                                                           & \multicolumn{2}{c}{$l=3$}                                                                            \\ 
\hline
s& $Re(\omega)$&$Im(\omega)$&$Re(\omega)$&$Im(\omega)$&$Re(\omega)$&$Im(\omega)$\\
\hline
\multirow{2}{*}{0.99\quad}       & 0.3762055912                                      & -0.0900896009                                     & 0.6237563062                                      & -0.0895373909                                     & 0.8720041949                                      & -0.0893846794                                      \\
                            & 0.3742670229                                      & -0.0900696201                                     & 0.6232727045                                      & -0.0895182148                                     & 0.8718206816                                      & -0.0893779664                                      \\
\multirow{2}{*}{0.9\quad}        & 0.3637066222                                      & -0.0947427505                                     & 0.6009488609                                      & -0.0942904001                                     & 0.8393317935                                      & -0.0941634988                                      \\
                            & 0.3618775351                                      & -0.0947062218                                     & 0.6004948731                                      & -0.0942713464                                     & 0.8391598515                                      & -0.0941570756                                      \\
\multirow{2}{*}{0.7\quad}        & 0.3408752613                                      & -0.0986496604                                     & 0.5624223918                                      & -0.0979828596                                     & 0.7852194051                                      & -0.0977944238                                      \\
                            & 0.3393237655                                      & -0.0987271148                                     & 0.5620307364                                      & -0.0979828452                                     & 0.7850707716                                      & -0.0977934267                                      \\
\multirow{2}{*}{0.5\quad}        & 0.3235342925                                      & -0.0993515868                                     & 0.533818086                                       & -0.0985724807                                     & 0.7452888813                                      & -0.0983503413                                      \\
                            & 0.3219858221                                      & -0.0995455106                                     & 0.5334311042                                      & -0.0985914448                                     & 0.7451428561                                      & -0.098354732                                       \\
\multirow{2}{*}{0.3\quad}        & 0.3096415907                                      & -0.0989749919                                     & 0.5110162987                                      & -0.0981311196                                     & 0.7135039562                                      & -0.0978893749                                      \\
                            & 0.307999452                                       & -0.0992457629                                     & 0.5106143307                                      & -0.0981635738                                     & 0.7133535414                                      & -0.0978977124                                      \\
\multirow{2}{*}{0.1\quad}        & 0.2980681332                                      & -0.0981534347                                     & 0.4920504262                                      & -0.0972678241                                     & 0.6870780311                                      & -0.0970134153                                      \\
                            & 0.2963068621                                      & -0.0984754655                                     & 0.491628113                                       & -0.097310014                                      & 0.6869212869                                      & -0.0970246966                                      \\
\multirow{2}{*}{0\quad}          & 0.2929361333                                      & -0.0976599889                                     & 0.4836438722                                      & -0.096758776                                      & 0.6753662325                                      & -0.0964996277                                      \\
                            & 0.2911141164                                      & -0.0980013631                                     & 0.4832110304                                      & -0.0968048549                                     & 0.675206178                                       & -0.0965121143                                      \\
\multirow{2}{*}{-0.1\quad}       & 0.2881615316                                      & -0.0971353427                                     & 0.4758233999                                      & -0.0962210405                                     & 0.6644712038                                      & -0.0959579129                                      \\
                            & 0.2862799244                                      & -0.0974930215                                     & 0.4753801758                                      & -0.0962705126                                     & 0.6643078766                                      & -0.0959714677                                      \\
\multirow{2}{*}{-0.3\quad}       & 0.2795122233                                      & -0.0960340955                                     & 0.4616560916                                      & -0.0950993038                                     & 0.6447339818                                      & -0.0948299372                                      \\
                            & 0.2775174544                                      & -0.0964175448                                     & 0.4611930519                                      & -0.0951543929                                     & 0.6445644025                                      & -0.0948453003                                      \\
\multirow{2}{*}{-0.5\quad}       & 0.2718459161                                      & -0.0949062659                                     & 0.4490967969                                      & -0.0939565158                                     & 0.6272360885                                      & -0.0936825807                                      \\
                            & 0.2697477256                                      & -0.0953089565                                     & 0.4486155668                                      & -0.0940160429                                     & 0.6270607761                                      & -0.093699416                                       \\
\multirow{2}{*}{-0.7\quad}       & 0.2649696261                                      & -0.0937815642                                     & 0.4378289057                                      & -0.0928207752                                     & 0.6115362336                                      & -0.0925434544                                      \\
                            & 0.2627780753                                      & -0.0941990015                                     & 0.4373312051                                      & -0.092883877                                      & 0.6113557438                                      & -0.0925615095                                      \\
\multirow{2}{*}{-0.9\quad}       & 0.2587422561                                      & -0.0926759376                                     & 0.4276214399                                      & -0.0917069907                                     & 0.5973126872                                      & -0.0914271526                                      \\
                            & 0.2564668676                                      & -0.0931049246                                     & 0.4271089136                                      & -0.0917730128                                     & 0.597127553                                       & -0.0914462318                                      \\
\multirow{2}{*}{-1\quad}         & 0.2558376364                                      & -0.0921331191                                     & 0.4228593181                                      & -0.0911609447                                     & 0.5906764879                                      & -0.0908801044                                      \\
                            & 0.2535236413                                      & -0.0925669597                                     & 0.4223399545                                      & -0.0912282303                                     & 0.5904892182                                      & -0.0908996369                                      \\
\multirow{2}{*}{-1.1\quad}       & 0.2530574278                                      & -0.0915978709                                     & 0.4183004849                                      & -0.0906229436                                     & 0.5843232974                                      & -0.0903412442                                      \\
                            & 0.2507068851                                      & -0.0920360624                                     & 0.4177746428                                      & -0.0906913813                                     & 0.5841340084                                      & -0.0903611949                                      \\
\hline\hline
\end{tabular}
\end{adjustbox}
\end{center}
\end{table}


\chapter{Superradiant stability of Mutated Reissner-Nordstr\"{o}m black holes}\blfootnote{\begin{flushleft} The work presented in this chapter is based on “Superradiant stability of mutated Reissner–Nordström black holes”, \textbf{Avijit Chowdhury} and Narayan Banerjee, Gen.\ Rel.\ Grav.\  \textbf{51}, 99 (2019).\end{flushleft}}  
\chaptermark{Superradiant stability of Mutated RN black holes}
\label{chap3}

\section{Introduction}
The long-awaited detection of gravity waves came from violent phenomena of the merger of two compact objects~\cite{ligo_PRL_2016,abott_PRL_2017, abott_GWTC_2018} and paved the way of a gravity wave astronomy. But radiations can emerge from black holes also from more sedate perturbations to start with, some of which may finally result in other kinds of spectacular events like a `black hole bomb'~\cite{press_Nature_1972}.
As discussed in section~\ref{ch1:sec_superrad}, when a bosonic wave of frequency less than a critical value, impinges upon a rotating black hole, the reflected wave gets amplified by extracting energy from the rotation of the black hole~\cite{zeldovich_JETPL_1971,zeldovich_JETP_1972,vilenkin_PLB_1978}. This critical frequency is known as the superradiance frequency. For Reissner-Nordstr\"{o}m (RN) black holes, such an amplification is possible for charged bosonic waves~\cite{bekenstein_PRD_1973}. If this superradiantly amplified wave is reflected back into  the black hole by a potential barrier, the reflected wave is further amplified. Repeated reflections of the amplified wave ultimately result in superradiant instability of the black hole, popularly known as the ``black hole bomb''~\cite{press_Nature_1972}. The mass ($\mu$) of the incident bosonic field needs to be more than the mode frequency ($\mu^2>\omega^2$) for producing a local minimum in the effective potential resulting in bound states of the bosonic wave and thus an instability~\cite{damour_LNC_1976, cardoso_PRD_2004, cardoso_PRD_2004_err}.

Reissner-Nordstr\"{o}m black holes are known to be superradiantly stable against perturbation by massive charged scalar fields~\cite{hod_PLB_2012,hod_PLB_2013,hod_PRD_2015,huang_EPJC_2016} in the entire parameter space and does not lead to a black hole bomb, whereas, rotating black holes (Kerr black holes) are not stable against perturbations by massive bosonic fields~\cite{vilenkin_PLB_1978,press_Nature_1972, press_APJ_1973,zouros_AP_1979,detweiler_PRD_1980,dolan_PRD_2007,hod_PRD_2010,beyer_JMP_2011,myung_PRD_2011,hod_PLB_2012_1,brito_PRD_2013,dolan_PRD_2013,witek_PRD_2013, cardoso_GRG_2013,okawa_PRD_2014,arderucio_ARXIV_2014}. Again, extremal brane-world Reissner-Nordström black holes are superradiantly unstable against massive charged scalar perturbation~\cite{zhang_ARXIV_2013}.

Kerr-Newman black holes are also superradiantly unstable against perturbation by massive charged scalar fields in a range of frequencies. {The range is given by $\xi(\mu, q)<\omega<\omega_c<\mu$ or $\xi(\mu, q)<\omega<\mu<\omega_c$ (see Refs.~\cite{huang_PRD_2016,huang_PRD_2018}), where $\xi(\mu,q)\equiv \frac{e q}{4 M}+\sqrt{\frac{\mu^2}{2}+\frac{e^2 q^2}{16 M^2}}$ and $\omega_c=m \Omega_H + q \Phi_H$ is the superradiance frequency. $\Omega_H$ and $\Phi_H$ are the horizon angular velocity and the electric potential at the horizon respectively and $m$ is the azimuthal harmonic index. }$M$ and $e$ are the mass and electric charge of the black hole, whereas, $\mu$ and $q$ are the mass and electric charge of the perturbing scalar field. 

At the superradiance frequency $\omega_c$, the imaginary part of $\omega$ vanishes and thus the modes do not grow or decay in time. These long-lived modes are dubbed as scalar clouds. Herdeiro and Radu~\cite{herdeiro_PRL_2014} studied the superradiance of massive scalar fields in the background of a rotating black hole. They obtained a numerical solution describing a rotating black hole with a complex scalar field. Gravitational wave signals from ultra-light bosonic clouds around rotating black holes have been studied in detail in Ref.~\cite{brito_PRL_2017, brito_PRD_2017, arvanitaki_PRD_2015, arvanitaki_PRD_2017}. Scalar clouds in the context of no-hair theorems has been studied in detail in~\cite{garcia_PRD_2019}.

Charged spherical black holes in string theory are also superradiantly stable against massive charged scalar perturbation~\cite{li_PRD_2013}. The recent work by Tokgoz~\cite{tokgoz_ARXIV_2019} on dilaton black holes is also referred to in this context. Superradiant stability of dilaton-axion black holes under scalar perturbation has been verified in~\cite{ghosh_EPL_2017}. Superradiance, in the context of scalar tensor gravity has been discussed by Cardoso, Carucci, Pani and Sotiriou~\cite{cardoso_PRL_2013, cardoso_PRD_2013}. 

In this chapter, we explore the existence of superradiance and the stability of the scalar-hairy Reissner-Nordstr\"{o}m (scalar-hairy RN)  black hole (Eq.~\eqref{ch1:eq_shRN}) against superradiance. We will particularly concentrate on the $s<-e^2$ regime, where the scalar-hairy RN black hole behaves as a ``mutated Reissner-Nordstr\"{o}m'' spacetime leading to an Einstein-Rosen bridge~\cite{rosen_PR_1935}. The idea is to check whether the superradiance condition and the bound state condition are satisfied simultaneously. We find that they do not, and thus such a system is stable against superradiance. This is firmly established for a large mass of the test field, and is very strongly indicated for smaller masses as well.

\section{Dynamics of massive charged scalar field}
As discussed in section~\ref{ch2:sec_msvsclrfld}, the radial dependence of a massive, charged test scalar field $\Phi$ (of mass $\mu$ and electric charge $q$) in the scalar-hairy RN background is governed by the radial Klein-Gordon equation~\eqref{ch2:eq_radialKG},
\begin{equation} \label{ch3:eq_radial}
\Delta \frac{d}{dr}\left(\Delta \frac{dR}{dr}\right)+U R=0,
\end{equation}
where $\Delta=r^2 f\left(r\right)$, $f\left( r \right)=\left(1-\frac{2M}{r}+\frac{e^2+s}{r^2}\right)$ and $U=\left( \omega r^2-e q r \right)^2-\Delta \left[ \mu ^2 r^2+l \left( l+1 \right)\right]$.
The position of the inner and outer horizons are given by the roots of $\Delta$,
\begin{equation}
r_\mp=M \mp \sqrt{M^2-e^2-s}.
\end{equation}
In terms of the tortoise coordinate $r_{*}$ (defined by $dr_{*}=dr/f\left( r\right)$), mapping the semi infinite region $\left[r_{+},\infty\right)$ to $(-\infty,\infty)$, Eq.~(\ref{ch3:eq_radial}) can be recast as
\begin{equation}\label{ch3:eq_trts}
\frac{d^2 \zeta}{d r_{*}^2}+W\left(\omega, r\right) \zeta=0,
\end{equation}
where $\zeta=r R \hspace{0.02\textwidth}\mbox{and}\hspace{0.02\textwidth} W\left(\omega, r\right)=\frac{U}{r^4}-\frac{\Delta}{r^3}\frac{d}{dr}\left(\frac{\Delta}{r^2}\right).$
{ In general, $\omega$ is a complex quantity with the real part $\omega_{Re}$ corresponding to the actual frequency of the wave motion and the imaginary part $\omega_{Im}$ taking care of the damping. At a linearized level, superradiant instabilities are related to the perturbations of the black hole which grows exponentially in time and corresponds to $\omega_{Im}>0$ with instability timescale $\tau\equiv 1/\omega_{Im}$.
In case of superradiant modes, this always occurs when the real part of the frequency satisfies the superradiance condition~\cite{brito_Springer_2015}. Hence, superradiant instability sets in through the real frequency modes (see Refs.~\cite{damour_LNC_1976,zouros_AP_1979,hartle_CMP_1974}), and thus we only consider modes with $\mid\omega_{Im}\mid<<\omega_{Re}$}.

For the scattering problem, with $\omega^2>\mu^2$, the physical boundary condition corresponds to an incident wave of amplitude $\mathcal{I}$ from infinity giving rise to a reflected wave of amplitude $\mathcal{R}$ near infinity and a transmitted wave of amplitude $\mathcal{T}$ at the horizon,
\begin{equation}\label{ch3:eq_BC_superrad}
\zeta \sim 
\begin{cases}
\mathcal{I}e^{-i\sqrt{\omega^2-\mu^2} r_*}+\mathcal{R}e^{i\sqrt{\omega^2-\mu^2} r_*} &\mbox{as} \quad r_*\rightarrow \infty\\
 \mathcal{T}e^{-i \left(\omega-\frac{e q}{r_{+}} \right) r_*}&\mbox{as} \quad  r_* \rightarrow -\infty. 
\end{cases}
\end{equation}
The invariance of the field equation under the transformation $t\rightarrow-t$ and $\omega\rightarrow-\omega$, leads to another linearly independent solution $\zeta^{*}$ which satisfies the complex conjugate boundary conditions. The Wronskian of the two solutions will be independent of $r_{*}$. Hence, the Wronskian evaluated at the horizon, $W_h= 2i\left(\omega-\frac{e q}{r_{+}} \right)|\mathcal{T}|^2$, must be equal to that evaluated near spatial infinity, $W_\infty=-2i\sqrt{\omega^2-\mu^2}\left(|\mathcal{R}|^2-|\mathcal{I}|^2\right)$. This yields,
\begin{equation}
|\mathcal{R}|^2=|\mathcal{I}|^2-\frac{\omega-\frac{e q}{r_+}}{\sqrt{\omega^2-\mu^2}}|\mathcal{T}|^2.
\end{equation}
The reflected wave is thus superradiantly amplified $\left( |\mathcal{R}|^2>|\mathcal{I}|^2 \right)$,  provided
\begin{equation}\label{ch3:eq_sprrdnc}
\omega<\frac{e q}{r_+}.
\end{equation}
However, when $\omega^2<\mu^2$, Eq.~(\ref{ch3:eq_trts}) results in bound states of the scalar field characterised by exponentially decaying modes near spatial infinity, 
\begin{equation}\label{ch3:eq_bndst1}
\zeta \sim 
 e^{-\sqrt{\mu^2 -\omega^2} r_*}\mbox{\hspace*{2mm} as \hspace*{2mm}} r_*\rightarrow \infty.
\end{equation}

As mentioned earlier, we are particularly interested in studying the superradiant stability of the scalar-hairy RN black hole with large negative values of the scalar charge,
\begin{equation}\label{ch3:eq_mtdcondtn}
s=-|s|\hspace{0.5cm} \mbox{and} \hspace{0.5cm} |s |> e^2.
\end{equation}
The assumption~(\ref{ch3:eq_mtdcondtn}) results in  $r_-<0$, which is unphysical and hence the mutated RN spacetime is characterised by only one event horizon at $r=r_{+}$.

\section{Existence of bound states}\label{section3}
In this section, we investigate the existence of bound states of the scalar field in the superradiant regime. We present the radial Klein-Gordon Eq.~(\ref{ch3:eq_radial}) in a Schr\"{o}dinger like form by defining a new radial function $\psi (r) =\sqrt{\Delta} R$, as
\begin{equation}
\frac{d^2 \psi}{dr^2}+\left(\omega^2-V \right)\psi=0,
\end{equation}
where
\begin{equation}\label{ch3:eq_pot}
V=\omega^2-\frac{U+M^2-e^2-s}{\Delta^2}
\end{equation}
and analyse the nature of the effective potential $V$ to check for the existence of a potential well outside the event horizon supporting meta-stable bound states in the superradiant regime. In the asymptotic limit,
\begin{equation}\label{ch3:eq_potlimit1}
V\left(r\rightarrow \infty\right)\rightarrow
 \mu ^2+\frac{2~a\left(\omega\right)}{r}+\mathcal{O}\left(\frac{1}{r^2}\right),
\end{equation}
where
\begin{equation}\label{ch3:eq_a}
a\left(\omega\right)=e q \omega +\mu ^2 M-2 M \omega ^2
\end{equation}
represents a convex parabola. The sign of $a\left(\omega\right)$ at the edges completely determines the asymptotic behaviour of the effective potential.

The superradiance condition $\left( \omega<e q/r_{+} \right)$ and the bound state condition $\left( \omega^2<\mu^2 \right)$ can be combined to yield
\begin{equation}\label{ch3:eq_omegalimit}
0\leq\omega<\mbox{min}\left\lbrace\tfrac{e q}{r_+},\mu \right\rbrace.
\end{equation}
 The left hand bound is required as $\omega$ is the frequency. The behaviour of $a\left(\omega\right)$ near the boundaries of~(\ref{ch3:eq_omegalimit}) can be summarized as
\begin{equation}
a\left(\omega\right)\rightarrow
\begin{cases}
 \mu^2 M >0 & \mbox{as} \quad \omega\rightarrow 0 \\
\mu\left(e q-M\mu \right)>\mu M\left(\frac{e q}{r_+}-\mu \right) >0 & \mbox{as} \quad \omega\rightarrow\mu \quad \mbox{for} \quad \mu<\tfrac{e q}{r_+}\\
M\mu^2+\frac{e^2 q^2}{r_+}\left(1-\frac{2 M}{r_+}\right)>0 & \mbox{as} \quad \omega\rightarrow\tfrac{e q}{r_+} \quad \mbox{for} \quad \tfrac{e q}{r_+}<\mu.
\end{cases}
\end{equation}
Thus,
\begin{equation}\label{ch3:eq_a>0}
a\left(\omega\right)>0
\end{equation}
in the entire range~(\ref{ch3:eq_omegalimit}). Further,
\begin{equation}\label{ch3:eq_potlimit2}
V\rightarrow
\begin{cases}
-\infty & \mbox{ as \hspace*{2mm}} r\rightarrow r_{+} \\
 \omega ^2-\frac{l(l+1)+1}{|s |-e^2}-\frac{M^2}{\left(|s |-e^2\right)^2}
 & \mbox{ as \hspace*{2mm}} r\rightarrow 0 \\
-\infty & \mbox{ as \hspace*{2mm}} r\rightarrow r_{-} .
\end{cases}
\end{equation}
From Eq.\eqref{ch3:eq_potlimit1} and  \eqref{ch3:eq_a>0}, we note that $V(r\rightarrow \infty)$ is positive definite and in light  of Eq.\eqref{ch3:eq_potlimit2} (namely, $V(r\rightarrow r_+)\rightarrow -\infty$), we infer that $V$ has at least one maximum outside the horizon $\left(r_{+}<r<\infty\right)$. Eq.\eqref{ch3:eq_potlimit2} also indicates that $V$ has another maximum in the unphysical region, $r_{-}<r<r_{+}$.

The derivative of the effective potential is given by 
\begin{equation}
\begin{split}
V^{'}=&-\tfrac{2}{\Delta^3}\left[a\left(\omega\right)r^4+\left[-2M^2 \mu^2-e^2 q^2+\left(|s|-e^2\right)\left(\mu^2 -2 \omega^2 \right)+2 M e q \omega+l\left(l+1 \right)\right]r^3 \right.\\
&\left.+3\left[-M \mu^2 \left(|s|-e^2 \right)+eq\omega\left(|s|-e^2  \right)-M l\left(l+1\right)\right]r^2 +\left[l \left(l+1\right)\left(e^2+2 M^2-|s|\right)\right.\right.\\
&\left.\left.-2 \left(M^2+|s|-e^2 \right) -e^2 q^2 \left(|s|-e^2\right)-\mu ^2 \left(|s|-e^2\right)^2 \right]r \right.\\
& \left.+2 M \left(M^2+|s|-e^2 \right) +l (l+1) M \left(|s|-e^2\right)\right],
\end{split}
\end{equation}
which in the asymptotic limit reduces to $V^{'}(r\rightarrow\infty)\rightarrow 0^-$, suggesting the absence of any potential well as $r\rightarrow\infty$.

If we define $z=r-r_{-}\left(=r+|r_- |\right)$, and write $V^{'}$ as a function of $z$, one has
\begin{equation}\label{ch3:eq_V'(z)}
V^{'}\left(z \right)=-\frac{\left(a z^4+bz^3+c z^2+d z+g \right)2}{\Delta^3},
\end{equation}
where
\begin{align}
\begin{split}
b={ }& -\mu ^2 \left(e^2+2 M^2+4 M |r_{-}|-|s|\right)+2 \omega ^2 \left(e^2+4 M |r_{-}|-|s|\right) \\& -e^2 q^2+2 e q \omega  (M-2 |r_{-}|)+l (l+1),
\end{split}\\
\begin{split}
c={ }& 3|r_{-}|^3\left( \tfrac{e q}{|r_{-}|}+\omega \right) \left( \tfrac{e q}{|r_{-}|}+2\omega \right)-3\mu ^2 |r_{-}|^2 \left(r_{+}-M \right)-3l \left(l+1 \right) \left(r_{+}-M \right),\label{ch3:eq_c}
\end{split}\\
\begin{split}
d={ }& 2 \mu ^2 |r_{-}|^2 \left(M^2+|s|-e^2\right)-e^2 q^2 |r_{-}|\left(r_{+}+3 |r_{-}|\right)\\
&-2 e q |r_{-}|^2 \omega  \left(3 r_{+}+2 |r_{-}|-3 M \right)-2 |r_{-}|^3 \omega ^2 \left(3 r_{+}-4 M \right)\\
&+2 \left(l \left(l+1 \right)-1\right) \left(M^2+|s|-e^2 \right),
\end{split}\\
\begin{split}
g={ }& 2 |r_{-}|^4 (r_{+}-M) \left(\tfrac{e q}{|r_{-}|}+\omega \right)^2+2 (r_{+}-M)^3.\label{ch3:eq_e}
\end{split}
\end{align}

If ${z_1,z_2,z_3,z_4}$ are the roots of the equation $V^{'}(z)=0$, then Vieta's formulas~\cite{barnard_child_1959} give
\begin{eqnarray}
&z_1+z_2+z_3+z_4=-\frac{b}{a}\label{ch3:eq_z1+z2+z3+z4},\\
&z_1z_2+z_1z_3+z_1z_4+z_2z_3+z_2z_4+z_3z_4=\frac{c}{a}\label{ch3:eq_z1z2+},\\
&z_1z_2z_3+z_1z_2z_4+z_1z_3z_4+z_2z_3z_4=-\frac{d}{a}\label{ch3:eq_z1z2z3+},\\
\mbox{and} &z_1z_2z_3z_4=\frac{g}{a}.\label{ch3:eq_z1z2z3z4}
\end{eqnarray}
The existence of at least one maximum of the effective potential outside the horizon guarantees that $V^{'}(z)$ has at least one positive root $\left(z_1\mbox{, say}\right)$. Similarly, the potential maximum in the unphysical region $r_{-}<r<r_{+}$ suggests another positive root of $V^{'}(z)$ $\left(z_2\mbox{, say}\right)$ with
\begin{equation}\label{ch3:eq_z1z2>0}
z_1>z_2>0.
\end{equation}
Eq.~(\ref{ch3:eq_e}) shows 
\begin{equation}
g>0
\end{equation}
which together with Eqs.~(\ref{ch3:eq_a>0}) and~(\ref{ch3:eq_z1z2z3z4}) implies
\begin{equation}\label{ch3:eq_z1z2z3z4>0}
z_1z_2z_3z_4>0.
\end{equation}
Combining the inequalities (\ref{ch3:eq_z1z2>0}) and~(\ref{ch3:eq_z1z2z3z4>0}) one can conclude that $z_3$ and $z_4$ must be of the same sign,\begin{equation}\label{ch3:eq_z3z4>0}
z_3z_4>0.
\end{equation}

For a potential well to exist outside the horizon, beyond $z_1$, the roots $z_3$ and $z_4$ must be real and positive with
\begin{equation}\label{ch3:eq_z3,z4>z1}
z_3,z_4>z_1.
\end{equation}
The inequality~(\ref{ch3:eq_z3,z4>z1}) in conjunction with Eqs.~(\ref{ch3:eq_a>0}),(\ref{ch3:eq_z1+z2+z3+z4}),(\ref{ch3:eq_z1z2+}),(\ref{ch3:eq_z1z2z3+}),(\ref{ch3:eq_z1z2>0}) and~(\ref{ch3:eq_z3z4>0}) implies
\begin{equation}\label{c>0b<0}
b\leq 0, \quad c\geq 0 \quad \mbox{and} \quad d\leq 0.
\end{equation}
If we drop the last term in the expression of $c$ in Eq.~(\ref{ch3:eq_c}) (which is at most zero) and define
\begin{equation}
\tilde{c}=3|r_{-}|^3\left( \tfrac{e q}{|r_{-}|}+\omega \right) \left( \tfrac{e q}{|r_{-}|}+2\omega \right)-3\mu ^2 |r_{-}|^2 \left(r_{+}-M \right),
\end{equation}
then $\tilde{c}$ represents a concave parabola which crosses the $\omega$-axis at 
\begin{eqnarray}
&\omega_1=\frac{-3eq-\sqrt{e^2q^2+4|r_{-}|\left(r_{+}+|r_{-}|\right) \mu^2}}{4|r_{-}|} < 0\\
\mbox{and} &\omega_2=\frac{-3eq+\sqrt{e^2q^2+4|r_{-}|\left(r_{+}+|r_{-}|\right) \mu^2}}{4|r_{-}|}.
\end{eqnarray}

As $\omega_1$ is a negative definite quantity and we are looking for the possibility of an enhanced radiation, we shall work with $\omega_2$. We separate the parameter space of the massive charged test scalar field in three regions based on the mass of the test field.
\begin{enumerate}[label={\bf Region \Roman*}] 
\item \label{regionI}{\bf  :}
\begin{equation}
\mu\geq \mu_1 \left(>\frac{e q}{r_{+}} \right),
\label{ch3:eq_massivelimit}
\end{equation}
where
\begin{equation}
\mu_1=\frac{e q}{r_{+}}\sqrt{\frac{2\left(r_{+} + 2|r_{-} |\right)}{|r_{-} |}}.
\end{equation}
For such values of the field mass, one finds
\begin{equation}
\omega_2\geq \tfrac{e q}{r_{+}}>0
\end{equation}
and since the concave parabola $\tilde{c}$ crosses the $\omega$-axis at $\omega_1(<0)$ and $\omega_2$, for any $\omega$ in the region $0\leq\omega<\frac{e q}{r_+}$, one has
\begin{equation}\label{ch3:eq_c<0}
c\leq \tilde{c}<0.
\end{equation}
The inequality~(\ref{ch3:eq_c<0}) contradicts condition (\ref{c>0b<0}). Hence, there is no potential well in the physical region $r>r_{+}$. Thus for $\mu$ satisfying the condition~(\ref{ch3:eq_massivelimit}), there is no bound state and so the mutated black hole is stable against superradiance.

\item \label{regionII}{\bf  :}\\
This region is defined as
\begin{equation}\label{ch3:eq_w2=0}
\left(\frac{e q}{r_+} <\right)\mu_2<\mu<\mu_1,
\end{equation}
where
\begin{equation}
\mu_2=e q\sqrt{\frac{2}{|r_{-} |\left(r_{+}+|r_{-} |\right)}}.
\end{equation}
Now, $\omega_2$ lies in the range
\begin{equation}
0<\omega_2<\tfrac{e q}{r_{+}},
\end{equation}
and there exists an $\omega$, in the range $0\leq\omega<\omega_2$, for which $\tilde{c}< 0$, in contradiction with~(\ref{c>0b<0}) and as before, the spacetime is superradiantly stable.

\item \label{regionIII}{\bf  :} 

\begin{equation}\label{ch3:eq_w2<0}
0<\mu\leq \mu_2,
\end{equation}
For $\mu$ in this range, $\omega_2\leq 0$ resulting in $\tilde{c}>0$ for any $\omega>0$. The coefficients $b$ and $d$ can also be shown to be negative for $l=0$ in this range. To further ascertain the absence of potential well in this region, we analyse the signature of the discriminant of Eq.\eqref{ch3:eq_V'(z)}.
\end{enumerate}

It is well known that a negative discriminant of quartic polynomial implies the existence of two real and two complex roots (see Ref.~\cite{rees_JSTOR_1922}). Table~\ref{ch3:table1} summarizes the signature of the discriminant $(D)$ of $V^{'}(z)$ and the nature of $z_3$ and $z_4$ at some values of $\mu$ for $l=0$ at the boundaries of the superradiant regime. The expression for $D$ is,
\begin{align}\label{ch3:eq_dscrmnt}
D& =-2 a b d \left(96 a g^2+40 c^2 g-9 c d^2\right)+b^2 \left(144 a c g^2-6 a d^2 g-4 c^3 g+c^2 d^2\right)\nonumber \\
& \quad +a \left(-128 a c^2 g^2+144 a c d^2 g+a \left(256 a g^3-27 d^4\right)+16 c^4 g-4 c^3 d^2\right)\nonumber\\
& \quad -27 b^4 g^2+b^3 \left(18 c d g-4 d^3\right).
\end{align}

\begin{table}[htbp!]
\centering
\caption{Table showing the signature of the discriminant $D$ in the superradiant regime at some discrete values of $\mu$.}
\begin{adjustbox}{width=\textwidth}
\begin{tabular}{|c|c|c|c|c|c|}
\hline
\rule[-1.5ex]{0pt}{2.7ex}$\mu\rightarrow$    & \multicolumn{2}{c|}{\begin{tabular}[c]{@{}c@{}}$\mu_1$\end{tabular}} & \multicolumn{2}{c|}{$\mu_2$} & 0 \\ \hline
\rule[-1.5ex]{0pt}{2.7ex}$\omega\rightarrow$ & 0 & ${e q}/{r_{+}}$ & 0 & ${e q}/{r_{+}}$ & 0  \\ \hline
\rule[-1.5ex]{0pt}{2.7ex}Sign$(D)$ & $<0$ (for $|s|-e^2\leq 4.414 M^2$) & $<0$ & $<0$ & $<0$ & $<0$ \\ \hline
\rule[-1.5ex]{0pt}{2.7ex}Nature of $z_3,z_4$ & \thead{Complex  (for $|s|-e^2\leq 4.414 M^2$);\\ Real, negative otherwise} & Complex & Complex & Complex & Complex \\
\hline
\end{tabular}
\end{adjustbox}
\label{ch3:table1}
\end{table} 
We observe from table~\ref{ch3:table1} that at the boundaries of~(\ref{ch3:eq_w2=0}) and~(\ref{ch3:eq_w2<0}),  $z_3$ and $z_4$ are complex in the superradiant regime. Thus, there exists no potential well and hence the spacetime is expected to be superradiantly stable, irrespective of the mass of the test field.
 \begin{figure}[!htbp]
\centering
\includegraphics[width=1\textwidth, keepaspectratio]{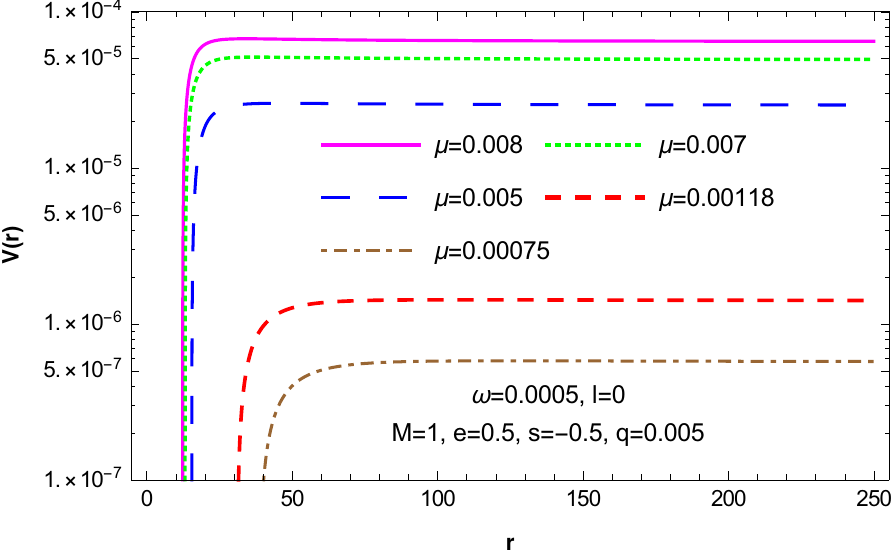}
\caption{Variation of the effective potential $V(r)$ with $r$ in logarithmic scale for different values of the field mass $\mu$.}
\label{ch3:fig:1}
\end{figure}
Fig.\ref{ch3:fig:1} shows the potential profile of a mutated RN black hole of unit mass with electric charge $e=0.5$ and scalar charge $s=-0.5$ for different values of the field mass for the $l=0$ mode in logarithmic scale. The event horizon of the black hole is at $r_+=2.118$. The electric charge of the test field $q=0.005$ which yields a superradiance frequency $e q/r_+=0.00118$. The boundaries of~\ref{regionI} and~\ref{regionII} are at $\mu_1=0.00745$ and $\mu_2=0.00688$. We observe that no potential well exist outside the event horizon when $\mu$ lies in~\ref{regionI} ($\mu>\mu_1$). This observation also holds when $\mu$ lies in~\ref{regionII} ($\mu_2<\mu<\mu_1$) with $\omega_2(=0.00024)<\omega(=0.0005)$ and even for $\mu<\mu_2$.

\section{Summary and Discussion}\label{section4}
In the present work, we investigated the possibility of the existence of superradiance and the superradiant stability of a charged spherically symmetric black hole with scalar hair, particularly for $s<-e^2$, representing a mutated RN spacetime. The quasinormal modes of the mutated RN spacetime against perturbation by massive (and massless) charged (and uncharged) scalar fields have been rigorously studied in Ref.~\cite{chowdhury_EPJC_2018}. Those quasinormal frequencies were obtained with negative imaginary part implying that the perturbations decay in time and the corresponding modes are stable.

We considered modes with $Re(\omega)\gg |Im(\omega)|$ and evaluated the superradiance condition for the mutated RN spacetime (see~(\ref{ch3:eq_sprrdnc})). The superradiance frequency is found to be consistent with that of a standard RN black hole~\cite{bekenstein_PRD_1973,hod_PLB_2012,hod_PLB_2013,hod_PRD_2015,huang_EPJC_2016}.

We observe that in the superradiant regime, if the mass of the test field lies in~\ref{regionI}, there is no potential well outside the horizon and hence no bound state resonance of the massive scalar fields. The spacetime is thus superradiantly stable in the  range~(\ref{ch3:eq_sprrdnc}). If the mass of the test scalar lies in~\ref{regionII}, the stability can be conclusively proved in the range of frequencies, $0\leq\omega<\omega_2$ with $0<\omega_2<e q/r_+$.

In~\ref{regionIII} where $\mu<\mu_2$, Descartes' rule of signs eliminates the possibility of negative real roots of $V^{'}(z)$, so we checked the signature of the discriminant ($D$) of $V^{'}(z)$ for the lowest angular momentum state (see~(\ref{ch3:eq_dscrmnt})). We note that as $\mu\rightarrow \mu_2$, $D<0$ at both ends of the superradiant regime~(\ref{ch3:eq_sprrdnc}) suggesting the existence of two complex roots ($z_3,z_4$). This confirms the superradiant stability of the spacetime. We observe that for vanishingly small field masses in the superradiant regime, $D<0$. This is expected, as for very small values of the field mass the potential well will almost cease to exist and the spacetime will be superradiantly stable. Further, as $\mu\rightarrow\mu_1$ and $\omega\rightarrow 0$, $z_3,z_4$ are complex for $|s|-e^2\leq 4.414M^2$ and real negative otherwise. In either situation, there will be no potential well outside the horizon and the spacetime will be superradiantly stable.

We conclude that the mutated RN black hole is superradiantly stable. For a large mass of the test field, this stability is proved. As with smaller masses, the strength of the well should decrease, the stability is expected to be ensured. However, such a comprehensive proof could not be provided for smaller masses. But the signature analysis of the discriminant quite strongly indicates a stable superradiance rather than a black hole bomb, for smaller masses as well. The potential profile shown in Fig.\ref{ch3:fig:1} for different values of the field mass also supports this conclusion.

Albeit a mutated RN spacetime is qualitatively different from an RN black hole, the results in connection with superradiance is very much similar in the two cases~\cite{hod_PLB_2012,hod_PLB_2013,hod_PRD_2015,huang_EPJC_2016}. Calculations for the case,  $0<s+e^2<M^2$ are omitted, as it would effectively resemble an RN black hole (see metric\eqref{ch1:eq_shRN}) and no characteristic difference from the RN black hole is expected.


\chapter{Hawking emission of charged particles from a charged spherical black hole with scalar hair} \blfootnote{\begin{flushleft} The work presented in this chapter is based on “Hawking emission of charged particles from an electrically charged spherical black hole with scalar hair”, \textbf{Avijit Chowdhury}, Eur.\ Phys.\ J.\ C \textbf{ 79}, 928 (2019).\end{flushleft}}   
\chaptermark{Hawking emission of charged particles \ldots}
\label{chap4}

\section{Introduction}
In 1974, Hawking~\cite{hawking_CMP_1975} showed that ``black holes'' are not ``entirely black''; they spontaneously emit particles at a temperature proportional to their surface gravity. Since then various derivations of Hawking radiation have been proposed. Parikh and Wilczek~\cite{parikh_PRL_2000} used the semi-classical tunneling formalism to study Hawking emission of massless uncharged particles  from the Schwarzschild and Reissner-Nordstr\"{o}m (RN)  black holes. According to this formalism, the black hole loses energy due to radiation and thereby decreases its horizon radius. Thus, the outgoing particle creates a potential barrier by itself~\cite{parikh_IJMPD_2004, parikh_GRG_2004,  parikh_MARCELGROSSMANN10}. Using WKB approximation one can evaluate the emission rate from the imaginary part of the action of the outgoing particles. The tunneling method has since been used to study Hawking radiation from a wide variety of black holes. Zhang and Zhao~\cite{zhang_JHEP_2005} extended the  tunneling method to study the emission of massive charged particles from Reissner-Nordstr\"{o}m black holes. Later, Jiang and Wu~\cite{jiang_PLB_2006} used tunneling formalism to study Hawking radiation of charged particles from Reissner-Nordstr\"{o}m-de Sitter black hole. Tunneling of charged particles in the modified RN black hole was studied by Liu~\cite{liu_IJTP_2014}. Jiang,Wu and Cai~\cite{jiang_PRD_2006} studied the tunneling of particles from Kerr and Kerr-Newman black holes. Jiang, Yang and Wu~\cite{jiang_IJTP_2006} also studied the tunneling of charged particles from Reissner-Nordstr\"{o}m black holes of arbitrary  dimensions. Jiang and Wu~\cite{jiang_PLB_2008} studied the tunneling of charged particles from Reissner-Nordstr\"{o}m-de Sitter black holes with a global monopole. Sarkar and Kothawala~\cite{sarkar_PLB_2008} gave a generalized treatment of Hawking radiation as tunneling for asymptotically flat, spherically symmetric black holes. Tunneling mechanism has also been extensively used to study Hawking radiation from numerous other black holes~\cite{wu_JHEP_2006, li_PLB_2006, chen_PLB_2008, chen_CQG_2008}). Tunneling of Dirac particles from black rings was studied by Jiang~\cite{jiang_PRD_2008}. Jiang, Chen and Wu~\cite{jiang_JCAP_2013} also studied the tunneling of massive particles from the cosmological horizon of a Schwarzschild-de Sitter black hole.The general outcome of all these investigations is the non-thermality of the Hawking emission spectrum. For a detailed review of the tunneling mechanism, we refer to the work of Vanzo~\cite{vanzo_CQG_2011}.

The question of information loss has also been addressed using the tunneling formalism. Using logarithmic correction to the Bekenstein-Hawking entropy, Chen and Shao~\cite{chen_PLB_2009} showed that in a black hole evaporation process unitarity is preserved and the information loss paradox can be successfully resolved. Singleton \textit{et al.}~\cite{singleton_JHEP_2010} showed that if the back reaction and the quantum corrections are taken into account, the information can be carried away by the correlations of the outgoing radiation during complete evaporation of the black hole and the information loss paradox can be resolved. Sakalli \textit{et al.}~\cite{sakalli_ASS_2012}  studied hawking radiation in  linear dilaton black holes and showed that no  information is lost during complete evaporation of the black hole. 

In the present chapter, we use the tunneling formalism to study Hawking radiation of charged particles from an asymptotically flat static spherically symmetric charged black hole, endowed with a scalar hair, dubbed as the scalar-hairy RN black hole~\cite{astorino_PRD_2013}. The primary objective of the present work is to  study the effect of the scalar hair on the transmission rate of charged Hawking quanta. Based on the non-negativity of mutual information between consecutive Hawking emissions~\cite{zhang_PLB_2009, kim_PLB_2014}, we also find the dependence of the charge to mass ratio of the emitted particles on the scalar charge $s$ in each step.

We observe that the change in entropy of the scalar-hairy RN black hole due to the emission of charged particles contains an energy-dependent contribution which vanishes during emission of uncharged particles and also in the absence of the scalar hair. The tunneling rate  for the emission of charged particles from the scalar-hairy RN black hole matches smoothly with that of the standard RN black hole, in the absence of the scalar hair.
\section[Painl\'{e}ve coordinates]{The scalar-hairy-Reissner-Nordstr\"{o}m black hole in Painl\'{e}ve coordinates}
As discussed in section~\ref{ch1:sec_shRN}, scalar-hairy RN spacetime is described by the line element,
\begin{equation}\label{ch4:eq_metric}
ds^2=-f\left(r \right)dt_R^2+{f\left(r \right)}^{-1}dr^2+r^2 \left( d\theta^2+\sin^2{\theta} d\phi^2 \right),
\end{equation}
with 
\begin{equation}
\label{ch4:eq_f(r)}
f\left( r \right)=\left(1-\frac{2M}{r}+\frac{e^2+s}{r^2}\right),
\end{equation}
where $M$ and $e$ are respectively the mass parameter and electric charge of the black hole. The scalar-hairy RN spacetime, Eq.\ \eqref{ch4:eq_metric}, is characterised by an event horizon at $r_+$ and a Cauchy horizon at $r_{-}$ where
\begin{equation}\label{ch4:eq_r+-}
r_\pm=M \pm \sqrt{M^2-e^2-s}~.
\end{equation}
The spacetime suffers from coordinate singularities both at the event horizon at $r_+$ and at the inner horizon at $r_-$~. To eliminate these singularities at $r_{\pm}$, we introduce the generalized Painlev\'{e} transformation,
 \begin{equation}\label{ch4:eq_painleve_trans}
 \begin{split}
 t&= t_R+2\sqrt{2Mr-e^2-s}+M \ln\left( \frac{r-\sqrt{2Mr-e^2-s}}{r+\sqrt{2Mr-e^2-s}} \right) \\
& +\frac{e^2+s-2M^2}{\sqrt{M^2-e^2-s}} \tanh^{-1}\left(\frac{\sqrt{M^2-e^2-s}\sqrt{2Mr-e^2-s}}{Mr-e^2-s}\right),
\end{split}
\end{equation}  
which transforms the scalar-hairy RN metric (Eq.\ \eqref{ch4:eq_metric}) to
\begin{equation}\label{ch4:eq_painleve_metric}
ds^2= -f(r)dt^2+2\sqrt{1-f(r)} dt dr+dr^2+r^2 d\theta^2+r^2\sin^2{\theta} d\phi^2 .
\end{equation}
The line element in Eq.\ \eqref{ch4:eq_painleve_metric} highlights the stationary but non-static character of the spacetime. The constant time slices are flat Euclidean by construction. Furthermore, the generator of  time-translation is a killing vector.  As $f(r)$ vanishes asymptotically, these coordinates are similar to static coordinates to an asymptotic observer. 
{ Another important property of the Painl\'{e}ve coordinates is related to the synchronisation of coordinate clocks.
According to Landau's theory of coordinate clock synchronisation~\cite{landau_book}, the coordinate time difference between two simultaneous events taking place at different space points in a spacetime decomposed in (3+1) dimensions is given by~\cite{zhang_chinese,zhang_JHEP_2005}
\begin{equation}\label{ch4:eq_deltaT}
\Delta T=-\int \frac{g_{0i}}{g_{00}}dx^i \qquad (i=1,2,3)~.
\end{equation}
If the simultaneity of the coordinate clocks can be transmitted from one point to another and is independent of the integration path, then
\begin{equation}\label{ch4:eq_clock_sync}
\frac{\partial}{\partial x^i}\left(- \frac{g_{0j}}{g_{00}}\right)=\frac{\partial}{\partial x^j}\left(- \frac{g_{0i}}{g_{00}}\right) \qquad (i,j=1,2,3)~.
\end{equation}
The Painl\'{e}ve line element~\eqref{ch4:eq_painleve_metric} satisfies the condition~\eqref{ch4:eq_clock_sync}, and hence simultaneity of events can be transmitted from one place to another though the metric is essentially non-diagonal.}
We will use this line element to study the emission of massive charged particles from the scalar-hairy RN black hole  using the semi-classical tunneling formalism~\cite{parikh_PRL_2000, parikh_IJMPD_2004, zhang_JHEP_2005}.

\section{Tunneling rate of charged particles}\label{ch4:sec_tunneling rate}
The tunneling of particles across a potential barrier being an instantaneous phenomenon, the metric must obey Landau's condition of coordinate clock synchronisation. Two events occur simultaneously during tunneling, a radially moving particle tunnels into the barrier while another particle tunnels out of the barrier. { Thus, in terms of} Landau's theory of coordinate clock synchronisation, the coordinate time difference between two simultaneous events, occurring at two different space points is given by,
\begin{equation}
dt=-\frac{g_{01}}{g_{00}}dr_c~, \quad \left(\mbox{for ~} d \theta=d \phi =0 \right),
\end{equation}
where $r_c$ is the position of the tunneling particle.
Following~\cite{zhang_JHEP_2005, zhang_PLB_2005, jiang_PRD_2006,  jiang_PLB_2006, ali_IJTP_2014, ali_IJTP_2008}, we consider the outgoing charged particle to be represented by a de Broglie s-wave whose phase velocity $v_p$ is related to its group velocity $v_g$ as,
\begin{equation}\label{ch4:eq_vp_vg}
v_p=\frac{v_g}{2}~.
\end{equation}
The group velocity of a de Broglie s-wave representing an outgoing charged Hawking quanta is given by,
\begin{equation}\label{ch4:eq_groupvel}
v_g=\frac{dr_c}{dt}=-\frac{g_{00}}{g_{01}}=\frac{e^2-2 M r+r^2+s}{r \sqrt{-e^2+2 M r-s}}~,
\end{equation}
which results in a phase velocity of
\begin{equation}\label{ch4:eq_vp}
\dot{r}=v_p=\frac{e^2-2 M r+r^2+s}{2r \sqrt{-e^2+2 M r-s}} ~.
\end{equation}
The electric charge of the black hole gives rise to an electromagnetic field  $F_{\mu\nu}$, given by the vector potential $A_\mu=-\delta^0_\mu \left(e/r \right)$. The Lagrangian of this electromagnetic field is $L_e=-\frac{1}{4} F_{\mu\nu} F^{\mu\nu}$. However, $L_e$ being independent of the corresponding generalized coordinates $A_\mu=\left( A_t,0,0,0 \right)$, there exists a gauge freedom in the choice of $A_t$. To eliminate this freedom, we write the action of an outgoing charged massive particle as~\cite{landau_book, zhang_JHEP_2005} 
\begin{equation}
\mathcal{A} = \int^{t_f}_{t_i} \left(L-P_{A_t}\dot{A_t}\right) dt~,
\end{equation}
where $P_{A_t}$ is the canonical momentum conjugate to the generalized coordinate $A_t$ and $L$ is the total Lagrangian of the matter-gravity system.
When a particle of charge $q$ and mass $\omega$ tunnels out of the event horizon the electric charge of the black hole changes from $e\rightarrow e-q$ and the ADM mass changes from $M_{ADM}\rightarrow M_{ADM}-\omega$. The imaginary part of the corresponding action can thus be written as
\begin{equation}\label{ch4:eq_Ims}
\begin{split}
Im \hspace{2pt} \mathcal{A} & = Im \left\lbrace \int^{r_f}_{r_i} \left( P_r \dot{r}-P_{A_t}\dot{A_t}\right)\frac{dr}{\dot{r}} \right\rbrace \\
& = Im \left\lbrace \int^{r_f}_{r_i} \left[ \int^{\left( P_r P_{A_t} \right)}_{\left(0,0\right)} \left( d P'_r \dot{r}-\dot{A_t} dP'_{A_t} \right) \right] \frac{dr}{\dot{r}} \right\rbrace,
\end{split}
\end{equation}
where $r_i$ and $r_f$ are the position of the event horizon before and after the tunneling of the charged Hawking quanta. Using Hamilton's equations,
\begin{eqnarray}
\dot{r} &=& \frac{d H}{d P_r}\bigg\vert_{\left( r; A_t,P_{A_{t}} \right)}\\
\mbox{and} \quad \dot{A_t} &=& \frac{d H}{d P_{A_{t}}}\bigg\vert_{\left( A_{t}; r,P_r \right)},
\end{eqnarray}
with
\begin{eqnarray}
& dH\vert_{\left( r; A_t,P_{A_{t}} \right)} = d \left(M_{ADM}-\omega'\right)=-d\omega' \\
\mbox{and} \quad & dH\vert_{\left( A_{t}; r,P_r \right)} = \frac{e-q'}{r}d(e-q')=-\frac{e-q'}{r} dq',
\end{eqnarray}
we rewrite the Eq.\ \eqref{ch4:eq_Ims} as
\begin{equation}\label{ch4:eq_Ims1}
Im \hspace{2pt} \mathcal{A}=-Im \left[ \int^{r_f}_{r_i}  \left( \int^{\omega}_0  d \omega' - \int^q_0 \frac{e-q'}{r} d q'\right)\frac{dr}{\dot{r}} \right].
\end{equation}

The ADM mass $M_{ADM}$ is related to the  mass parameter $M$ as
\begin{equation}\label{ch4:eq_MADM}
M_{ADM}=\frac{M}{1+s/e^2}~.
\end{equation} 
{Henceforth, we will restrict ourselves to the regime $s>-e^2$ such that  $M_{ADM}$ is always positive definite.} For the standard RN black hole $(s=0)$, it is equal to the mass parameter $M.$ However, in the limit of $e\rightarrow	0$ with a finite non-zero $s$, $M_{ADM}$ vanishes. 

Substituting Eq.\ \eqref{ch4:eq_MADM} in the expression of $\dot{r}$ in Eq.\ \eqref{ch4:eq_vp}  with $M_{ADM}\rightarrow M_{ADM}-\omega'$ and $e\rightarrow e-q'$ we get
\begin{equation}\label{ch4:eq_vp_final}
\dot{r}=\frac{-2 r (M_{ADM}-\omega' ) \left(\frac{s}{(e-q')^2}+1\right)+(e-q')^2+r^2+s}{2 r \sqrt{2 r (M_{ADM}-\omega' ) \left(\frac{s}{(e-q')^2}+1\right)-(e-q')^2-s}}~.
\end{equation}
Eq.\ \eqref{ch4:eq_Ims1} in conjunction with Eq.\ \eqref{ch4:eq_vp_final} yields
\begin{equation}\label{ch4:eq_ImA}
\begin{split}
Im \hspace{2pt} \mathcal{A} &=-Im \Bigg[ \int^{r_f}_{r_i} \left( \int^\omega_0 d \omega' - \int^q_0 \frac{e-q'}{r} d q'\right) \\
& \times \frac{2 r \sqrt{2 r (M_{ADM}-\omega' ) \left(\frac{s}{(e-q')^2}+1\right)-(e-q')^2-s}}{\left(r-r'_+\right) \left(r-r'_-\right)} dr \Bigg]~,
\end{split}
\end{equation}
where
\begin{eqnarray}
\begin{split}
&r'_{\pm} = (M_{ADM}-\omega' ) \left(\frac{s}{(e-q')^2}+1\right)\\ 
 & \pm \sqrt{(M_{ADM}-\omega' )^2 \left(\frac{s}{(e-q')^2}+1\right)^2-(e-q')^2-s}~,
\end{split}\\
\label{ch4:eq_ri}
r_i = M_{ADM} \left(\frac{s}{e^2}+1\right)+\sqrt{M_{ADM}^2 \left(\frac{s}{e^2}+1\right)^2-e^2-s}~, \\
\begin{split}
&r_f = (M_{ADM}-\omega ) \left(\frac{s}{(e-q)^2}+1\right)\\ 
 &+\sqrt{(M_{ADM}-\omega )^2 \left(\frac{s}{(e-q)^2}+1\right)^2-(e-q)^2-s}~.
\end{split}
\end{eqnarray}
Interchanging the order of integrations in \eqref{ch4:eq_ImA}, we note that the $r$ - integral has a pole at $r'_+$. Deforming the contour around this pole we obtain,
\begin{equation}\label{integral_inexact}
Im\hspace{2pt} \mathcal{A} =\pi \left( \int^{\omega}_0 \frac{2 {r'_+}^2}{r'_+-r'_-} d\omega'- \int^q_0 \frac{2 r'_+\left(e-q'\right)}{r'_+-r'_-} dq'\right).
\end{equation} 

The First law of black hole mechanics for a scalar-hairy RN black hole is
\begin{equation}\label{ch4:eq_1st law}
d M_{ADM}=\frac{\kappa}{8 \pi \left(1+s/e^2 \right)}d A +\Phi \left[ 1+\frac{s}{{r_{-}}^2}\right] de ~,
\end{equation}
where $A$ is the area of the event horizon, $\kappa$ is the surface gravity and $\Phi$ is the electric potential at the horizon,
\begin{equation}
A= 4 \pi {r_+}^2~, \quad \kappa=\frac{r_{+}-r_{-}}{2 {r_{+}}^2}~, \quad \Phi=\frac{e}{r_{+}}~ .
\end{equation}

Using the First law \eqref{ch4:eq_1st law}, the integral in \eqref{integral_inexact} can be written as,
\begin{equation}
Im\hspace{2pt} \mathcal{A} =-\frac{\pi}{2} \int^{\left(\omega,q\right)}_{\left(0,0\right)}d\left(\frac{{r'_{+}}^2}{1+\frac{s}{\left(e-q' \right)^2}}\right) -\frac{\Delta S_{\omega}}{2}~,
\end{equation}
where
\begin{equation}\label{ch4:eq_Somega}
\Delta S_{\omega} =-2 \pi s \int^q_0 \frac{\left( e-q' \right)}{r'_-{^2}} \frac{r'_+ + r'_-}{r'_+-r'_-}dq'.
\end{equation}
Thus, we get 
\begin{eqnarray}\label{ch4:eq_ImS_corr}
Im\hspace{2pt} \mathcal{A} &=&  -\frac{\pi}{2} \left[ \frac{r_f^2}{1+\frac{s}{\left(e-q\right)^2}} - \frac{r_i^2}{1+\frac{s}{e^2}} \right]-\frac{\Delta S_{\omega}}{2}\\
&=&-\frac{1}{2}\left( \Delta S_{BH}+\Delta S_{\omega}\right),
\end{eqnarray}
where 
\begin{equation}\label{ch4:eq_Sbh}
\Delta S_{BH}=\left( \pi\frac{r_f^2}{1+\frac{s}{\left(e-q\right)^2}} -\pi \frac{r_i^2}{1+\frac{s}{e^2}} \right)~,
\end{equation}
is the change in the Bekenstein-Hawking entropy~\cite{hawking_PRL_1971, bekenstein_PRD_1973} of the scalar-hairy RN black hole, $S_{BH}=A/(4 \tilde{G})=\frac{A}{4 (1+s/e^2)}$.
The tunneling rate is given by 
\begin{equation}\label{ch4:eq_tunneling_prob}
\Gamma=e^{-2 Im \hspace{2pt} \mathcal{A}}=e^{ \Delta S_{charged} }~,
\end{equation}
where 
\begin{equation}\label{ch4:eq_Stotal}
\Delta S_{charged}=\Delta S_{BH}+\Delta S_{\omega}~,
\end{equation}
is the total change in entropy of the scalar-hairy RN black hole due to the emission of massive charged particle.

For the emission of uncharged particles, $\Delta S_{\omega}=0$ and we get the tunneling rate as
\begin{equation}\label{ch4:eq_tunneling_prob_massless_uncharged}
\begin{split}
\Gamma&=e^{\Delta S_{BH}} \quad \mbox{with} \\ 
\Delta S_{BH}&=\frac{\pi}{1+s/e^2}\Bigg[\Bigg( (M_{ADM}-\omega ) \left(\frac{s}{e^2}+1\right)\\ 
 & \left.+\sqrt{(M_{ADM}-\omega )^2 \left(\frac{s}{e^2}+1\right)^2-e^2-s}\right)^2  \\
 &\left. - \left(M_{ADM}\left(\frac{s}{e^2}+1\right)+\sqrt{M_{ADM}^2\left(\frac{s}{e^2}+1\right)^2-e^2-s}  \right)^2\right].
\end{split}
\end{equation}
In case of vanishing scalar hair, $s=0$, we recover the tunneling rate for standard Reissner-Nordstr\"{o}m black hole~\cite{parikh_PRL_2000, zhang_JHEP_2005}.

Expanding both the terms on the RHS of Eq.\ \eqref{ch4:eq_ImS_corr} to leading orders in $\omega$ and $q$,

\begin{equation}
\begin{split}
\Delta S_{BH}&=-\frac{2 \pi \left( M_{ADM} \left(\frac{s}{e^2}+1\right)+\sqrt{M_{ADM}^2 \left(\frac{s}{e^2}+1\right)^2-e^2-s} \right)^2}{\sqrt{M_{ADM}^2 \left(\frac{s}{e^2}+1\right)^2-e^2-s}}\\
& \times \Bigg[ \omega -\frac{e q}{M_{ADM} \left(\frac{s}{e^2}+1\right)+\sqrt{M_{ADM}^2 \left(\frac{s}{e^2}+1\right)^2-e^2-s}} \\
&-\frac{q s M_{ADM} }{e\left(e^2+s \right)} \Bigg]~,
\end{split}
\end{equation}
\begin{equation}
\begin{split}
\Delta S_{\omega}&= -2 \pi\frac{q s M_{ADM}}{e\left(e^2+s \right)\sqrt{M_{ADM}^2 \left(\frac{s}{e^2}+1\right)^2-e^2-s}} \\
& \times  \left( M_{ADM} \left(\frac{s}{e^2}+1\right)+\sqrt{M_{ADM}^2 \left(\frac{s}{e^2}+1\right)^2-e^2-s} \right)^2,
\end{split}
\end{equation}
we get the tunneling rate~\eqref{ch4:eq_tunneling_prob} as
\begin{equation}\label{ch4:eq_thermal_prob}
\Gamma \sim e^{ -\beta\left(\omega-\omega_0\right) } ~,
\end{equation}
where $\beta=1/T_{BH}$ is the inverse of the black hole temperature, 
\begin{equation}
\beta=\frac{2 \pi \left( M_{ADM} \left(\frac{s}{e^2}+1\right)+\sqrt{M_{ADM}^2 \left(\frac{s}{e^2}+1\right)^2-e^2-s} \right)^2}{\sqrt{M_{ADM}^2 \left(\frac{s}{e^2}+1\right)^2-e^2-s}}
\end{equation}
and
\begin{equation}
\omega_0=\frac{e q}{ M_{ADM} \left(\frac{s}{e^2}+1\right)+\sqrt{M_{ADM} \left(\frac{s}{e^2}+1\right)^2-e^2-s}}~.
\end{equation}
As Eq.\ \eqref{ch4:eq_thermal_prob} includes only the leading order, Hawking radiation has an approximately thermal spectrum.

\begin{figure*}[!htbp]\centering
\includegraphics[width=1\textwidth, height=0.38\textwidth]{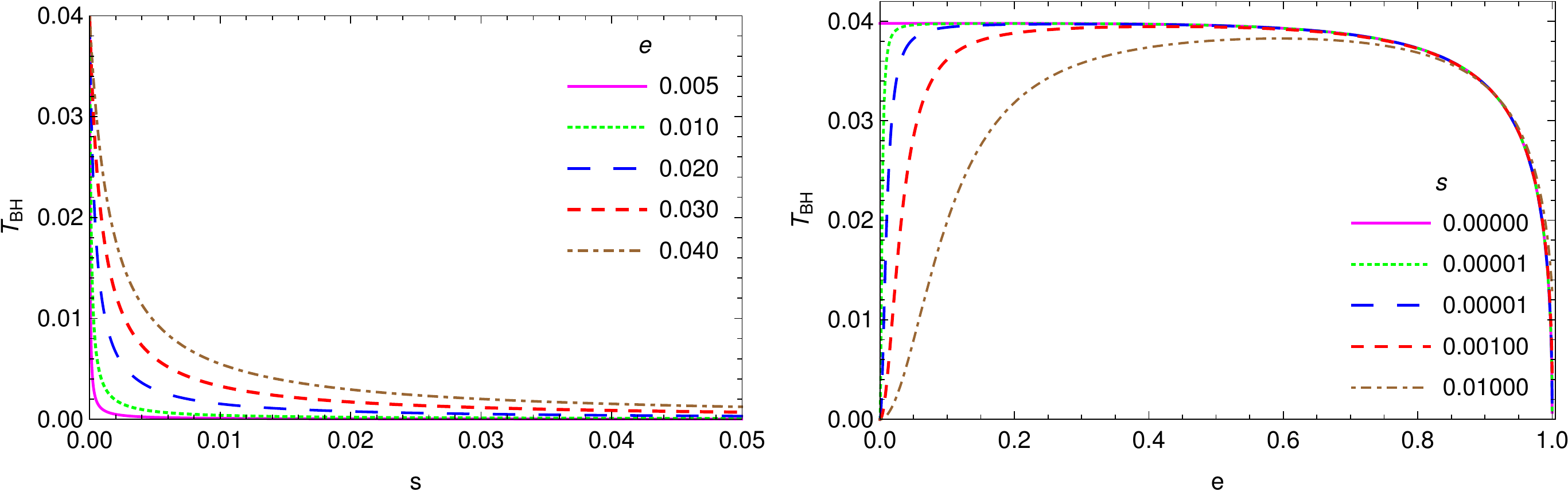}
\caption{{\bf (Left)} Plot of the Hawking temperature $T_{BH}$ of the scalar-hairy RN black hole with respect to the scalar charge $s$ for different values of the electric charge $e$  with $M_{ADM}=1$~. {\bf (Right)} Plot of the Hawking temperature $T_{BH}$ of the scalar-hairy RN black hole with respect to the electric charge $e$ for different values of the scalar charge $s$  with $M_{ADM}=1$~.}\label{ch4:fig_1}
\end{figure*}

Fig.\ref{ch4:fig_1}  shows the plot of the black hole temperature  with respect to the scalar and electric charges.

From Fig.\ref{ch4:fig_1} we observe that for a constant electric charge,  the black hole temperature $T_{BH}$  decreases very rapidly with $s$ but reaches a plateau quickly. For non-zero scalar charge, the black hole temperature rises with the electric charge, reaches a plateau and sharply falls to zero as the extremal value of the electric charge, $e_{ext}={\sqrt{\sqrt{M_{ADM}^2 \left(M_{ADM}^2+4 s\right)}+M_{ADM}^2}}/{\sqrt{2}}$~, is approached.  At sufficiently small values of $e$, $T_{BH}$ becomes vanishingly small.
\section{Charge-mass ratio bound from Mutual information of successively emitted charged Hawking quanta} \label{ch4:sec_charge-mass_ratio}
As already seen in Sec.\ref{ch4:sec_tunneling rate}, the Hawking radiation spectrum is not strictly thermal,  the emission in each step depends on the previous one. A quantity of importance in such a scenario is the ``Mutual Information", as defined in Refs.~\cite{zhang_PLB_2009, kim_PLB_2014},
\begin{equation}
\begin{split}
S_{MI}=  \Delta S_{charged}\left(M_{ADM},e;\omega_2,q_2|\omega_1,q_1\right)-\Delta S_{charged}\left(M_{ADM},e;\omega_2,q_2\right),
\end{split}
\end{equation}
where $\omega_1$, $\omega_2$and $q_1$, $q_2$ are the mass and charge of two consecutive emissions, $\Delta S_{charged}\left(M_{ADM},e;\omega_2,q_2|\omega_1,q_1\right)$ is the change in entropy of the black hole due to the emission of a particle of mass $\omega_2$ and charge $q_2$ considering a previous emission of mass $\omega_1$ and charge $q_1$.
$S_{MI}$ gives a measure of the correlation between two consecutive emission. The conservation of energy in the tunneling method automatically ensures the conservation of information~\cite{zhang_PLB_2009}. The non-negativity of mutual information during emission of two successive Hawking quanta gives rise to bounds on the charge to mass ratio of the emitted particles.

Far away from extremality, in the non superradiant regime, in the limit of large  black hole mass  and charge, $M_{ADM}\gg \left\{\omega_1, \omega_2\right\}$; $e\gg \left\{q1,q2\right\}$ and $M_{ADM}\gg e,s$, the non-negativity condition yields,
\begin{equation}
\begin{split}
 \frac{q_i}{\omega_i}\leq  \frac{1}{e^3}\left[ \sqrt{2}\sqrt{2s^2\left( M_{ADM}-\omega_i \right)^2+e^4 \left( s+e^2  \right)}-2s \left( M_{ADM}-\omega_i \right) \right],
\end{split}
\end{equation}
which can be written as
\begin{equation}\label{ch4:eq_q/w}
\frac{q_i}{\omega_i}\leq-\frac{2 s M_{ADM}}{e^3}+\sqrt{ \left( \frac{2 s M_{ADM}}{e^3} \right)^2+2\left(1+s/e^2\right) }~,
\end{equation}
where, for convenience, we have chosen $\omega_1=\omega_2=\omega_i$ and $q_1=q_2=q_i$ .
In the absence of the scalar hair ($s=0$),  Eq.\ \eqref{ch4:eq_q/w} reduces to $q_i/\omega_i\leq \sqrt{2}$~, which matches exactly with the charge-mass ratio for the RN black hole, obtained in Ref.~\cite{kim_PLB_2014}.
In order to obtain Eq.\ \eqref{ch4:eq_q/w}, we expanded the integrand in Eq.\ \eqref{ch4:eq_Somega} to leading orders in $\omega$ and $q'$  as the integration is otherwise difficult to carry out.
\begin{figure*}[!htp] \centering
\includegraphics[width=1\textwidth, height=0.38\textwidth]{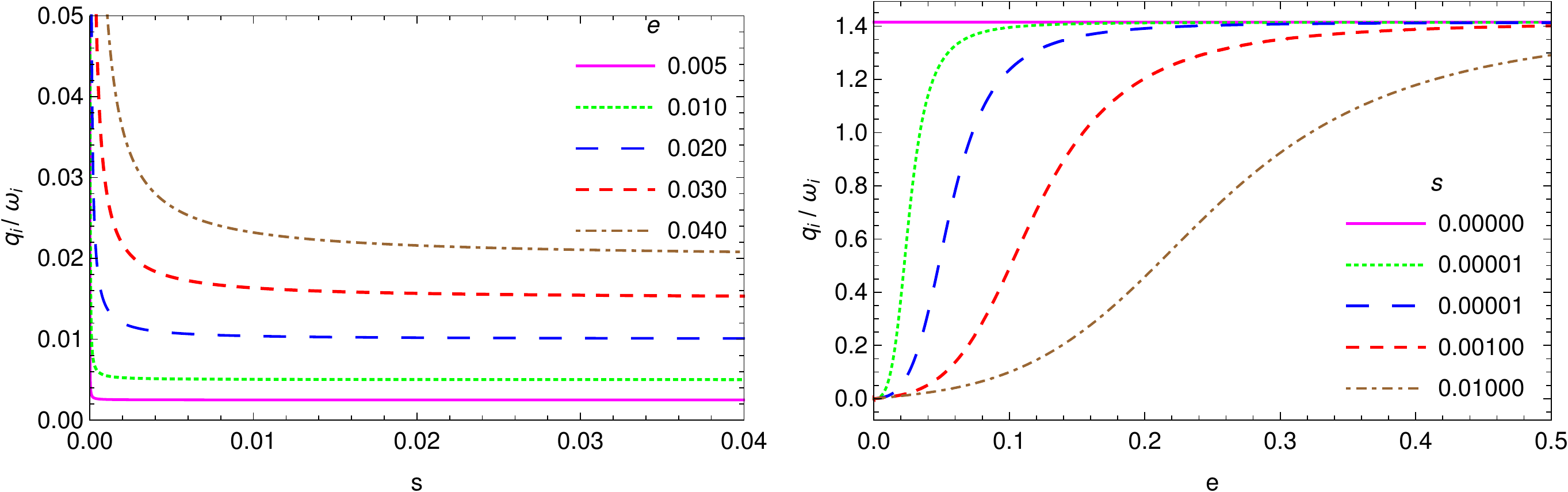}
\caption{{\bf (Left)} Plot of the maximum charge-mass ratio with $s$ for different values of the electric charge $e$ with $M_{ADM}=1$~. {\bf (Right)} Plot of the maximum charge-mass ratio with $e$ for different values of the scalar charge $s$ with $M_{ADM}=1$~.}\label{ch4:fig_2}
\end{figure*}

Fig.\ref{ch4:fig_2} shows the variation of the upper bound on the charge-mass ratio with the scalar and electric charges of the black hole. We observe that at a fixed electric charge of the black hole, for smaller values of the scalar charge, the upper bound on the charge-mass ratio decreases with $s$~; however, the rate of fall decreases with the increase in the scalar charge. On the other hand, for non-zero scalar charge, the maximum possible charge-mass ratio increases with the electric charge of the black hole  and approaches that for the standard RN black hole.
\section{Summary and Discussion}\label{ch4:sec_summary}
In this chapter, we studied the problem of Hawking emission of massive charged particles from a spherically symmetric charged black hole endowed with a scalar hair using the semi-classical tunneling formalism. The black hole is very similar to the standard Reissner-Nordstr\"{o}m black hole  but with an additional scalar hair~\citep{astorino_PRD_2013, chowdhury_EPJC_2018, chowdhury_GERG_2019}. The presence of the scalar hair gives rise to an effective Newtonian constant, $\tilde{G}=G\left( 1+s/e^2 \right)$ and an ADM mass, different from the mass parameter $M$ (see Eq.\ \eqref{ch4:eq_MADM}).
It is interesting to note that in the limit of vanishing electric charge,  the ADM mass of the black hole goes to zero which is due to the fact that the effective Newtonian constant $\tilde{G}$ blows up as $e\rightarrow0$. This suggests that the scalar-hairy RN black hole cannot radiate away its electric charge and settle down to a scalar-hairy (electrically) uncharged distribution. This is consistent with the following thermodynamic consideration. We note that for any positive value of the scalar charge, the black hole temperature becomes vanishingly small at sufficiently small values of the black hole electric charge (see Fig.\ref{ch4:fig_1}) and thus according to the Third Law of black hole thermodynamics~\cite{israel_PRL_1986}, it is impossible for the scalar-hairy RN black hole to radiate away its electric charge in any finite number of steps.
  
The total change in entropy of the scalar-hairy RN black hole due to the emission of the massive charged particle contains a $\omega$-dependent contribution due to the scalar charge (see Eqs. \eqref{ch4:eq_Somega},~\eqref{ch4:eq_Sbh} and \eqref{ch4:eq_Stotal}). For vanishing scalar charge $s$, the emission rate, (see Eq.\ \eqref{ch4:eq_tunneling_prob}), matches smoothly with that of the standard Reissner-Nordstr\"{o}m black hole~\cite{zhang_JHEP_2005}.
We also note that the Hawking emission spectrum deviates from pure thermality. This is consistent with the generic results found in the investigations~\cite{parikh_PRL_2000, parikh_IJMPD_2004, parikh_GRG_2004, parikh_MARCELGROSSMANN10, zhang_JHEP_2005, jiang_PLB_2006, liu_IJTP_2014, jiang_PRD_2006, jiang_IJTP_2006, jiang_PLB_2008, sarkar_PLB_2008, wu_JHEP_2006, li_PLB_2006, chen_PLB_2008, chen_CQG_2008, jiang_PRD_2008, jiang_JCAP_2013, zhang_PLB_2005} and the recent findings that the quantum process involved in Hawking radiation is unitary, meaning a pure state to pure state transition~\cite{chen_PLB_2009,singleton_JHEP_2010,sakalli_ASS_2012,saini_PRL_2015, arpit_EPJC_2019}.

It is important to note that using the modified geodesic equation instead of the Eq.\ \eqref{ch4:eq_vp_vg} as suggested by Pu and Han~\cite{pu_IJTP_2017} also yields the same expression of the tunneling rate.

We also studied the mutual information stored in consecutive Hawking emission. Demanding non-negativity of the mutual information, we observed that the maximum allowed charge-mass ratio of the emitted particles decreases with the scalar charge of the black hole. 




\chapter{Greybody factor and sparsity of Hawking radiation from a charged spherical black hole with scalar hair} \blfootnote{\begin{flushleft} The work presented in this chapter is based on “Greybody factor and sparsity of Hawking radiation from a charged spherical black hole with scalar hair”, \textbf{Avijit Chowdhury} and Narayan Banerjee, Phys.\ Lett.\ B \textbf{805}, 135417 (2020).\end{flushleft}} 
\label{chap5}
\chaptermark{Greybody factor and sparsity of Hawking radiation \ldots}

\section{Introduction}
To an asymptotic observer, the Hawking emission spectrum is not an exact black-body like Planckian distribution. The geometry interpolating between the black hole horizon and the asymptotic observer allows only a fraction of the emitted radiation to reach the asymptotic observer. This deviation of the Hawking emission spectrum from the perfect black body spectrum is described in terms of the greybody factor~\cite{page_PRD_1976,page_PRD2_1976}. Another important aspect of the Hawking radiation flow is its sparsity~\cite{gray_CQG_2016, hod_PLB_2016, hod_EPJC_2015, miao_PLB_2017,schuster_thesis, bibhas_IJMPA_2017}. The Hawking emission is known to be extremely sparse, that is the average time gap between emission of successive Hawking quanta is large compared to the characteristic time-scale of individual Hawking emission.

In Chapter~\ref{chap4}, we studied the Hawking emission of charged particles from the scalar-hairy RN black hole. We observed that the total change in entropy of the black hole due to the emission of massive charged particles contains an additional frequency-dependent contribution due to the black hole scalar charge. We also observed the nontrivial dependence of the black hole temperature on the scalar charge and the lowering of the maximum allowed charge-mass ratio of the emitted particles with the black hole scalar charge. So, one naturally asks, how the scalar hair affects the black hole greybody factor and the sparsity of the Hawking radiation ? 

Earlier, there have been attempts to study the sparsity and greybody factor of Hawking radiation from higher dimensional ($D>4$) black holes with scalar hair (of a different nature than that considered in this work, see Ref.~\cite{miao_PLB_2017} for example), however, the (3+1) dimensional scalar-hairy RN black hole is still unexplored. This work attempts to fill this gap and provide a cohesive understanding of the behaviour of the black hole greybody factor and the sparsity of Hawking radiation flow from the scalar-hairy RN black hole.

In this chapter, we specifically consider the emission of massless uncharged scalar particles from the scalar-hairy RN black hole. We find that unlike the electric charge which reduces the greybody factor, the scalar charge enhances the same. The scalar charge also reduces the sparsity of the Hawking cascade whereas the electric charge enhances it. {The greybody factor increases with the (ADM) mass of the black hole and the sparsity of the Hawking emission cascade decreases.
} 
We recollect from Chapter~\ref{chap4} that presence of the scalar hair modifies the ADM mass of the black hole,
\begin{equation}\label{ch5:eq_MADM}
M_{ADM}=\frac{M}{1+s/e^2}~,
\end{equation}
from the standard Reissner-Nordstr\"{o}m case $\left(M_{ADM}=M\right)$ and gives rise to a Hawking temperature of 
\begin{equation}\label{ch5:eq_TBH}
T_{BH}=\frac{r_+ - r_-}{4 \pi r_+^2}=\frac{\sqrt{M_{ADM}^2 \left(\frac{s}{e^2}+1\right)^2-e^2-s}}{2 \pi  \left(\sqrt{M_{ADM}^2 \left(\frac{s}{e^2}+1\right)^2-e^2-s}+M_{ADM} \left(\frac{s}{e^2}+1\right)\right)^2}~.
\end{equation}


\section{Hawking emission of uncharged particles}\label{sec:hawking_rad}
The energy emitted per unit time by a black hole at temperature $T_{BH}$ with frequency $\omega$ in the momentum interval $d^3\overrightarrow{k}$
is given by~\cite{miao_PLB_2017,gray_CQG_2016}
\begin{equation}
\frac{dE(\omega)}{dt}= \sum_l T_l (\omega)\frac{\omega}{e^{\omega/T_{BH}}-1} \hat{k} \cdot  \hat{n}~  \frac{d^3 k ~dA}{(2\pi)^3},
\end{equation}
where $\hat{n}$ is the unit normal to the surface element $dA$, $l$ is the angular momentum quantum number and $T_l(\omega)$ is the frequency dependent greybody factor. {$T_l(\omega)=1$ signifies perfect black body spectrum.} For massless particles $\big|\overrightarrow{k}\big|=\omega$. Integrating over the finite surface area $A$, we get the total power of Hawking radiation  as,
\begin{equation}\label{ch5:eq_P}
P=\sum_l \int_{0}^{\infty} P_l\left(\omega\right) d\omega,
\end{equation}
where
\begin{equation}\label{ch5:eq_Pl}
P_l\left(\omega\right)=\frac{A}{8\pi^2} T_l(\omega)\frac{\omega^3}{e^{\omega/T_{BH}}-1} 
\end{equation}
is the power emitted per unit frequency in the $l^{th}$ mode. The area $A$ is usually a multiple of the horizon area. For Schwarzschild black hole, $A$ is taken to be  $(27/4 )$ times the horizon area. This value of the effective surface area ensures that in the limit of vanishing electric charge of the black hole, the low-frequency results smoothly match with the high-frequency results. Here, we will consider the area $A$ to be the horizon area $A_H$, where $A_H= 4 \pi r_+^2$, as this does not affect the qualitative behaviour of the results.

The excitation of massless uncharged scalar fields around the scalar-hairy RN black hole is governed by the Klein-Gordon equation,
\begin{equation}\label{ch5:eq_KG}
\square \Phi _{l m}\left(t,r,\theta,\phi\right)=0
\end{equation} 
Decomposing the scalar field as
\begin{equation}
\Phi_{lm}\left(t,r,\theta,\phi\right)=e^{-i\omega t}Y^m_l\left(\theta,\phi\right)R_{lm}\left(r\right),
\end{equation}
{where $Y^m_l\left(\theta,\phi\right)$ are the spherical harmonics and $R_{lm}\left(r\right)$ is the radial component of $\Phi$,} we get the radial Klein-Gordon equation in the tortoise coordinate $r_*$~\cite{chowdhury_EPJC_2018, chowdhury_GERG_2019}, as,
\begin{equation}\label{ch5:eq_KG}
\frac{d^2 R_{lm}(r)}{d r_*^2}+\left( \omega^2-V_{eff}(r) \right)R_{lm}(r)=0,
\end{equation}
where
\begin{equation}\label{ch5:eq_Veff_unchrgd}
V_{eff}=f(r)\left(-\frac{2\left( e^2+s \right)}{r^4} +\frac{2M}{r^3}+\frac{l(l+1)}{r^2}\right)
\end{equation}
is the effective potential, $f\left( r \right)=\left(1-\frac{2M}{r}+\frac{e^2+s}{r^2}\right)$, $\omega$ is the conserved frequency, $l$ is the spherical harmonic index and $m$ ($-l\leq m\leq l$) is the azimuthal harmonic index.
{The radial coordinate $r$ is an intrinsic function of the tortoise coordinate, $r_*$,  defined by  $dr_*=dr/f(r)$, mapping the semi-infinite region $\left[ \left.r_+,\infty \right)\right.$ to $(-\infty, \infty)$.}
The effective potential $V_{eff}$, vanishes both at the horizon and near spatial infinity. Thus, the solution at the horizon consists only of ingoing modes whereas near infinity, it consists of both ingoing and outgoing modes. 

A fraction of the radiation emitted by the black hole is reflected back by the effective potential while the remaining is transmitted out. The greybody factor measures the transmission probability of the outgoing Hawking quanta to reach future infinity without being back-scattered by this effective potential.

\subsection{Bounds on the greybody factor} \label{subsec:bound}
There are various methods in literature~\cite{cardoso_PRL_2006,cardoso_JHEP_2006,harmark_ATMP_2010,grain_PRD_2005,fernando_GERG_2005,ida_PRD_2003,cvetic_FDP_2000,cvetic_PRD_1998,klebanov_NPB_1997,neitzke_ARXIV_2003,motl_ATMP_2003} to estimate the greybody factors, however,  derivation of exact analytical expression of greybody factor is limited only to very few cases. Following Refs.~\cite{visser_PRA_1999, boonserm_AP_2008, boonserm_thesis,boonserm_PRD_2008, ngampitipan_JPCS_2013,boonserm_JHEP_2014, miao_PLB_2017}, in this work, we will provide rigorous  bounds on the greybody factors of the scalar-hairy RN black hole. {A brief description of the method to put general bounds on the transmission probability is given in Appendix~\ref{AppendixB}.}
The general bounds on the greybody factor, as proposed by Visser~\cite{visser_PRA_1999} is given by,
\begin{equation}
\label{ch5:eq_bound}
T_l (\omega)\geq \sech^2\left\{ \int_{-\infty}^{\infty} \vartheta dr_{*} \right\}
\end{equation}
where \begin{equation}\label{ch5:eq_bound1}
\vartheta=\frac{\sqrt{\left[h'(r)\right]^2+\left[ \omega^2-V_{eff}-h(r)^2 \right]^2}}{2h(r)}.
\end{equation}
The arbitrary function $h(r)$ has to be positive definite everywhere and satisfy the boundary condition, $h(\infty)=h(r_+)=\omega$ for the bound~\eqref{ch5:eq_bound} to hold. A particularly simple choice of $h(r)$ for the present case is 
\begin{equation}\label{ch5:eq_h_unchrgd}
h(r)=\omega.
\end{equation}
Substituting Eq.\eqref{ch5:eq_h_unchrgd} in Eq.\eqref{ch5:eq_bound1} and using the definition of $r_*$, we get
\begin{equation}\label{ch5:eq_bound2}
\int_{-\infty}^{\infty} \vartheta dr_*=\int_{r_+}^{\infty} \frac{V_{eff}}{2 \omega f(r)}  dr.
\end{equation}
Eq.\eqref{ch5:eq_bound} in conjunction with  Eqs.\eqref{ch1:eq_f(r)},\eqref{ch1:eq_r_+-} \eqref{ch5:eq_Veff_unchrgd} and \eqref{ch5:eq_bound2} yields a relatively simple expression for the lower bound of the greybody factor,
\begin{equation}\label{ch5:eq_tl_unchrgd_rh}
T_l(\omega)\geq \sech^2\left\{\frac{1+2l(l+1)}{4r_+ \omega}-\frac{e^2+s}{12r_+^3 \omega} \right\}.
\end{equation}
Using the expression of the horizon radius, $r_+$ (Eq\eqref{ch1:eq_r_+-}), in terms of the ADM mass  (Eq.\eqref{ch5:eq_MADM}), we finally obtain, 
		\begin{equation}\label{ch5:eq_tl_unchrgd_Madm}
	\begin{split}	
	T_l(\omega)\geq \sech^2&\left\lbrace\frac{s M_{ADM}}{2 e^2 \omega  \left(\sqrt{ M_{ADM} ^2 \left(\frac{s}{e^2}+1\right)^2-e^2-s}+ M_{ADM}  \left(\frac{s}{e^2}+1\right)\right)^2}\right.\\
	&\left.-\frac{e^2+s}{3\omega \left(\sqrt{ M_{ADM} ^2 \left(\frac{s}{e^2}+1\right)^2-e^2-s}+ M_{ADM}  \left(\frac{s}{e^2}+1\right)\right)^3}\right.\\	
	&\left.+\frac{ l (l+1) M_{ADM} }{2 \omega  \left(\sqrt{ M_{ADM} ^2 \left(\frac{s}{e^2}+1\right)^2-e^2-s}+ M_{ADM}  \left(\frac{s}{e^2}+1\right)\right)}\right\rbrace .
	\end{split}
	\end{equation}
	
To better understand the dependence of the greybody factor and the power spectrum on the black hole scalar and electric charges and the black hole mass, we plot the lower bound of the greybody factor~\eqref{ch5:eq_tl_unchrgd_Madm} and the power spectrum\eqref{ch5:eq_Pl}  for different values of $s$, $e$ and $M_{ADM}$.

\begin{figure}[!h]
\centering
\includegraphics[width=0.99\linewidth, height=0.45\linewidth, keepaspectratio]{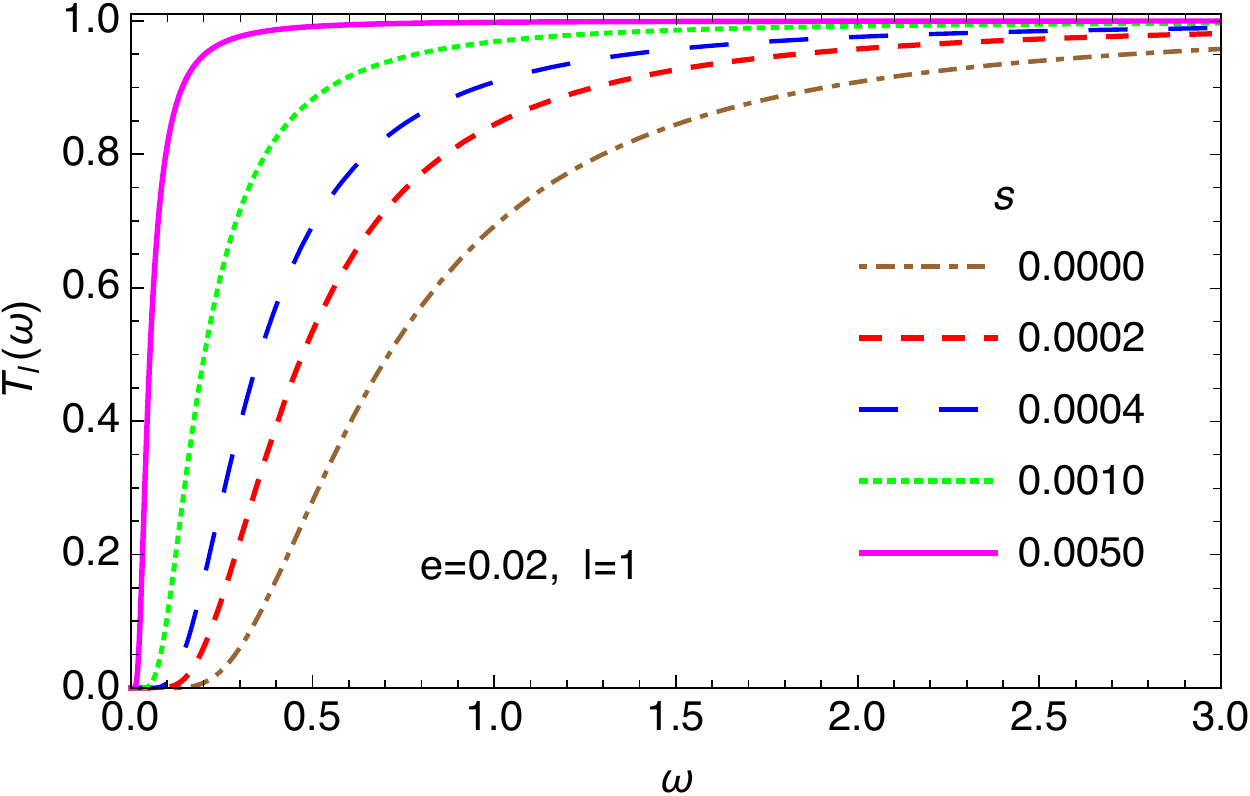}\\
\includegraphics[width=0.99\linewidth, height=0.45\linewidth, keepaspectratio]{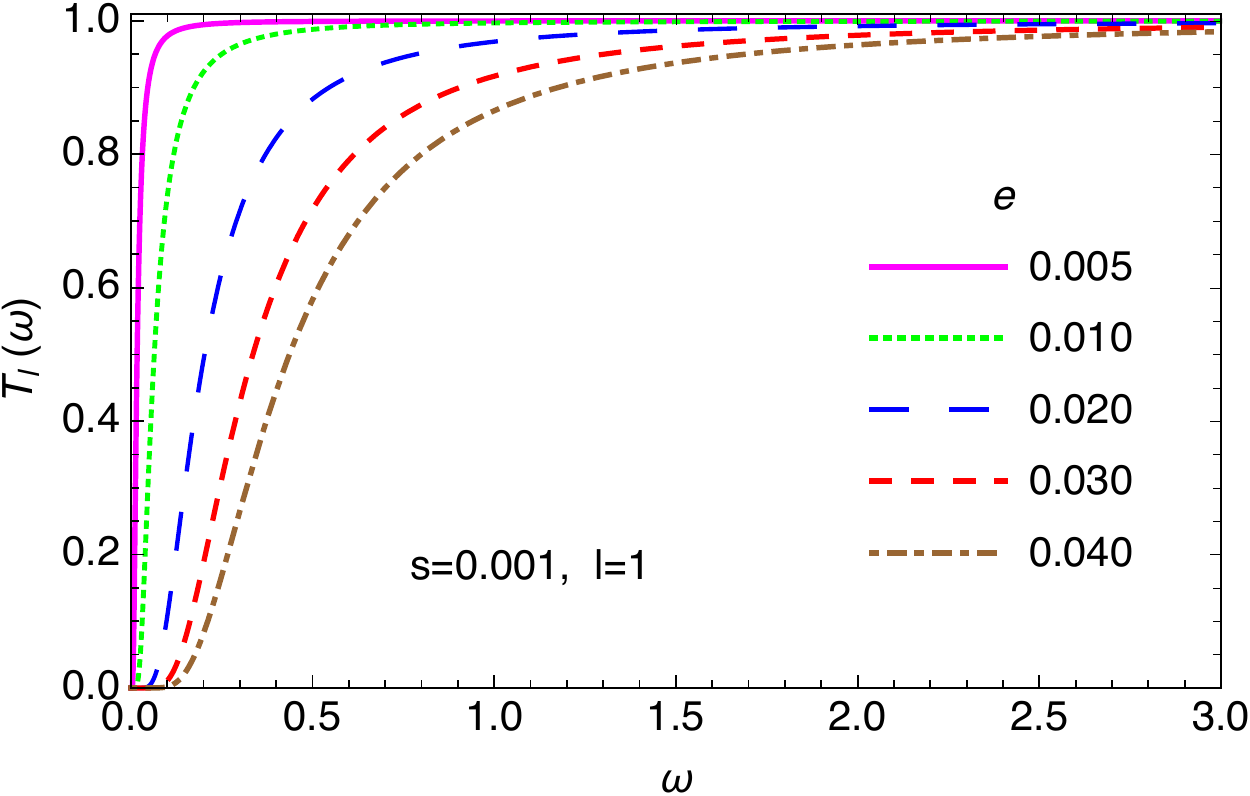}\\
\includegraphics[width=0.99\linewidth, height=0.45\linewidth, keepaspectratio]{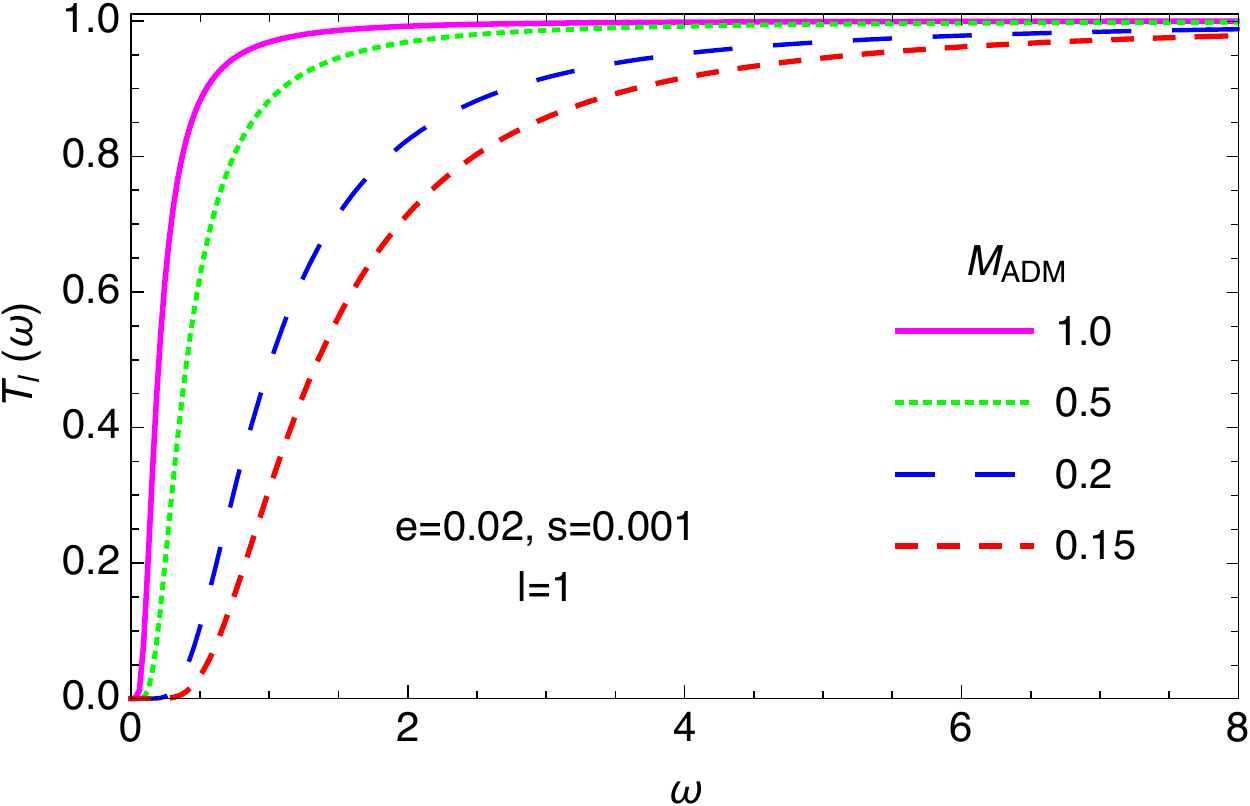}
\caption{{\bf (Top)} Plot of lower bound of the greybody factor $T_l(\omega)$ with $\omega$ for different values of the scalar charge $s$ with $e=0.02$, $l=1$ and $M_{ADM}=1$. {\bf (Middle)} Plot of lower bound of the greybody factor $T_l(\omega)$ with $\omega$ for different values of the electric charge $e$ with $s=0.001$, $l=1$ and $M_{ADM}=1$. {\bf (Bottom)} Plot of lower bound of the greybody factor $T_l(\omega)$ with $\omega$ for different values of  $M_{ADM}$ with $e=0.02$,  $s=0.001$ and $l=1$.}\label{ch5:fig:1}
\end{figure}
\begin{figure}[!h]
\centering
\includegraphics[width=0.99\linewidth, height=0.45\linewidth, keepaspectratio]{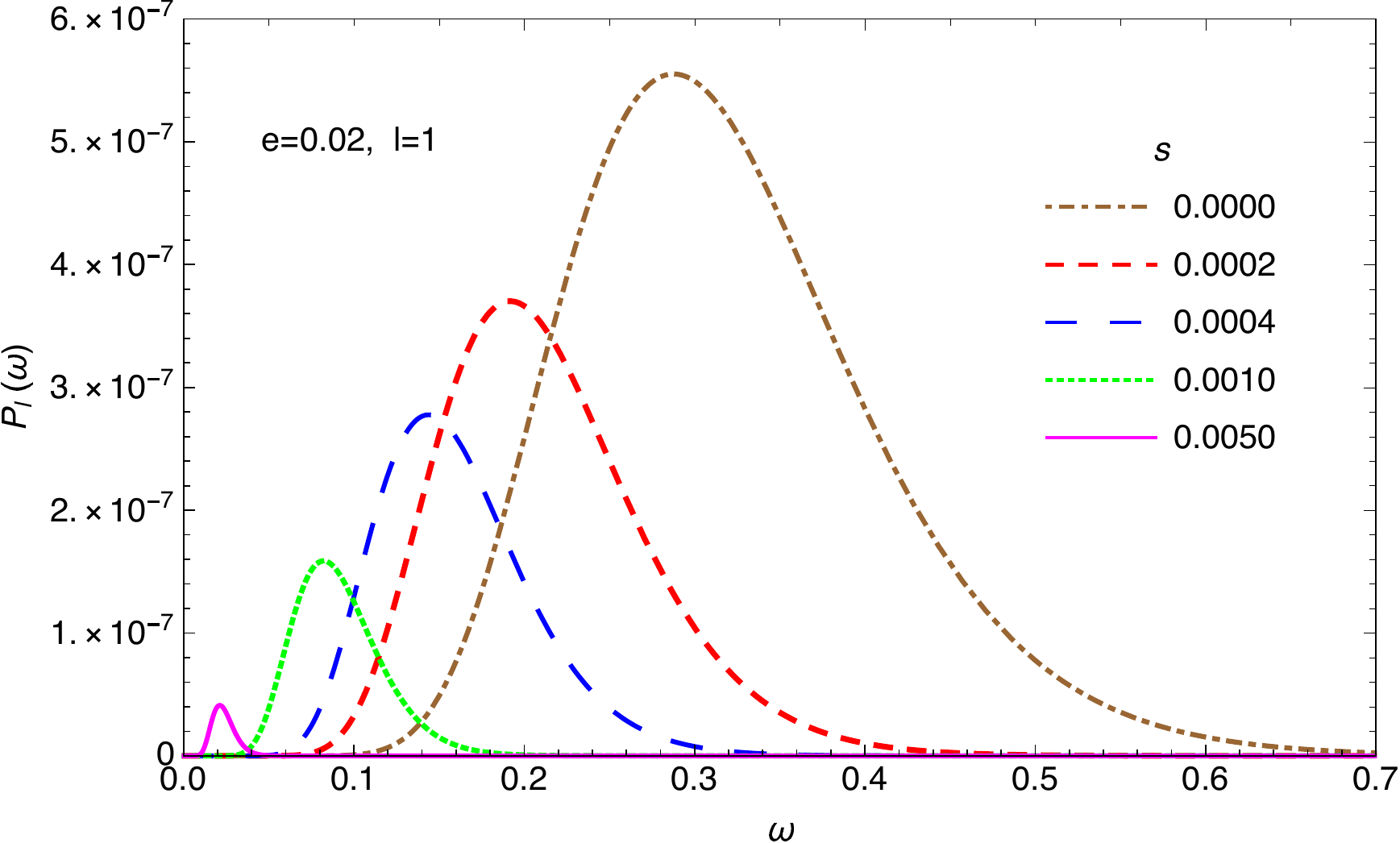}\\
\includegraphics[width=0.99\linewidth, height=0.45\linewidth, keepaspectratio]{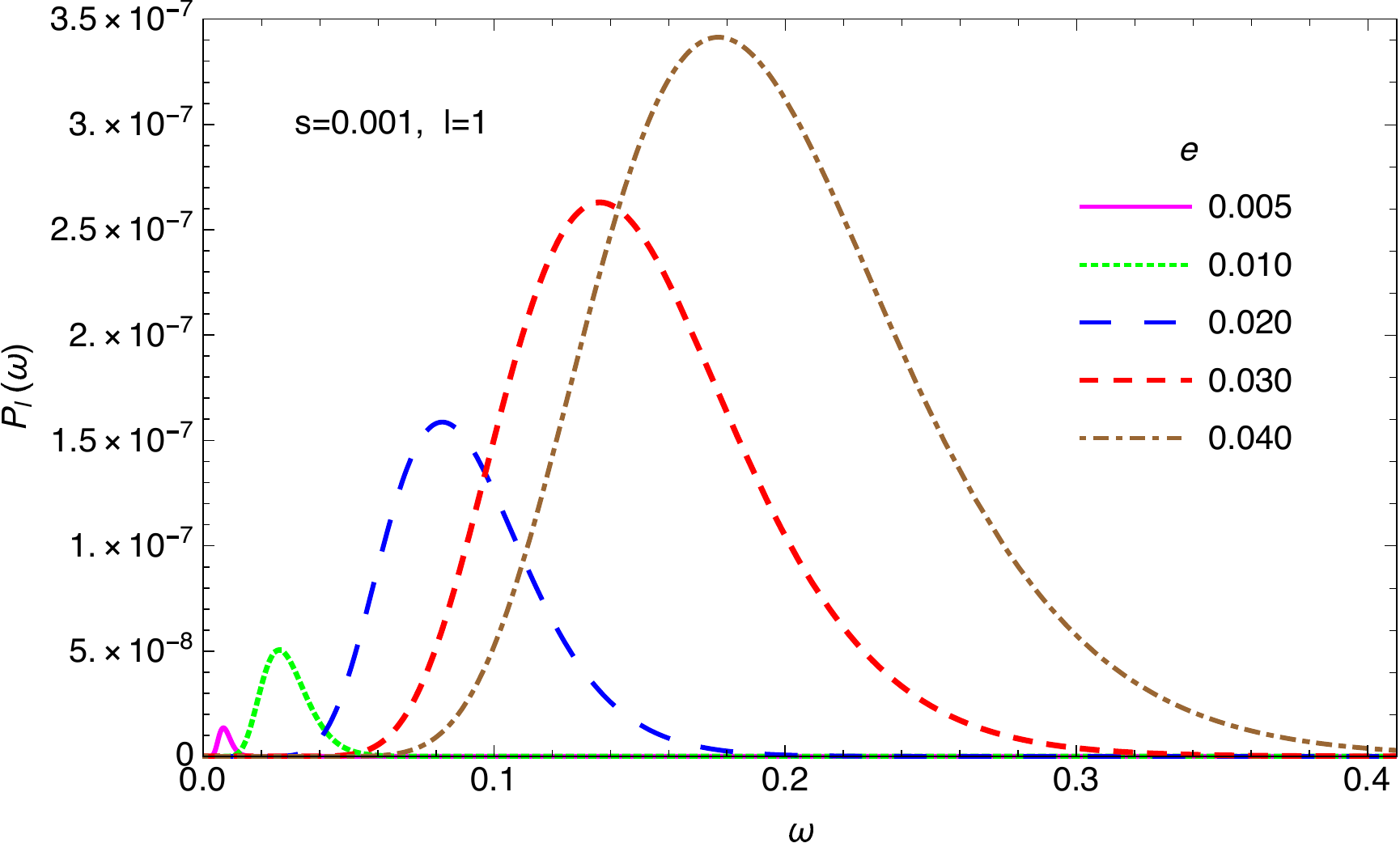}\\
\includegraphics[width=0.99\linewidth, height=0.45\linewidth, keepaspectratio]{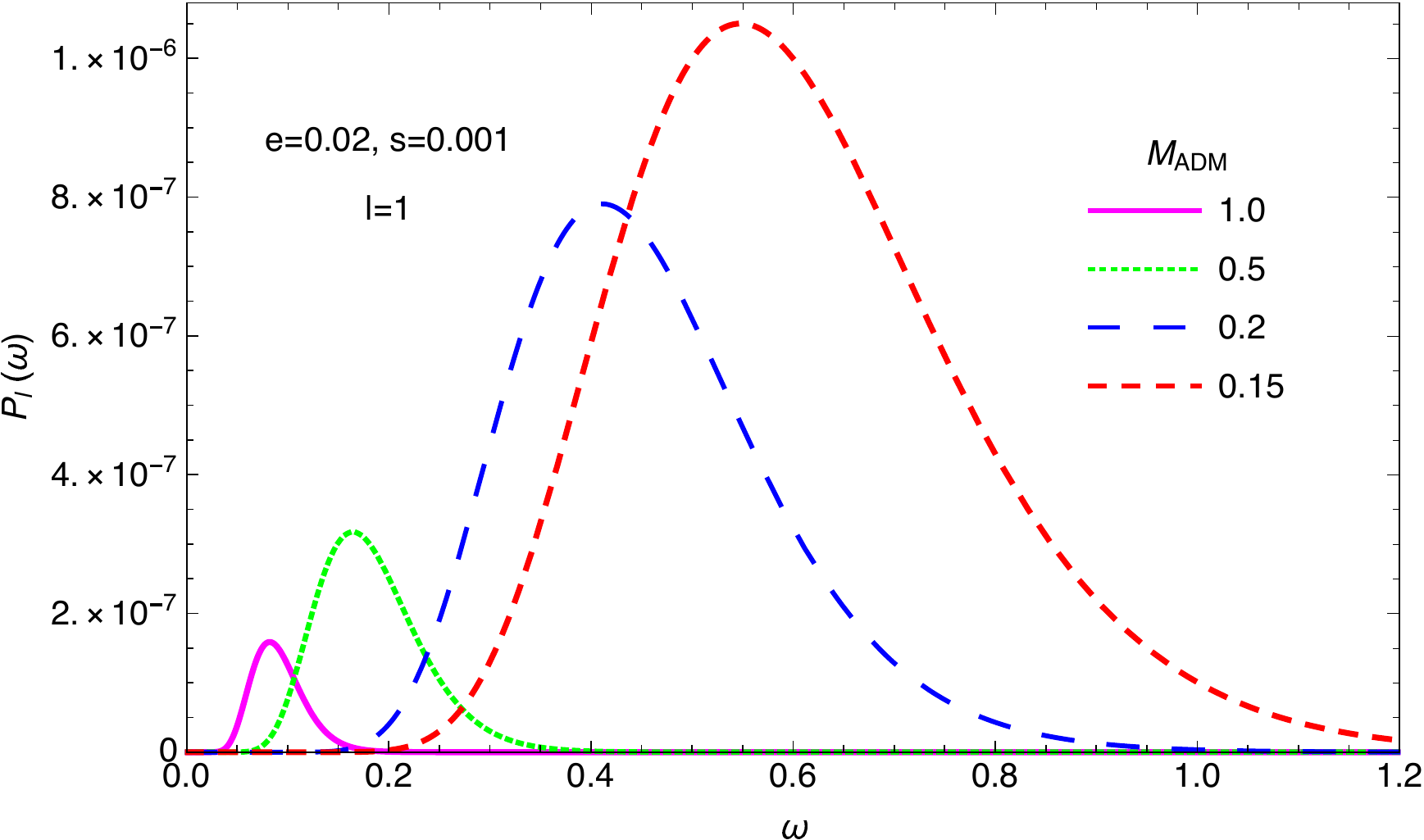}
\caption{{\bf (Top)} Plot of the power spectrum $P_l(\omega)$ with $\omega$ for different values of the scalar charge $s$ with $e=0.02$, $l=1$ and $M_{ADM}=1$. {\bf (Middle)} Plot of the power spectrum $P_l(\omega)$ with $\omega$ for different values of the electric charge $e$ with $s=0.001$, $l=1$ and $M_{ADM}=1$. {\bf (Bottom)} Plot of the power spectrum $P_l(\omega)$ with $\omega$ for different values of  $M_{ADM}$ with $e=0.02$,  $s=0.001$ and $l=1$.}\label{ch5:fig:2}
\end{figure}
We observe from the top and middle panels of Fig.\ref{ch5:fig:1}, that for a fixed ADM mass and spheroidal harmonic index, the lower bound of the greybody factor increases with the increase in the black hole scalar charge for a given value of the black hole electric charge whereas,  for fixed non zero values of the scalar charge, it decreases with increasing values of the electric charge.  The bottom panel of Fig.\ref{ch5:fig:1} shows that for fixed values of the black hole electric and scalar charges, the lower bound of the greybody factor decreases with the decrease in the black hole ADM mass.  In all cases, however, the far away observer misses more of the lower frequency contribution than the higher frequencies.  From the top and middle panels of Fig.\ref{ch5:fig:2}, we see that for a fixed ADM mass and spheroidal harmonic index, the peak of the power spectrum decreases and shifts towards lower frequencies as the scalar charge increases for a given value of the electric charge of the black hole, whereas, for a given non-zero scalar charge, it increases and shifts towards higher frequencies with the increase in  the electric charge. The bottom panel of Fig.\ref{ch5:fig:1} shows that for fixed values of the black hole electric and scalar charges, the peak of the power spectrum increases and shifts towards higher frequencies with the decrease in the black hole ADM mass.

\subsection{Sparsity of Hawking radiation}\label{subsec:sparsity}

As discussed in section~\ref{ch1:sec_sparsity}, the sparsity of Hawking radiation can be quantitatively defined using the dimensionless parameter $\eta$ as,
\begin{equation}\label{ch5:eq_eta_unchrgd}
\eta=\frac{\tau_{gap}}{\tau_{emission}},
\end{equation}
as a figure of merit. $\tau_{gap}$ is the average time interval between the emission of two successive Hawking quanta,
\begin{equation}\label{ch5:eq_tgap_unchrgd}
\tau_{gap}=\frac{\omega_{peak}}{P},
\end{equation}
where $P$ is defined in Eq.\eqref{ch5:eq_P} and $\omega_{peak} $ is the frequency at which the peak of the power spectrum~(Eq.\eqref{ch5:eq_Pl}) occurs, considering complete transmission,  \textit{i.e.,}  the position of the maximum of $\omega^3/\left(e^{\omega/T_{BH}}-1\right)$,
\begin{equation}\label{ch5:eq_wpeak_unchrgd}
\omega_{peak}=T_{BH} \left[ 3+ W \left(-3 e^{-3}\right)\right],
\end{equation}
where $W(x)$ is the Lambert $W$-function, defined as \\$W(x) e^{W(x)}=1$. $\tau_{emission}$  is the characteristic time for the emission of individual Hawking quantum and is bounded from below by the localisation time-scale, $\tau_{localisation}$, the characteristic time taken by the emitted wave field  with frequency $\omega_{peak}$ to complete one cycle of oscillation,
\begin{equation}\label{ch5:eq_temm_unchrgd}
\tau_{emission} \geq \tau_{localisation}=\frac{2 \pi}{\omega_{peak}}.
\end{equation}

Thus, $\eta\gg 1$ implies that the time gap between the emission of successive Hawking quanta is large compared to the time taken for the emission of individual Hawking quantum, suggesting an extremely sparse Hawking cascade. On the other hand, $\eta\ll 1$ suggests that the Hawking radiation flow is almost continuous. Table \ref{ch5:table_1}, \ref{ch5:table_2} and \ref{ch5:table_3} show the numerical values of  $\eta_{max}=\tau_{gap}/\tau_{localisation}$, $\left(\eta\leq\eta_{max}\right)$ for different values of the scalar and electric charges and the ADM mass.
\begin{table}[!h]
\centering
\caption{Numerical values of the dimensionless parameter $\eta_{max}=\tau_{gap}/\tau_{localisation}$ for the $l=1$ mode with $e=0.02$ and $M_{ADM}=1$ for different values of the scalar charge $s$ }
\label{ch5:table_1}
\begin{tabular}{cccccc}
\hline\hline  
\rule[-1ex]{0pt}{2.5ex} $s$ & 0.0000 & 0.0002 & 0.0004 & 0.0010 & 0.0050 \\ 
\rule[-1ex]{0pt}{2.5ex} $\eta_{max}$ & 16931.0 & 16927.7 & 16926.1 & 16924.0 & 16921.9 \\
\hline \hline 
\end{tabular} 
\end{table}
\begin{table}[!h]
\centering
\caption{Numerical values of the dimensionless parameter $\eta_{max}=\tau_{gap}/\tau_{localisation}$ for the $l=1$ mode with $s=0.001$ and $M_{ADM}=1$ for different values of the electric charge $e$ }
\label{ch5:table_2}
\begin{tabular}{cccccc}
\hline \hline 
\rule[-1ex]{0pt}{2.5ex} $e$ & 0.005 & 0.010 & 0.020 & 0.030 & 0.040 \\  
\rule[-1ex]{0pt}{2.5ex} $\eta_{max}$ & 16921.2 & 16921.4 & 16924.0 & 16931.6 & 16945.3 \\ 
\hline \hline 
\end{tabular} 
\end{table}
\begin{table}[!h]
\centering
\caption{Numerical values of the dimensionless parameter $\eta_{max}=\tau_{gap}/\tau_{localisation}$ for the $l=1$ mode with $e=0.02$ and $s=0.001$ for different values of $M_{ADM}$ }
\label{ch5:table_3}
\begin{tabular}{cccccc}
	\hline \hline 
	\rule[-1ex]{0pt}{2.5ex} $M_{ADM}$ & 1.00 & 0.80 & 0.50 & 0.20 & 0.15 \\  
	\rule[-1ex]{0pt}{2.5ex} $\eta_{max}$ & 16924 & 16925.6 & 16932.4 & 16991.5 & 17046.6 \\ 
	\hline \hline 
\end{tabular} 
\end{table}
The high values of the dimensionless parameter suggest that the Hawking cascade of massless uncharged scalar quanta from the scalar-hairy RN black hole is extremely sparse. Table~\ref{ch5:table_1} and Table~\ref{ch5:table_2} show that with the increase of scalar charge $s$, the sparsity decreases as opposed to the enhancement of the sparsity of the Hawking cascade with the electric charge $e$. Table~\ref{ch5:table_3} shows that for fixed $e$ and $s$, the sparsity of the Hawking radiation cascade increases with the decrease in the ADM mass of the black hole. Though these variations are steady and monotonic, the variations are rather small.


\section{Summary and Discussion }\label{ch5:sec_sum}

Despite the rather modest appearance of the scalar hair in the scalar-hairy-Reissner-Nordstr\"{o}m black hole (see Eq.\eqref{ch1:eq_shRN}) recent studies~\cite{astorino_PRD_2013, chowdhury_EPJC_2018, chowdhury_EPJC_2019} have shown that the presence of this additional scalar field greatly modifies the known physics of the standard Reissner-Nordstr\"{o}m black hole.  In the present study, we analyzed the greybody factor and  also estimated the dimensionless parameter, $\eta=\tau_{gap}/\tau_{emission}$, to study the sparsity of Hawking radiation from the scalar-hairy-Reissner-Nordstr\"{o}m black hole. We considered the emission of massless uncharged scalar quanta for this purpose.

We find that the black hole scalar and the electric charges oppositely affect the greybody factor.  Increasing the scalar charge increases the greybody factor, whereas, increasing the electric charge has the effect of lowering the greybody factor.  For a given electric charge, the total Hawking radiation power emitted in each mode decreases with the scalar charge. The peak of the power spectrum also diminishes and shifts towards lower frequencies. This in-turn reduces the sparsity of Hawking radiation flow with the scalar charge. However,  increasing the electric charge increases the power of Hawking radiation. The peak value of the emitted power spectrum also increases and shifts towards higher frequencies and the Hawking flux becomes even more sparse.  
 
As the black hole continues to Hawking radiate, its ADM mass decreases. The lowering of the ADM mass, on one hand, raises the black hole temperature $T_{BH}$ (see Eq.\eqref{ch5:eq_TBH}) and enhances the Hawking emission power, while on the other hand, it reduces the grey body factor and  increases the sparsity of the Hawking radiation cascade. However, for the emission of uncharged particles, the ADM mass of the black hole cannot reduce to zero since it is bounded by the mass of the extremal black hole, $M_{ADM}\geq \frac{e^2}{\sqrt{e^2+s}}$. 

It is also interesting to note that if quantum gravity effects result in the formation of a black hole remnant at the end stages of Hawking radiation~\cite{chen_PR_2015, ong_JHEP_2018, ong_PLB_2018, gohar_PRD_2018, gohar_IJMPD_2018}, then, Eq.\eqref{ch5:eq_MADM} suggests that for non-zero scalar charge, the remnant has to be electrically charged. A detailed study of such a remnant is outside the scope of this thesis.

Both the greybody factor and the sparsity are considered to be important characteristics of Hawking radiation, which should help understanding the mechanism of the quantum processes involved. As the modes of scalar fields are instrumental in the formulation of Hawking radiation, these investigations are likely to be extremely relevant.


\chapter{Summary and Conclusions}
\label{chap6}
Observation of gravitational waves by the LIGO-Virgo collaboration \cite{LIGO_GW151266, ligo_PRL_2016, LIGO_GW170104, abbott_APJ_2017, ligo_PRL_2017_1, abott_PRL_2017, ligo_PRX_2019}, supplemented by the findings of the Event Horizon Telescope collaboration~\cite{eht_ApJ_2019, eht_ApJ_2019_1, zhu_ApJ_2018} has proved with certainty that black holes are not merely mathematical constructs but actual physical entities. The unique properties of a black hole are due to the existence of the event horizon, which acts as a one-way membrane, causally separating its interior from the rest of the universe. A black hole is believed to be characterised only by a few parameters, its mass, electric charge and angular momentum~\cite{ruffini_PT_1971}. This is the so-called `No-hair conjecture'. Thus, no matter how a black hole is formed or what information one throws in it, once it is inside the black hole, it is lost irreversibly; the exterior geometry does not convey much information except for the mass, electric charge and angular momentum of the black hole. Since the idea of the No-hair conjecture came into being, there were attempts to find `hairy' black hole solutions. A `hair' in this context refers to a black hole parameter, other than its mass, electric charge and angular momentum that can be measured by an exterior observer. The success of scalar fields in fundamental particle physics and cosmology (see Ref.~\cite{saha_2018, faraoni_book} and references therein), presented them as a viable black hole-hair candidate. A particularly promising class of scalar-hairy black holes are the Bocharova–Bronnikov–Melnikov–Bekenstein (BBMB)~\cite{bocharova_1970,bekenstein_AP_1974,bekenstein_AP_1975} black hole where a massless scalar field is conformally coupled to gravity. However, as discussed in Chapter~\ref{chap1}, the BBMB black hole solutions are plagued with many problems. In 2013, Astorino~\cite{astorino_PRD_2013} provided a charged generalization of BBMB black holes that is devoid of the anomalies of the general BBMB solutions. In the present thesis, we considered the static, spherically symmetric, electrically charged black hole with scalar hair~\eqref{ch1:eq_shRN}, put forward by Astorino.  An interesting feature of this spacetime as discussed in section~\ref{ch1:sec_shRN} is that when the scalar charge $s$ is smaller than the negative of the square of the black hole's electric charge $e$, the spacetime effectively behaves as a mutated Reissner-Nordstr\"{o}m black hole, leading to an Einstein-Rosen bridge or a wormhole~\cite{rosen_PR_1935}. The motivation of the present thesis is to look at the effect of the additional scalar field on the quasinormal mode spectrum, superradiant stability and Hawking radiation from a scalar-hairy Reissner-Nordstr\"{o}m black hole.

In Chapter~\ref{chap2} we explored the effect of this primary scalar hair on the quasinormal mode spectrum due to perturbation by scalar and Dirac fields. We also studied the impact of the mass of the scalar field on the QNM frequencies.  As discussed in section~\ref{ch2:sec_sum}, we found that in the mutated RN regime ($s<-e^2$), the damping rate (imaginary part of the quasinormal frequency) varies monotonically with the black hole's electric charge of the black hole as opposed to the existence of a peak for the pure Reissner-Nordstr\"{o}m case. In almost all cases, both the frequency and the damping rate decrease with the magnitude of the negative scalar charge. For a massive scalar field, the damping rate falls off sharply compared to the massless case, whereas the real frequency falls off at a much slower pace. For  electrically charged fields, the oscillation frequency and the damping rate is more for higher values of the field charge.

For a massive electrically charged field incident on an electrically charged black hole, there is a distinct possibility that the spacetime might be superradiantly unstable. The idea is that a charged/rotating black hole can, under certain specific conditions, superradiantly amplify an incident wave. If such an amplified wave is confined and reflected back into the black hole, its amplitude will increase iteratively, ultimately leading to an exponential amplification and hence to an instability. In Chapter~\ref{chap3}, we investigated the superradiant stability of the scalar-hairy Reissner-Nordstr\"{o}m black hole, mainly concentrating on the mutated Reissner-Nordstr\"{o}m regime. The motivation is to check whether the superradiant modes of the incident scalar wave is compatible with the bound-state condition. We verified that in a mutated Reissner-Nordstr\"{o}m black hole, the superradiance condition \eqref{ch3:eq_sprrdnc} and the bound-state condition{} are incompatible. The same is also true for $s>-e^2$ regime. Thus, much like a standard Reissner-Nordstr\"{o}m black hole, the scalar-hairy Reissner-Nordstr\"{o}m black hole is superradiantly stable against perturbation by massive charged scalar fields for all values of the scalar charge.


In Chapter~\ref{chap4}, we studied Hawking radiation of massive charged scalar particles from a scalar-hairy Reissner-Nordstr\"{o}m black hole. We used the tunnelling method following Parikh and Wilczek~\cite{parikh_PRL_2000} to calculate the Hawking emission rate. The tunnelling method depends on the conservation of energy and highlights the non-thermality of Hawking radiation. We observed that the emission rate of the scalar-hairy RN black hole differs considerably from that of a standard Reissner-Nordstr\"{o}m black hole. The total change in entropy of the scalar-hairy RN black hole due to the emission of the massive charged particle contains an energy-dependent contribution due to the scalar charge. We also evaluated the upper bound of the charge-mass ratio.

A strongly gravitating object such as a black hole distorts the spacetime around it. The spacetime around a black bole acts as a potential barrier, allowing only a part of the Hawking radiation to reach an asymptotic observer. The greybody factor measures the fraction of this transmission  probability. Thus, it encodes information regarding the spacetime geometry surrounding a black hole. In Chapter~\ref{chap5}, we studied the effect of the scalar hair on the grey body factor and also the sparsity of Hawking emission of massless uncharged scalar particles from the scalar-hairy RN black hole. We observed that unlike the electric charge, the scalar charge increases the greybody factor and reduces the sparsity of the Hawking emission cascade. As the black hole continues to Hawking radiate, its ADM mass decreases which in turn lowers the greybody factor and enhances the sparsity of Hawking radiation.

{In the present thesis, we concentrated solely on spherically symmetric black hole solution with scalar hair, however, it will be interesting to explore the effect of scalar hair in a rotating black hole solution.}



\begin{appendices} 

\chapter{Penrose diagrams for scalar-hairy Reissner-Nordstr\"{o}m} 
\label{AppendixA}
As mentioned in Sec.~\ref{ch1:sec_shRN}, for $s>-e^2$, the line element of the scalar-hairy Reissner-Nordström~\eqref{ch1:eq_shRN} black hole is similar to that of a Reissner-Nordstr\"{o}m black hole of effective electric charge $Q^2=e^2+s$ and mass $M$,
\begin{equation}\label{apA:eq_shRN}
ds^2=-f(r)dt^2+\frac{dr^2}{f(r)}+r^2\left(d\theta^2 +\sin^2\theta d\phi^2\right)~,
\end{equation}
with
\begin{equation}\label{apA:eq_f(r)}
f(r)=1-\frac{2M}{r}+\frac{Q^2}{r^2}=\frac{1}{r^2}\left(r-r_+\right)\left(r-r_-\right)~.
\end{equation}
where,
\begin{equation}\label{apA:eq_rpm}
r_\pm=M\pm \sqrt{M^2-Q^2}
\end{equation}
The Reissner-Nordström metric has a true curvature singularity at $r=0$ and coordinate singularities at $r=r_\pm$, depending on the relative values of $M^2$ and $Q^2$. The surface $r=r_\pm$ represents the position of the outer and the inner horizons respectively. 

\section*{Sub-extremal Reissner-Nordström solution}
\begin{figure}[!t]
\begin{center}
  \includegraphics[width=0.9\textwidth]{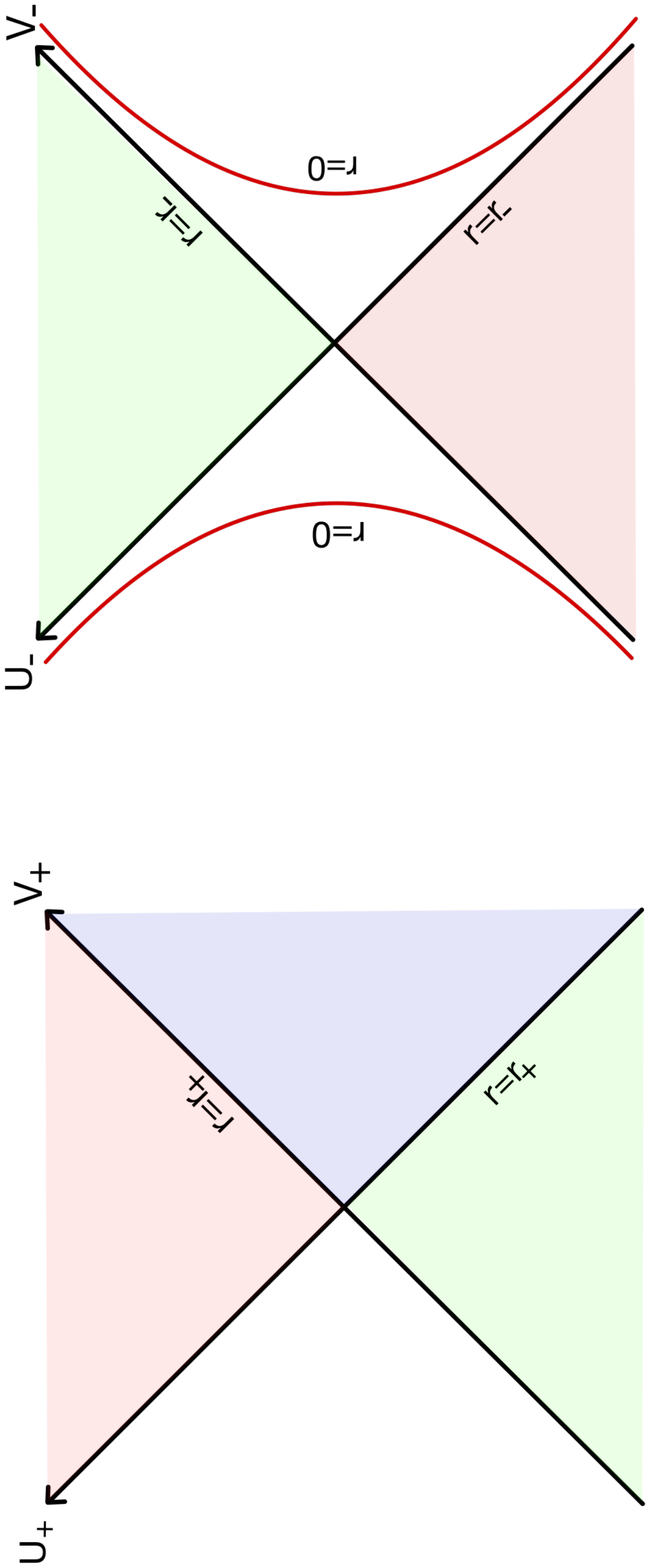}
\caption{$U_+, V_+$ and $U_- , V_-$ coordinate planes} 
\label{apA:fig_1}
\end{center}
\end{figure}

\begin{figure}[!htbp]
\begin{center}
  \includegraphics[width=0.8\textwidth, keepaspectratio]{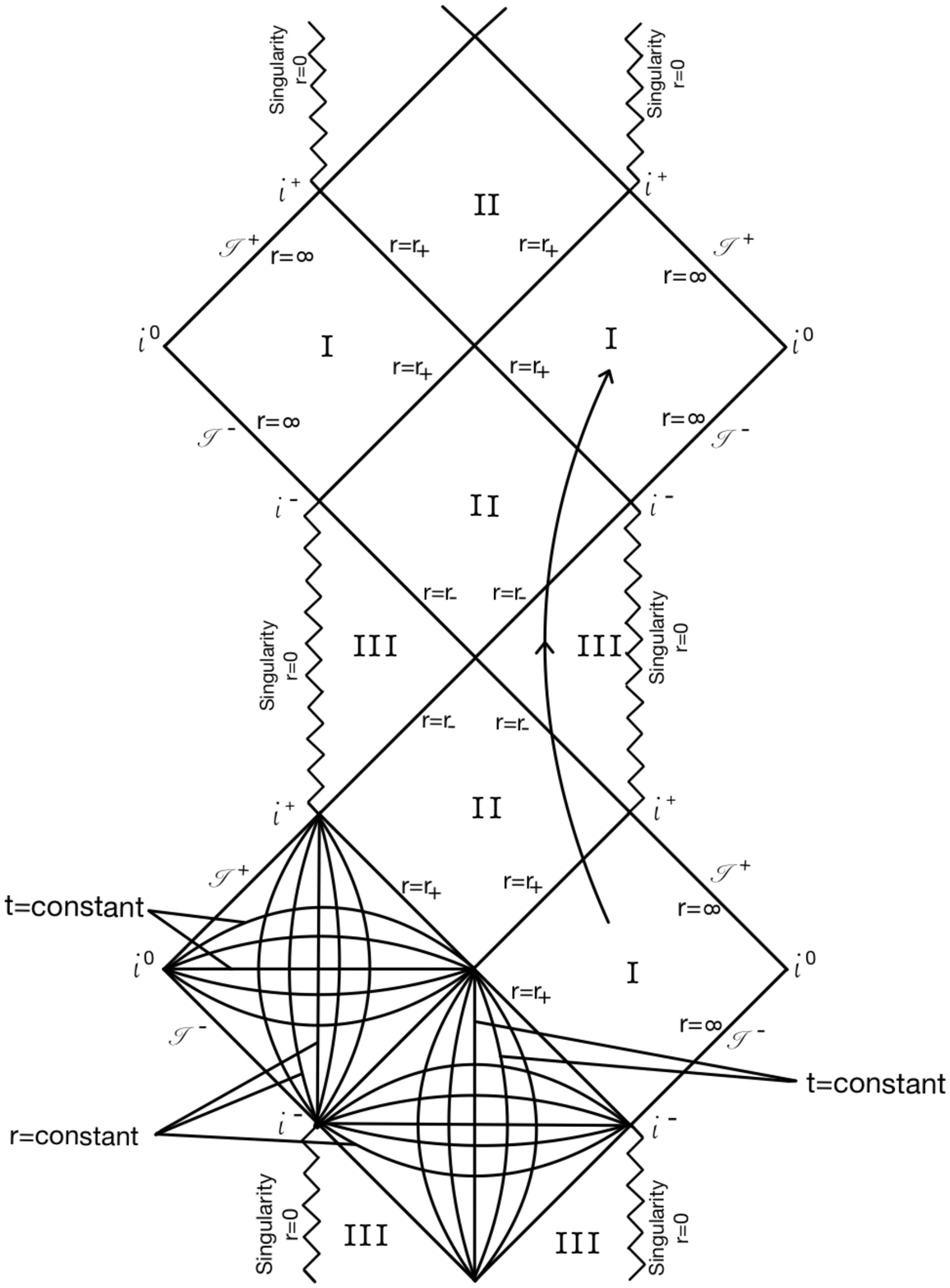}
\caption{Penrose diagram for a Reissner-Nordström black hole with $Q^2<M^2$} 
\label{apA:fig_2}
\end{center}
\end{figure}

For $Q^2<M^2$, $r_\pm$ are real, however, the singularities at both these locations are removable by appropriate choice coordinates. 
We start by defining the tortoise coordinate,
\begin{equation}
dr_*=\frac{dr}{f(r)}~,
\end{equation}
which gives
\begin{equation}
r_*=r+\frac{1}{2 k_+} \log \left| \frac{r-r_+}{r_+}\right|+\frac{1}{2 k_-} \log \left| \frac{r-r_-}{r_-}\right|
\end{equation}
with
\begin{equation}
k_\pm=\frac{r_\pm - r_\mp}{2r^2_\pm}~.
\end{equation}
 $k_\pm$ represents the surface gravity on the two horizons $\left(r=r_\pm\right)$.
 The tortoise coordinate $r_*$ maps the  region $r\in (r_+, \infty)$ to $r_* \in \left(-\infty, \infty\right)$. 
 Using the pair of null coordinates,
\begin{equation}
v=t+r_* \quad \mbox{and} \quad u=t-r_*~,
\end{equation}
one defines the Kruskal-like coordinates
\begin{equation}
U_\pm=-e^{k_\pm u} \quad \mbox{and} \quad V_\pm=\pm e^{k_\pm v}~.
\end{equation}
The coordinates $v$ and $u$ are known as the ingoing and outgoing Eddington-Finkelstein coordinates.\\
The coordinates $U_+$ and $V_+$ have the property,
\begin{equation}
U_+V_+=-e^{2 k_+ r_*} = -\frac{r-r_+}{r_+} \left(\frac{r_-}{r-r_-}\right)^{r^2_-/r^2_+}e^{2 k_+ r}.
\end{equation}
It may be noted that in the limit of vanishing electric charge, $Q\rightarrow 0$, the coordinates $U_+$ and $V_+$ coincide with the standard Kruskal coordinates of the Schwarzschild black hole.\footnote{In case of the scalar-hairy Reissner-Nordström black hole  $Q\rightarrow 0$ implies $s\rightarrow -e^2$ which causes the scalar field to blow up and hence this limit is unattainable.} In these coordinates the Reissner-Nordström metric ~\eqref{apA:eq_shRN} takes the form,
\begin{equation}
\begin{split}
ds^2&=-f(r) du dv +r^2 d\Omega^2 \\
&= -\frac{r_+ r_-}{k^2_+ r^2}\left(\frac{r-r_-}{r_-}\right)^{1-r^2_-/r^2_+} e^{-2k_+ r} dU_+dV_+ +r^2 d\Omega^2
\end{split}
\end{equation}
where $d\Omega^2= d\theta^2 +\sin^2\theta d\phi^2$ and $ r=r\left(U_+,V_+\right)$. The region outside the Reissner-Nordström black hole, $r>r_+$, corresponds to $U_+<0$ and $V_+>0$. However, extending the Kruskal-like  coordinates to $U_+,V_+\in \mathbf{R}$, we note that the coordinates $U_+$ and $V_+$ extend only down to the inner horizon $r_-$ and not down to $r=0$. In fact, $U_+V_+\rightarrow\infty$ as $r\rightarrow r_-$. The outer horizon at $r=r_+$ corresponds to both the null surfaces $U_+=0$ and $V_+=0$, representing the white hole and black hole horizons respectively. To study the spacetime in the region $r\leq r_-$, we use the other set of Kruskal-like coordinates, $\left(U_-,V_-\right)$ such that
\begin{equation}
U_-V_-=e^{2 k_- r_*} = \frac{r-r_-}{r_-} \left(\frac{r_+}{r_+-r}\right)^{r^2_+/r^2_-}e^{2 k_- r}.
\end{equation}
As $r\rightarrow r_+$,  $U_-V_-\rightarrow \infty$. Thus, the coordinates $\left(U_-,V_-\right)$ covers only the interior of the black hole and not the outside. The inner horizon at $r=r_-$ corresponds to the two null hypersurfaces $U_-=0$ and $V_-=0$. The region, $r_-<r<r_+$~, is covered by both sets of coordinates with $U_+,V_+>0$ and $U_-,V_-<0$. The different region of the extended Reissner-Nordström black hole is depicted in Fig.~\ref{apA:fig_1}. 
The singularity at $r=0$ is represented by $U_-V_-=-1$. It is interesting to note that unlike the Schwarzschild case where the singularity is spacelike, the singularity in the Reissner-Nordström case is timelike and is in fact avoidable. A fortunate observer (uncharged particle) entering the event horizon into the region, $r_-<r<r_+$, i.e.,  $U_+,V_+>0$ or $U_-,V_-<0$ and following future directed time like trajectory may completely miss the singularity and land up in the region  $U_-,V_->0$ in finite proper time. The region $U_-,V_->0$ is indeed an unexpected part of the spacetime and is isomorphic to the region $U_+,V_+<0$. This suggests that the spacetime may be further extended and the entire process can be repeated infinitely. The Kruskal diagrams can be patched together to yield the Penrose diagram for the Reissner-Nordstrom black hole. 
The idea behind the Penrose diagram is to draw the entire spacetime is a finite  piece of paper. For that, one uses conformal transformations of the form
\begin{equation}
U_\pm =\tan \tilde{U}_\pm \quad \mbox{and} \quad V_\pm =\tan \tilde{V}_\pm~.
\end{equation}
Figure~\ref{apA:fig_2} shows the Penrose diagram for the maximally extended Reissner-Nordström black hole. Each point in the Penrose diagram represents a two-sphere since angular coordinates $\theta$ and $\phi$ has been suppressed.
To appreciate the elegance of the Penrose conformal diagrams, it is imperative to under stand the conformal infinities presented in the diagram:
\begin{itemize}
\item[]\hspace{0.3\linewidth}$i^+$ = future timelike infinity
\item[]\hspace{0.3\linewidth}$i^0$ = spacelike infinity
\item[]\hspace{0.3\linewidth}$i^-$ = past timelike infinity
\item[]\hspace{0.3\linewidth}$\mathscr{I}^+$ = future null infinity
\item[]\hspace{0.3\linewidth}$\mathscr{I}^-$ = past null infinity~.
\end{itemize}
The complete maximally extended Reissner-Nordström spacetime consists of an infinite number of asymptotically flat regions (I), connected by region (II) $r_-<r<r_+$  and region (III) $0<r<r_-$. For every exterior region (I) $r>r_+$, there exists another identical region which cannot be connected by timelike curves. The outer horizon at $r=r_+$ is the event horizon, since it forms the boundary of the causal past of future null infinity $\mathscr{I}^+$. The null surface at $r=r_-$ is the cauchy horizon. Starting from the initial data given on some spacelike hypersurface, the equations of motion can uniquely determine the behaviour of a field only upto the surface $r=r_-$. Beyond that, one has to specify the behaviour of the field at the singularity for unique time evolution.\\
As already noted, the singularity at $r=0$ is timelike. Hence, an uncharged particle from region I crossing the event horizon into region II following future directed timelike trajectory may pass through region III without hitting the singularity and re-emerge into another asymptotically flat region. 
\section*{Extremal Reissner-Nordström solution}
\begin{figure}[!t]
\begin{center}
  \includegraphics[width=0.55\textwidth, keepaspectratio]{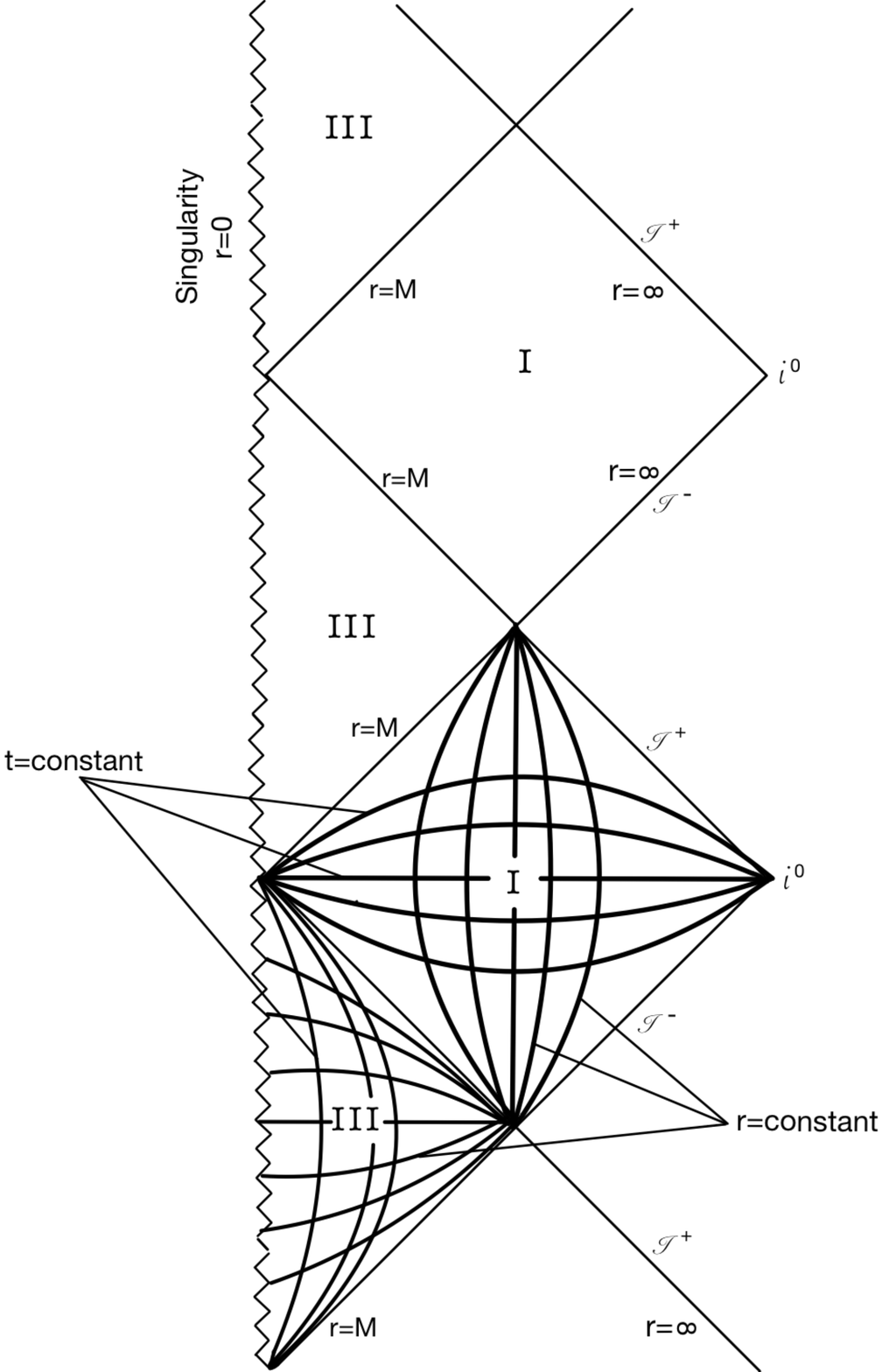}
\caption{Penrose diagram for an extremal Reissner-Nordström black hole} 
\label{apA:fig_3}
\end{center}
\end{figure}

Finally, for $Q^2=M^2$, the outer event horizon and the inner cauchy horizon merges such that
\begin{equation}
r_\pm=2M.
\end{equation}
	The spacetime in this case is referred to as the extremal Reissner-Nordström solution. The Penrose diagram for the extremal Reissner-Nordström solution is depicted in Fig.~\ref{apA:fig_3}. The singularity in this case is timelike as well and thus can be avoided and just as before, the spacetime can be further extended. The extremal Reissner-Nordström solution is an extremely interesting and curious solution of the Einstein-Maxwell field equations and play an important role in examining the role of black holes in quantum gravity. For detailed description of the extremal Reissner-Nordström black hole we refer to~\cite{griffiths}.
\section*{Super-extremal Reissner-Nordström solution}
\begin{figure}[!bbpt]
\begin{center}
  \includegraphics[width=0.3\textwidth, keepaspectratio]{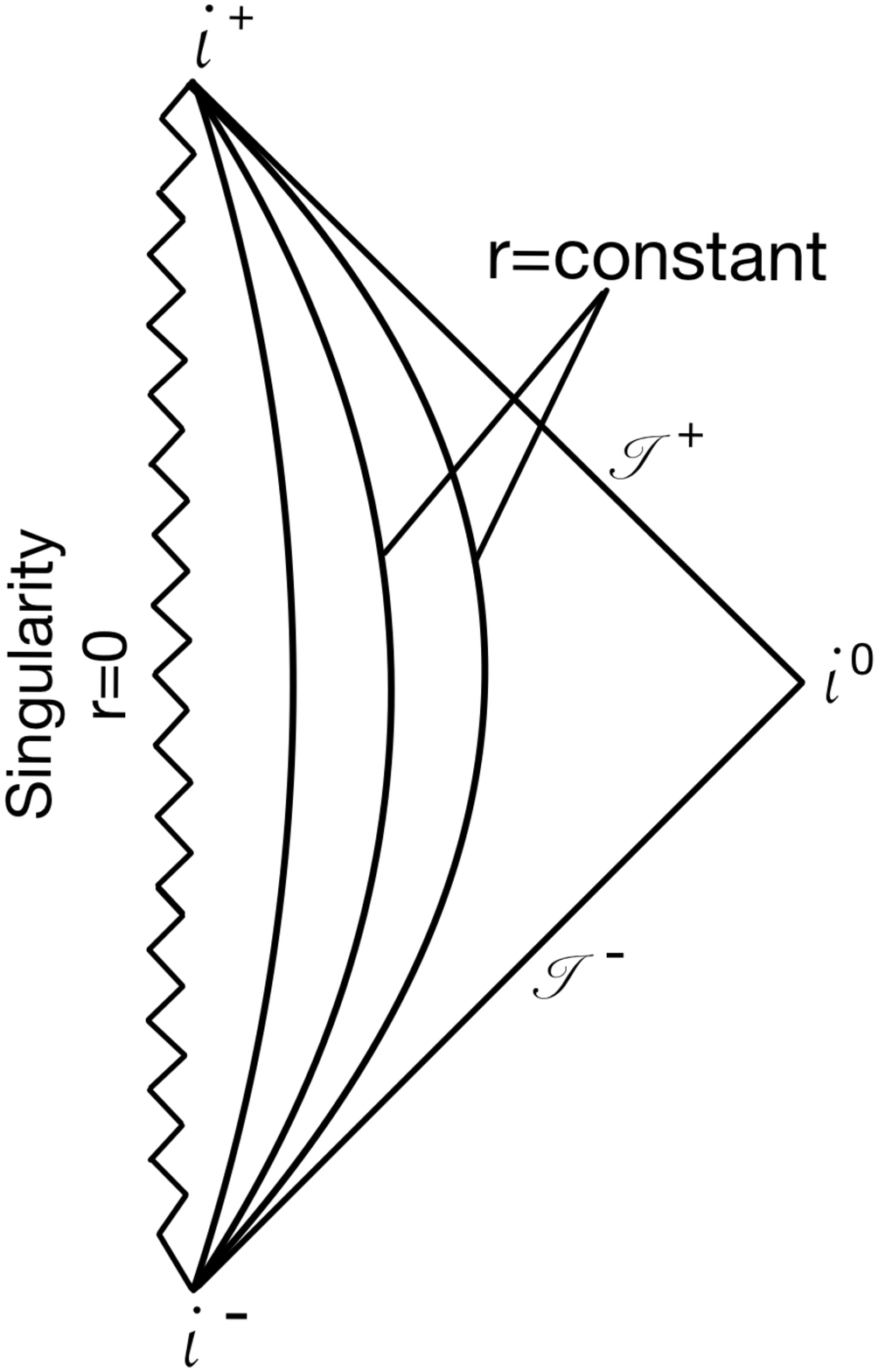}
\caption{Penrose diagram for a super-extremal Reissner-Nordström solution with $Q^2>M^2$} 
\label{apA:fig_4}
\end{center}
\end{figure}

For $Q^2>M^2$, $f(r)$ has no real roots, thus, the singularity is not cloaked by an event horizon and is globally naked. The metric is regular down to the singularity at $r=0$, which in this case, is a timelike line. Such naked singularities violates the cosmic censorship conjecture and are less likely to be the outcome of a gravitational collapse.	
The Penrose diagram for the superextreme Reissner-Nordström spacetime is shown in Fig.~\ref{apA:fig_4}.

For $s<-e^2$ $\left(Q^2<0\right)$, the scalar hairy Reissner-Nordström black hole leads to an Einstein-Rosen bridge~\cite{rosen_PR_1935}.

\chapter{Bounds on transmission and reflection probability} 
\label{AppendixB}
This appendix provides a brief description of the general methodology to derive bounds on the transmission and reflection probability as proposed by Visser in \cite{visser_PRA_1999}.
\section{Shabat-Zakharov systems}
In general, the one dimensional Schr\"{o}dinger equation with asymptotically constant potential,
\begin{equation}\label{apB:eq_KG_a}
-\frac{\hbar^2}{2m}\frac{d^2}{dx^2}\psi(x)+V(x)\psi(x)=E \psi(x) ;
\end{equation}
\begin{equation}
V\left(x\rightarrow \pm \infty \right)\rightarrow V_{\pm \infty}
\end{equation}
admits an exact set of solutions known as the Jost solutions ${J}_\pm \left(x \right)$~\cite{jost} that satisfy
\begin{eqnarray}
{J}_+ \left(x\rightarrow +\infty\right) \rightarrow \frac{\exp\left(+ i k_{+ \infty} x\right)}{\sqrt{k_{+\infty}}}~,\\
{J}_- \left(x\rightarrow -\infty\right) \rightarrow \frac{\exp\left(- i k_{- \infty} x\right)}{\sqrt{k_{-\infty}}}~,
\end{eqnarray}
and 
\begin{eqnarray}
{J}_+ \left(x\rightarrow -\infty\right) \rightarrow& \alpha \frac{\exp\left(+ i k_{- \infty} x\right)}{\sqrt{k_{-\infty}}} + \beta \frac{\exp\left(- i k_{- \infty} x\right)}{\sqrt{k_{-\infty}}}~, \\
{J}_- \left(x\rightarrow +\infty\right) \rightarrow& \alpha^* \frac{\exp\left(- i k_{+ \infty} x\right)}{\sqrt{k_{+\infty}}} + \beta^* \frac{\exp\left(+ i k_{+ \infty} x\right)}{\sqrt{k_{+\infty}}}~,
\end{eqnarray}
where $k=\frac{\sqrt{2m(E-V(x))}}{\hbar}$ and $k_{\pm\infty}=\frac{\sqrt{2m(E-V_{\pm\infty})}}{\hbar}$~.  $e^{i k x}$ corresponds to the modes moving in the positive $x$ direction (right moving), whereas, $e^{-i k x}$ corresponds to the modes moving in the negative $x$ direction (left moving). $\alpha$ and $\beta$ are the (right moving) Bogoliubov coefficients.
Considering an incoming flux of particles, moving in the positive $x$ direction from the left being partially reflected with amplitude $\tilde{r}$ and partially transmitted with amplitude 
$\tilde{t}$,
\begin{equation}
\tilde{r}=\frac{\beta}{\alpha}~, \quad \tilde{t}=\frac{1}{\alpha}.
\end{equation}
$\alpha^*$ and $\beta^*$ are the left moving  Bogoliubov coefficients. They are the complex conjugates of the corresponding right moving coefficients $\alpha$ and $\beta$. The idea is to derive bounds on the $|\alpha|$ and $|\beta|$ which will in-turn provide general bounds on the reflection probability and the transmission probability,
\begin{equation}
\tilde{R}=|\tilde{r}|^2;\qquad \tilde{T}=|\tilde{t}|^2.
\end{equation}
To derive the bounds, one rewrites the second order Schr\"{o}dinger equation as two coupled first order differential equations (Shabat-Zakharov system~\cite{shabat}). One starts by defining an auxiliary function $\phi\left(x\right)$, so that the radial function $\psi$ becomes
\begin{equation}\label{apB:eq_new_repr}
\psi\left(x\right)=a\left(x\right)\frac{e^{+i\phi}}{\sqrt{\phi'}}+b\left(x\right)\frac{e^{-i\phi}}{\sqrt{\phi'}}.
\end{equation}
$\phi\left(x\right)$ can be either real or imaginary but $\phi'\left(x\right)\neq 0$. The representation \eqref{apB:eq_new_repr} introduces additional degrees of freedom. One effectively replaces the complex function $\psi\left(x\right)$ with two complex coefficients $a \left(x\right)$, $b\left(x\right)$ and an auxiliary function $\phi\left(x\right)$. Thus, without any loss of generality, one can choose the coefficients $a\left(x\right)$ and $b\left(x\right)$ such that they reach a constant value at spatial infinity, which in-turn implies choosing $\phi'(x\rightarrow \pm \infty)\rightarrow k_{\pm \infty}$. To reduce the extra degrees of freedom, one imposes the gauge condition,
\begin{equation}\label{apB:eq_gauge}
\frac{d}{dx}\left(\frac{a}{\sqrt{\phi'}}\right)e^{+i\phi}+\frac{d}{dx}\left(\frac{b}{\sqrt{\phi'}}\right)e^{-i\phi}=0.
\end{equation}
Subject to the above gauge condition, one obtains,
\begin{equation}
\frac{d\psi}{dx}=i \sqrt{\phi'}\left[a\left(x\right) e^{+i\phi}-b\left(x\right)e^{-i \phi}\right],
\end{equation}
and 
\begin{align}
\label{apB:eq_d2}
\frac{d^2\psi}{dx^2}&= \frac{d}{dx}\left(i \sqrt{\phi'}\left[a e^{+i\phi}-be^{-i \phi}\right]\right)\\
&= \begin{multlined}[t]
\frac{\left( i\phi'\right)^2}{\sqrt{\phi'}}\left(ae^{+i\phi}+be^{-i\phi}\right)+\frac{i\phi''}{\sqrt{\phi}}\left(ae^{+i\phi'}-be^{-i\phi}\right)\\
+i \phi'\left[ \frac{d}{dx}\left(\frac{a}{\sqrt{\phi'}}\right)e^{+i\phi}-\frac{d}{dx}\left(\frac{b}{\sqrt{\phi'}}\right)e^{-i\phi} \right]~.
\end{multlined}
\end{align}

Using the gauge condition~\eqref{apB:eq_gauge} in Eq.~\eqref{apB:eq_d2}, one gets,
\begin{equation}
\frac{d^2\psi}{dx^2}=-\frac{\phi'^2}{\sqrt{\phi'}}\left( a e^{+i\phi}+be^{-i\phi} \right) + \frac{2 i \phi'}{\sqrt{\phi'}}\frac{da}{dx}e^{+i\phi} -i\frac{\phi''}{\sqrt{\phi'}}be^{-i\phi} ~,
\label{apB:eq_d21}
\end{equation}
\begin{equation}
\label{apB:eq_d22}
\mbox{and }\frac{d^2\psi}{dx^2}=-\frac{\phi'^2}{\sqrt{\phi'}}\left( a e^{+i\phi}+be^{-i\phi} \right) - \frac{2 i \phi'}{\sqrt{\phi'}}\frac{db}{dx}e^{-i\phi} +i\frac{\phi''}{\sqrt{\phi'}}ae^{+i\phi} ~.
\end{equation}
Substituting Eq.~\eqref{apB:eq_d21} and Eq.~\eqref{apB:eq_d22} back in Eq.~\eqref{apB:eq_KG_a}, one arrives at the Shabat-Zakharov system of coupled first order differential equations,
\begin{eqnarray}
\label{apB:eq_sch_a}
\frac{da}{dx}=\frac{1}{2\phi'}\left\lbrace \phi'' b e^{-2i\phi}+i\left[k^2(x)-{\phi'}^2\right]\left[a+b e^{-2i\phi}\right]\right\rbrace~,\\
\frac{db}{dx}=\frac{1}{2\phi'}\left\lbrace \phi'' a e^{+2i\phi}-i\left[k^2(x)-{\phi'}^2\right]\left[b+a e^{+2i\phi}\right]\right\rbrace~.
\end{eqnarray}

\section{Derivation of bounds}
To put bounds on the Bogoliubov coefficients we assume $\phi\left(x\right)$ to be real. The probability currentis given by,
\begin{equation}
\mathcal{J}=Im\left\lbrace \psi^*\frac{d \psi}{d x} \right\rbrace=\left\lbrace |a|^2-|b|^2 \right\rbrace.
\end{equation}
Since we are  using the one-dimensional Jost solution~\cite{jost}, as $x\rightarrow +\infty$, the wave function is purely right moving and normalised to unity. Thus, we get,
\begin{equation}\label{apB:eq_prob_cons}
|a|^2-|b|^2=1 \qquad \forall x~.
\end{equation}
 Equation~\eqref{apB:eq_prob_cons} is useful in interpreting $a\left(x\right)$ and $b\left(x\right)$ as positio-dependent Bogoliubov coefficients.
 Using the fact that,
 \begin{equation}
\frac{d|a|}{dx}=\frac{1}{2|a|}\left(a^*\frac{da}{dx}+a\frac{da^*}{dx}\right)
\end{equation}
and Eq.~\eqref{apB:eq_sch_a} we get,
\begin{equation}
\frac{d|a|}{dx}=\frac{1}{2|a|}\frac{1}{2\phi'} Re\left(\left\lbrace\phi''+i\left[k^2\left(x\right)-\left(\phi'\right)^2\right]\right\rbrace \left[a^*be^{-2i\phi}\right]\right)
\end{equation}
 Now, for any two complex numbers $A$ and $B$, we know, $Re(AB)\leq|A||B|$, thus,
 \begin{equation}
\frac{d|a|}{dx}\leq\frac{\sqrt{\left(\phi''\right)^2+\left[k^2\left(x\right)-\left(\phi'\right)^2\right]^2}}{2|\phi'|}|b|.
\end{equation}
Defining, 
\begin{equation}
\vartheta=\frac{\sqrt{\left(\phi''\right)^2+\left[k^2\left(x\right)-\left(\phi'\right)^2\right]^2}}{2|\phi'|}
\end{equation}
and using the conservation law~\eqref{apB:eq_prob_cons}, we obtain,
\begin{equation}\label{apB:eq_diff_limit}
\frac{d|a|}{dx}\leq\vartheta\sqrt{|a|^2-1}.
\end{equation}
Integrating the above equation, we get,
\begin{equation}
\left\lbrace\cosh^{-1}|a|\right\rbrace_{x_i}^{x_f}\leq\int_{x_i}^{x_f}\vartheta \,dx
\end{equation}
Choosing the limit of integration as ${x_i}\rightarrow-\infty$ and ${x_f}\rightarrow+\infty$, $a(x)$ and  $b(x)$ tend to the Bogoliubov coefficients $\alpha$ and $\beta$ with
\begin{equation}
\label{apB:eq_alpha}
|\alpha|\leq\cosh\left(\int_{-\infty}^{\infty}\vartheta \,dx\right), 
\end{equation}
and hence,
\begin{equation}
\label{apB:eq_beta}
|\beta|\leq\sinh\left(\int_{-\infty}^{\infty}\vartheta \,dx\right) .
\end{equation}
Equations~\eqref{apB:eq_alpha} and \eqref{apB:eq_beta} yields the bounds on the transmission and reflection probabilities as,
\begin{equation}
T\geq \sech^2\left(\int_{-\infty}^{+\infty}\vartheta dx\right)
\end{equation}
and
\begin{equation}
R\leq \tanh^2\left(\int_{-\infty}^{+\infty}\vartheta dx\right).
\end{equation}

%

\end{appendices}

\bibliographystyle{mybst}
\addcontentsline{toc}{chapter}{Bibliography}
\bibliography{thesis}


\end{document}